\pdfoutput=1
\documentclass[preprint,aps,showpacs,showkeys]{revtex4}
\usepackage{supertabular}
\usepackage{amsfonts}
\usepackage[pdftex]{graphicx}
\usepackage{amsmath}
\usepackage{amsfonts}
\usepackage{amssymb} 
\usepackage{subfig}
\usepackage{amsxtra} 
\DeclareRobustCommand{\spm}{\mathrel{\text{\framebox[0.9\width]{$\pm$}}}}
\DeclareRobustCommand{\smp}{\mathrel{\text{\framebox[0.9\width]{$\mp$}}}}

\DeclareMathOperator{\dia}{dia}
\newcommand\splus{\boxplus}
\newcommand\sminus{\boxminus}

\newcommand{\be}{\begin{equation}}
\newcommand{\ee}{\end{equation}}
\newcommand{\bn}{\begin{eqnarray}}
\newcommand{\en}{\end{eqnarray}}

\begin{document}
\thispagestyle{empty} 
\title{The symmetry of $4 \times 4$ mass matrices predicted by the {\it spin-charge-family} theory
--- $SU(2) \times SU(2) \times U(1)$ ---  remains in all  loop corrections~%
\footnote{This is the part of the talk presented by N.S. Manko\v c Bor\v stnik at the  $21^{st}$
Workshop "What Comes Beyond the Standard Models", Bled, 23 June - 1 of July, 2018, and 
published in the Proceedings to this workshop.}}

%
\author{A. Hernandez-Galeana${}^2$ and N.S. Manko\v c Bor\v stnik${}^1$\\
${}^1$University of Ljubljana,  \\
Jadranska 19, SI-1000 Ljubljana, Slovenia\\
${}^2$Departamento de F\'{\i}sica,   ESFM - Instituto Polit\'ecnico Nacional. \\
              U. P. "Adolfo L\'opez Mateos". C. P. 07738, Ciudad de M\'exico, M\'exico
}

%

\begin{abstract}
The  {\it spin-charge-family} theory~\cite{IARD,ND2017,NH2017,JMP2015,%
norma2014MatterAntimatter,JMP,NBled2013,NBled2012,norma92,norma93,norma94,%
pikanorma,portoroz03,norma95,gmdn07,gn,gn2013,gn2015,NPLB,N2014scalarprop} predicts
the existence of the fourth family to the observed three. The $4 \times 4$ mass matrices ---
determined by the nonzero vacuum expectation values of the two triplet scalars, the gauge fields
of the two groups of $\widetilde{SU}(2)$ determining family quantum numbers, and by the 
contributions of the dynamical fields of the two scalar triplets and the three scalar singlets  
with the family members quantum numbers ($\tau^{\alpha}=(Q, Q',Y')$) --- manifest the symmetry 
$\widetilde{SU}(2) \times \widetilde{SU}(2) \times U(1)$. 
All scalars carry the weak and the hyper charge of the {\it standard model} higgs field
 ($\pm \frac{1}{2},\mp  \frac{1}{2}$,  respectively).
It is demonstrated, using the massless spinor basis,  
that the symmetry of the $4\times4$ mass matrices  remains $SU(2) \times SU(2) \times U(1)$ 
in all  loop corrections, and it is discussed under which conditions this symmetry is kept under all  
corrections, that is with the corrections induced  by the repetition of the nonzero vacuum 
expectation values included.
\end{abstract}


\maketitle

 \section{Introduction}
 \label{introduction}

The {\it spin-charge-family} theory~\cite{IARD,ND2017,NH2017,JMP2015,%
norma2014MatterAntimatter,JMP,NBled2013,Bled2018NH,NBled2012,norma92,norma93,norma94,%
pikanorma,portoroz03,norma95,gmdn07,gn,gn2013,gn2015,NPLB,N2014scalarprop} predicts
before the electroweak break four - rather
than the observed three --- coupled massless families of quarks and leptons.

The $4 \times 4$ mass matrices of all the family members demonstrate in this theory the same
symmetry~\cite{IARD,norma2014MatterAntimatter,JMP2015,gn2013,gn2015}, determined by
the scalar fields originating in $d > (3+1)$: the two triplets --- the gauge fields of the two 
$\widetilde{SU}(2)$ family groups with the generators $\vec{\tilde{N}}_{L}$, 
$\vec{\tilde{\tau}}^{1}$, operating among families --- and the three singlets --- the gauge 
fields of the three charges ($\tau^{\alpha}=(Q, Q',Y')$)) ---
distinguishing among family members. All these scalar fields carry the weak and the hyper
charge as does the scalar higgs of the {\it standard model}: ($\pm \frac{1}{2}$ and
$\mp \frac{1}{2}$, respectively)~\cite{IARD,JMP2015,N2014scalarprop}. 
The loop corrections in all orders make each matrix element of mass matrices dependent 
on the quantum numbers of each of the family members.

Since there is no direct observations of the fourth family quarks with masses below $1$
TeV, while the fourth family quarks with masses above $1$ TeV would contribute according to the
{\it standard model} (the {\it standard model} Yukawa couplings of the quarks with the scalar
higgs is proportional to  $  \frac{m^{\alpha}_{4}}{v}$, where $m^{\alpha}_{4}$ is the
fourth family member ($\alpha=u,d$) mass and $v$ the vacuum expectation value of the
scalar higgs) to either the quark-gluon fusion production of the scalar field (the higgs) or
to the scalar field decay  
 too much in comparison with the observations, the high energy physicists do not expect the
existence of the fourth family members at all~\cite{AhmedAli,MatthiasNeubert}.

One of the authors (N.S.M.B) discusses in Refs.~(\cite{IARD
}, Sect.~4.2.) that  the {\it standard model} estimation with one higgs scalar might not
be the right way to evaluate whether the fourth family, coupled to the observed three,
does exist or not.
%
%
The $u_i$-quarks and $d_i$-quarks of an $i^{th}$ family, namely, if they couple with the opposite
 sign 
to the scalar fields carrying the family ($\tilde{A}, i$) quantum numbers 
and  have the same masses, do not contribute to either the quark-gluon fusion production
of the scalar fields with the family quantum numbers or to the decay of these scalars into two
photons.
The strong influence of the scalar fields carrying the family members quantum numbers to the
masses of the lower (observed) three families manifests in the huge differences in the  masses
of the family members, let say  $u_i$ and $d_i$, $i=(1,2,3)$, and families ($i$). For the fourth
family quarks, which are more and more decoupled from the observed three families the higher
are their masses~\cite{gn2015,gn2013}, the influence of the scalar fields carrying the family
members quantum numbers on their masses is in the {\it spin-charge-family} theory expected 
to be much weaker.
Correspondingly the $u_4$ and $d_4$ masses become closer to each other the higher are their
masses and the weaker are their couplings (the mixing matrix elements) to the lower three families.
For $u_4$-quarks and $d_4$-quarks with the similar masses the observations might consequently
not be in contradiction with the {\it spin-charge-family} theory prediction that there exists the
fourth family coupled to the observed three ~(\cite{NH2017notsent}, which is in preparation).
%


{\it We demonstrate in the main Sect.~\ref{main} that the symmetry $\widetilde{SU}(2) \times 
\widetilde{SU}(2) \times U(1)$, which the mass matrices demonstrate on the tree level, after the 
gauge scalar fields of the two $\widetilde{SU}(2)$ family groups triplets gain nonzero vacuum
expectation values, keeps the same in all loop corrections. 
We use the massless basis. We discuss also the conditions under which the mass matrix keeps
the symmetry of the tree level, as well as the break of the tree level symmetry.}


 In Sect.~\ref{SCFT} we present shortly the {\it spin-charge-family} theory  and its 
achievements so far. All the mathematical support appears in appendices.

Let  be in this introduction  stressed what supports the {\it spin-charge-family} theory to be
the right next step beyond the {\it standard model}. This  theory can not only explain --- while
starting from a very simple action in $d \ge (13+1)$, Eqs.~(\ref{wholeaction}
) in App.~\ref{SCFT}, with massless fermions (with the spin of the two kinds,
$\gamma^a$ and $\tilde{\gamma}^a$,  one kind taking care of the spin and of all
the charges of the family members (Eq.~(\ref{tau})), the second kind taking care of 
families  (Eqs.~(\ref{gammatildegamma}, \ref{sabtildesab})))  coupled only to the gravity
(through the vielbeins and the two kinds of the spin connections fields
$\omega_{ab\alpha} f^{\alpha}{}_{c}$ and $\tilde{\omega}_{ab\alpha} f^{\alpha}{}_{c}$,
the gauge fields of $S^{ab}$ and $\tilde{S}^{ab}$ (Eqs.~(\ref{wholeaction})), respectively --- 
all the assumptions of the {\it standard model}, but also answers several open questions
beyond the {\it standard model}. It offers the explanation for~\cite{JMP2015,%
norma2014MatterAntimatter,JMP,IARD,NBled2013,NBled2012,norma92,norma93,norma94,%
pikanorma,portoroz03,norma95,gmdn07,gn,gn2013,gn2015,NPLB,N2014scalarprop}:\\
\noindent
{\bf a.}~The appearance of all the charges of the left and right handed family members and for
 their families and their properties.  \\
{\bf b.}~The appearance of all the corresponding vector and scalar gauge fields and
 their properties (explaining the appearance of higgs and the Yukawa couplings). \\
{\bf c.}~The appearance and properties of the dark matter. \\
{\bf d.}~The appearance of the matter/antimatter asymmetry in the universe.\\
%

This theory predicts for the low energy regime:\\
{\bf i.}~$\;$The existence of the fourth family to the observed three.\\
{\bf ii.}~The existence of twice two triplets and three singlets of scalars, all with the
properties of the higgs with respect to the weak and hyper charges, what explains the origin of
the Yukawa couplings.\\
{\bf iii.}~There are several other predictions, not directly connected with the topic of this
paper.\\

The fact that the fourth family quarks have not yet been observed --- directly or indirectly ---
pushes the fourth family quarks masses to values higher than $1$ TeV.

Since the experimental accuracy of the $3\times 3$ submatrix of the $4 \times 4$ mixing
matrices is not yet high enough~\cite{data2016}, it is not yet possible to calculate the mixing 
matrix elements among the fourth family and the observed three~%
\footnote{The $3\times 3$ submatrix, if accurate, determines the $4\times 4$ unitary matrix 
uniquely.}. 
Correspondingly it is not possible yet to estimate masses of the fourth family members by 
fitting the experimental data to the free parameters of mass matrices, the number of which is 
limited by the symmetry $\widetilde{SU}(2) \times \widetilde{SU}(2) \times U(1)$, predicted 
by the {\it spin-charge-family}~\cite{gn2015,gn2013}.

If we assume the masses of the fourth family members, the matrix elements can be estimated
from the measured $3\times 3$ submatrix elements of the $4 \times 4$ 
matrix~\cite{gn2015,gn2013}~%
\footnote{While the fitting procedure is not influenced considerably by the accuracy of the 
measured masses of the lower three families, the accuracy of the measured values of the 
mixing matrices do influence, as expected, the fitting results very much.}.

The more effort and work is put into the {\it spin-charge-family} theory, the more explanations of
the observed phenomena and the more predictions for the future observations follow out of it.
Offering the explanation for so many observed phenomena --- keeping in mind that all the
explanations for the observed phenomena originate in a simple starting action --- qualifies
the {\it spin-charge-family} theory as the candidate for the next step beyond the {\it standard
model}.





The reader is kindly asked to learn more about the {\it spin-charge-family} theory in Refs.~\cite{%
ND2017,NH2017,JMP2015,IARD, norma2014MatterAntimatter,JMP} and the references
 therein. We shall point out sections in these references, which might be of particular help, when
 needed.

%
\section{The symmetry of the family members mass matrices}
\label{main}

The mass term $ \sum_{s=7,8}\;  \bar{\psi} \gamma^{s} p_{0s} \; \psi$, Eq.~(\ref{faction}),  
of the starting action, Eq.~(\ref{wholeaction}), manifests in the {\it spin-charge-family} theory~%
\cite{JMP2015,IARD,norma2014MatterAntimatter,JMP} the $\widetilde{SU}(2) \times $ 
$\widetilde{SU}(2)$ $\times U(1)$ symmetry. The infinitesimal generators of the two family groups
namely commute among themselves, $\{\vec{\tilde{N}}_{L}$, 
$\vec{\tilde{\tau}}^{\tilde{1}}\}_{-}=0$, Eq.~(\ref{tildetaucom}), and with all the infinitesimal 
generators of the family members groups, $\{\tilde{\tau}^{Ai}$, $\tau^{\alpha}\}_{-}=0$,
($\tau^{\alpha}=(Q, Q',Y')$), 
Eq.~(\ref{tautildetaucom}). After the scalar gauge fields, carrying the space index 
$(7,8)$, of the generators $\vec{\tilde{N}}_{L}$ and $\vec{\tilde{\tau}}^{\tilde{1}}$ of the two 
$\widetilde{SU}(2)$ groups gain nonzero vacuum expectation values, spinors (quarks and leptons), 
which interact with these scalar gauge fields, become massive. There are the scalar gauge fields,
carrying the space index $(7,8)$, of the group $U(1)$ with the infinitesimal generators 
$\tau^{\alpha}$ (=$(Q,Q',Y')$), which are responsible for the differences in mass matrices 
among the family members ($u^{i}, \nu^{i}, d^{i}, e^{i}, i (1,2,3,4)$, $i$ determines four 
families). Their couplings to the family members  depends strongly on the quantum numbers 
$(Q, Q', Y')$.

It is shown in this main section that the mass matrix elements of any family member keep the 
$\widetilde{SU}(2) \times \widetilde{SU}(2)$ $ \times U(1)$ symmetry of the tree
level in all  corrections (the loops one and the repetition of the nonzero vacuum expectation values),
provided that the scalar gauge fields of the $U(1)$ group have no 
nonzero vacuum expectation values. In the case that the scalar gauge fields of the $U(1)$ 
group have nonzero vacuum expectation values, the symmetry is changed, unless some 
of the scalar fields with the family quantum numbers have nonzero vacuum expectation values.
We comment on all these cases in what follows.

Let us first present the symmetry of the mass term in the starting action, 
Eq.~(\ref{wholeaction}). 

We point out that the symmetry $\widetilde{SU}(2) \times$ $ \widetilde{SU}(2)$ belongs to the two 
$\widetilde{SO}(4)$ groups --- to $\widetilde{SO}(4)_{\widetilde{SO}(3,1)}$ and to 
$\widetilde{SO}(4)_{\widetilde{SO}(4)}$.  The infinitesimal operators of the first and the second
$\widetilde{SO}(4)$ groups are, Eqs.~(\ref{so1+3tilde}, \ref{so42tilde}), 
\begin{eqnarray}
\label{tildeso3142}
\vec{\tilde{N}}_{+}(= \vec{\tilde{N}}_{L}): &=& \,\frac{1}{2} (\tilde{S}^{23} +
 i \tilde{S}^{01}, \tilde{S}^{31}+ i \tilde{S}^{02}, \tilde{S}^{12} + i \tilde{S}^{03} )\,,
\nonumber\\
 \vec{\tilde{\tau}}^{1}:&=&\frac{1}{2} (\tilde{S}^{58}-  \tilde{S}^{67}, \,\tilde{S}^{57} + 
 \tilde{S}^{68}, \,\tilde{S}^{56}-  \tilde{S}^{78} )\,,
\end{eqnarray}
respectively.
$U(1)$ contains the subgroup of the subgroup $SO(6)$ as well as the subgroup of $SO(4)$
 ($SO(6)$ and $SO(4)$ are together with $SO(3,1)$ the subgroups of the group $SO(13,1)$) 
with the infinitesimal operators equal to, Eq.~(\ref{so64tilde}),
 \begin{eqnarray}
 \label{so64}
\tau^{4}& = &-\frac{1}{3}(S^{9\;10} + S^{11\;12} + S^{13\;14})\,,\nonumber\\
 \vec{\tau}^{1}&=&\frac{1}{2} (S^{58}-  S^{67}, \,S^{57} + S^{68}, \,S^{56}-  S^{78} )\,,
\nonumber\\
 \vec{\tau}^{2}&=& \frac{1}{2} (S^{58}+  S^{67}, \,S^{57} - S^{68}, \,S^{56}+  S^{78} )\,.
 \end{eqnarray}
There are  additional subgroups $\widetilde{SU}(2) \times$ $ \widetilde{SU}(2)$, which belong to
 $\widetilde{SO}(4)_{\widetilde{SO}(3,1)}$ and $\widetilde{SO}(4)_{\widetilde{SO}(4)}$,
Eqs.~(\ref{so1+3tilde}, \ref{so42tilde}), the scalar gauge fields of which do not influence the 
masses of the four families to which the three observed families belong according to the 
predictions of the {\it spin-charge-family} theory~%
\footnote{The gauge scalar fields of these additional subgroups  
$\widetilde{SU}(2) \times$ $ \widetilde{SU}(2)$ influence the masses of the upper four families,
the stable one of which contribute to the dark matter.}.

All the  degrees of freedom and properties of spinors (of quarks and leptons)  and of gauge fields, 
demonstrated below, follow from the simple starting action, Eq.~(\ref{wholeaction}), after 
breaking the starting symmetry.

Let us rewrite formally the fermion part of the starting action, Eq.~(\ref{wholeaction}),  in the way
that it manifests, Eq.~(\ref{faction}), the kinetic and the interaction term in $d=(3+1)$ 
(the first line, $m=(0,1,2,3)$), the mass term (the second line, $s= (7,8)$) and the rest 
(the third line, $t=(5,6,9,10,\cdots,14)$).
\begin{eqnarray}
\label{faction}
{\mathcal L}_f &=&  \bar{\psi}\gamma^{m} (p_{m}- \sum_{A,i}\; g^{Ai}\tau^{Ai}
A^{Ai}_{m}) \psi + \nonumber\\
               & &  \{ \sum_{s=7,8}\;  \bar{\psi} \gamma^{s} p_{0s} \; \psi \} + \nonumber\\
& & \{ \sum_{t=5,6,9,\dots, 14}\;  \bar{\psi} \gamma^{t} p_{0t} \; \psi \}
\,,
\end{eqnarray}
where $p_{0s}$ $ =p_{s}  - \frac{1}{2}  S^{s' s"} \omega_{s' s" s} -
                    \frac{1}{2}  \tilde{S}^{ab}   \tilde{\omega}_{ab s}$,
         $p_{0t}$ $= p_{t}  - \frac{1}{2}  S^{t' t"} \omega_{t' t" t} -
                    \frac{1}{2}  \tilde{S}^{ab}   \tilde{\omega}_{ab t}$%
~\footnote{If there are no fermions present, then either $\omega_{ab c}$
or $\tilde{\omega}_{abc}$ are expressible by vielbeins $f^{\alpha}{}_{a}$%
~[\cite{ND2017,norma2014MatterAntimatter}, and the references therein]. We assume that
there are spinor fields which determine spin connection fields -- $\omega_{ab c}$ and
$\tilde{\omega}_{abc}$. In general one would have~\cite{JMP}: $p_{0a}$  $=
f^{\alpha}{}_{a} p_{0 \alpha} + \frac{1}{2E} \{p_{\alpha}, E f^{\alpha}{}_{a} \}_{-}$,
$p_{0 \alpha} =$  $p_{\alpha}  - \frac{1}{2}  S^{s' s"} \omega_{s' s" \alpha} -
                    \frac{1}{2}  \tilde{S}^{ab}   \tilde{\omega}_{ab \alpha}$. Since the term
$ \frac{1}{2E} \{p_{\alpha}, E f^{\alpha}{}_{a} \}_{-}$  does not influece the symmetry
of mass matrices, we do not treat it in this paper.},
with $ m \in (0,1,2,3)$, $s \in (7,8),\, (s',s") \in (5,6,7,8)$, $(a,b)$ (appearing in
 $\tilde{S}^{ab}$) run within  either $ (0,1,2,3)$ or $ (5,6,7,8)$, $t$ runs $ \in (5,\dots,14)$,
$(t',t")$ run either $ \in  (5,6,7,8)$ or $\in (9,10,\dots,14)$ 
\footnote{ We use units $\hbar=1=c$}.
The spinor function $\psi$ represents all family members, presented on 
Table~\ref{Table so13+1.}, of all the $2^{\frac{7+1}{2}-1}=8$ families, presented on 
Table~\ref{Table III.}. In this paper we pay attention on  the lower four families.

The first line of Eq.~(\ref{faction}) determines in $d=(3+1)$ the kinematics and dynamics of
spinor (fermion) fields, coupled to the vector gauge fields. The generators $\tau^{Ai} $ of the
charge groups are expressible  
in terms of $S^{ab}$ through the complex coefficients $c^{Ai}{ }_{ab}$ (the coefficients
$c^{Ai}{ }_{ab}$ of $\tau^{Ai}$ can be found in Eqs.~(\ref{so42}, \ref{so64})~%
\footnote{
Before the electroweak break there are the conserved (weak) charges  $\vec{\tau}^1$
(Eq.~(\ref{so42})),  $\vec{\tau}^{3} (Eq.~(\ref{so64}$) 
and $ Y:= \tau^{4} + \tau^{23}$  (Eqs.~(\ref{so42}, \ref{so64}) and the non conserved 
charge $Y':= -\tau^{4} \tan^2\vartheta_2 + \tau^{23}\,$, where $\vartheta_2$ is the angle 
of the break of $SU(2)_{II}$ from $SU(2)_{I} \times SU(2)_{II} \times $ $U(1)_{II}$ to
$SU(2)_{I} \times U(1)_{I}$. After the electroweak break the conserved charges are
$\vec{\tau}^{3}$ and $Q:= Y + \tau^{13}$, the non conserved charge is
$Q':= -Y \tan^2\vartheta_1 + \tau^{13}$, where
 $\vartheta_1$ is the electroweak angle.},
%
\begin{eqnarray}
\tau^{Ai} = \sum_{a,b} \;c^{Ai}{ }_{ab} \; S^{ab}\,,
\label{tau}
\end{eqnarray}
fulfilling the commutation relations
\begin{eqnarray}
\{\tau^{Ai}, \tau^{Bj}\}_- = i \delta^{AB} f^{Aijk} \tau^{Ak}\,.
\label{taucom}
\end{eqnarray}
They represent the colour ($\tau^{3i}$), the weak $(\tau^{1i})$ and the hyper ($Y$)
charges~%
\footnote{There are as well the $SU(2)_{II}$ ($\tau^{2i}$, Eq.~(\ref{so42})) and 
$U(1)_{II}$ ($\tau^{4}$, Eq.~(\ref{so64})) charges, the vector gauge fields of these last two 
groups gain masses when interacting with  the condensate,
Table~\ref{Table con.}~(\cite{IARD,JMP2015,norma2014MatterAntimatter} and the references
therein). The condensate leaves massless, besides the colour and gravity
gauge fields in $d=(3+1)$, the weak and the hyper charge vector gauge fields.}.
The corresponding vector gauge fields  $A^{Ai}_{m}$ are expressible with the spin connection 
fields $\omega_{st m}$, Eq.~(\ref{gaugevectorAiomega})~%
\footnote{Both fields, $ A^{Ai}_{m}$ and $\tilde{A}^{\tilde{A}i}_{m}$, are 
expressible with only the vielbeins, if there are no spinor fields present~\cite{ND2017}.}
\begin{eqnarray}
A^{Ai}_{m} = \sum_{s,t} \;c^{Ai}{ }_{st} \; \omega^{st}{}_{m}\,.
\label{AAiomega}
\end{eqnarray}

The second line of Eq.~(\ref{faction}) determines masses of each family member $(u^{i}, d^{i},
\nu^{i}, e^{i})$. The scalar gauge fields of the charges --- those of the family members, 
determined by $S^{ab}$ and those of the families, determined by $\tilde{S}^{ab}$ --- carry 
space index ($7,8$). Correspondingly the operators $\gamma^0 \gamma^s$, appearing in the 
mass term, transform the left handed members of any family into the right handed members 
of the same family, what can easily be seen in Table~\ref{Table so13+1.}. Operators $S^{ab}$
transform one family member of a particular family into the same  family member of another 
family.

Each scalar gauge fields (they are the gauge fields with space index $s\ge 5$) are as well 
expressible with the spin connections and vielbeins, Eq.~(\ref{gaugescalarAiomega})~%
\cite{ND2017}.  

The groups $SO(3,1)$, $SU(3)$, $SU(2)_{I}$, $SU(2)_{II}$ and $U(1)_{II}$ (all embedded 
into $SO(13+1)$) determine spin and charges of spinors, the groups 
$\widetilde{SU}(2)_{\widetilde{SO}(3,1)}$, Eqs~(\ref{tildeso3142}), 
$\widetilde{SU}(2)_{\widetilde{SO}(4)}$, Eqs.~(\ref{tildeso3142}), (embedded into 
$\widetilde{SO}(13+1)$)
 determine family quantum numbers%
%
%
%
%
%
%
%
%
%
%
%
~\footnote{$\widetilde{SU}(3)$  do not contribute to the families at low energies. We studied
such possibilities in a toy model, Ref.~\cite{familiesDNproc}.}.

The generators of these later groups are expressible by $\tilde{S}^{ab}$
\begin{eqnarray}
\tilde{\tau}^{Ai} = \sum_{a,b} \;c^{Ai}{ }_{ab} \; \tilde{S}^{ab}\,,
\label{tautilde}
\end{eqnarray}
fulfilling again the commutation relations
\begin{eqnarray}
\{\tilde{\tau}^{Ai},\tilde{\tau}^{Bj}\}_- = i \delta^{AB} f^{Aijk}  \tilde{\tau}^{Ak}\,,
\label{tildetaucom}
\end{eqnarray}
while
\begin{eqnarray}
\{\tau^{Ai},\tilde{\tau}^{Bj}\}_- =0\,.
\label{tautildetaucom}
\end{eqnarray}
%


The scalar gauge fields of the groups $\widetilde{SU}(2)_{I}$ 
($=\widetilde{SU}(2)_{\widetilde{SO}(3,1)}$ with generators $\vec{\tilde{N}}_{L}$,
 Eq.~(\ref{so1+3tilde})), $\widetilde{SU}(2)_{I}$ ($=\widetilde{SU}(2)_{\widetilde{SO}(4)}$, 
with generators $\vec{\tilde{\tau}}^{1}$, Eq.~(\ref{so42tilde})) and
$U(1)$ (with generators $(Q, Q', Y')$, Eq.~(\ref{YQY'Q'andtilde})) are presented in
 Eq.~(\ref{gaugescalarAiomega})~%
\footnote{All the scalar gauge fields, presented in Eq.~(\ref{gaugescalarAiomega}), are 
expressible with the vielbeins and spin connections with the space index $a\ge 5$,
 Ref.~\cite{ND2017}.}. 
The application of the generators
 $\vec{\tilde{\tau}}^{1}$, Eq.~(\ref{so42tilde}), $\vec{\tilde{N}}_{L}$, Eq.~(\ref{so1+3tilde}),
which distinguish among families and are the same for all the family members, is presented in
Eqs.~(\ref{snmb:gammatildegamma}, \ref{grapheigen}, \ref{tildetau1nl}). 

The application of  the family members generators ($Q, Q', Y'$) on the family members of 
any family is presented on Table~\ref{Table QYTAU.}.
The contribution of the scalar gauge fields to masses of different family members strongly 
depends on the quantum numbers $Q$, $Q'$ and $Y'$ as one can read from 
Table~\ref{Table QYTAU.}. In loop corrections the contribution of the scalar gauge fields of 
($Q, Q', Y'$)  is proportional  to the even power of these quantum numbers, while 
the nonzero vacuum expectation values of these scalar fields contribute in odd powers.
 \begin{table}
 \begin{center}
 \begin{tabular}{|r|r|r|r|r|r|r|r|r|r|r|r|}
 \hline
 $R$ &$Q_{\small{L,R}}$&$Y$&$\tau^{4}_{L,R}$&$\tau^{23}$&$Y'$& $Q'$& $L$ &$Y$&
$\tau^{13}$&$Y'$&$Q'$\\
 \hline
 $u^{i}_R$  & $ \frac{2}{3}$ & $ \frac{2}{3}$ & $ \frac{1}{6}$ &$\frac{1}{2}$&
$ \frac{1}{2}\,(1-\frac{1}{3} \tan^2 \vartheta_2)$ &$ -\frac{2}{3}\tan^2 \vartheta_1$&
 $u^{i}_L$  & $ \frac{1}{6}$&$\frac{1}{2}$& $-\frac{1}{6}\tan^2 \vartheta_2$& $ \frac{1}{2}
(1-\frac{1}{3} \tan^2 \vartheta_1) $
 \\
  $d^{i}_R$  & $-\frac{1}{3}$& $- \frac{1}{3}$&$\frac{1}{6}$&$-\frac{1}{2}$&
$ -\frac{1}{2}\,(1+\frac{1}{3} \tan^2 \vartheta_2)$&$\frac{1}{3}\tan^2 \vartheta_1$&
 $d^{i}_L$  & $  \frac{1}{6}$&$ -\frac{1}{2}$&$ - \frac{1}{6}\tan^2 \vartheta^2$&
$ -  \frac{1}{2}(1+ \frac{1}{3}\tan^2 \vartheta_1) $
 \\
 $\nu^{i}_R$& $0$ &$0$&$-\frac{1}{2}$  &$\frac{1}{2}$& $\frac{1}{2}\,
(1+  \tan^2 \vartheta_2)$ &
 $0$& $\nu^{i}_L$&$-\frac{1}{2}$& $\frac{1}{2}$&$\frac{1}{2}\,       
\tan^2 \vartheta_2 $ &
 $ \frac{1}{2}(1+ \tan^2 \vartheta_1)$
 \\
 $e_R$ & $-1$ &$-1$ &$-  \frac{1}{2}$&$-\frac{1}{2}$ & $\frac{1}{2}\,(-1+ \tan^2 \vartheta_2)$&
                                      $\tan^2 \vartheta_1$&
 $e_L$ & $-\frac{1}{2}$&$-\frac{1}{2}$ &$ \frac{1}{2}                  \tan^2 \vartheta_2 $ &
        $ -\frac{1}{2}(1-               \tan^2 \vartheta_1)$
 \\
 \hline
 \end{tabular}
 \end{center}
 \caption{\label{Table QYTAU.} The quantum numbers $Q, Y, \tau^4, Y', Q', \tau^{23}, \tau^{13}$,
Eq.~(\ref{YQY'Q'andtilde}), of  the family members $u^{i}_{L,R}, \nu^{i}_{L,R}$ of one family
(any one)~\cite{JMP} are presented. The left and right handed members of any family have the 
same $Q$ and $\tau^4$, the right handed members have $\tau^{13} =0$, and $\tau^{23}=  
\frac{1}{2}$ for $(u^{i}_{R}, \nu^{i}_{R})$ and $- \frac{1}{2}$ for $(d^{i}_{R}, e^{i}_{R})$,
while  the left handed members have  $\tau^{23} = 0$ and $\tau^{13} = \frac{1}{2}$ for 
 $(u^{i}_{L}, \nu^{i}_{L})$ and  $- \frac{1}{2}$ for $(d^{i}_{L}, e^{i}_{L})$. $\nu^{i}_{R}$
couples only to $A^{Y'}_{s}$ as seen from the table. 
}
 \end{table}

There are in the {\it spin-charge-family} theory $2^{\frac{(1+7)}{2} - 1}=8$ families~%
\footnote{In the break from $SO(13,1)$ to $SO(7,1) \times SO(6)$ only eight families remain 
massless, those for which the symmetry $\widetilde{SO}(7,1)$ remains. 
In Ref.~\cite{familiesDNproc} such kinds of breaks are discussed for a toy model.},
which split in two groups of four families, due to the break of the symmetry from $\widetilde{SO}
(7,1)$ into $\widetilde{SO}(3,1)$ $\times \widetilde{SO} (4)$. Each of these two groups manifests
$\widetilde{SU}(2)_{\widetilde{SO}(3,1)}\times$$\widetilde{SU}(2)_{\widetilde{SO}(4)}$
symmetry~\cite{JMP}. These decoupled twice four families are presented in Table~\ref{Table III.}.

The lowest of the upper four families, forming neutral clusters with respect to the electromagnetic
and colour charges, is the candidate to form the dark matter~\cite{gn}.

We discuss in this paper symmetry properties of the lower four families, presented in
Table~\ref{Table III.} in the first four lines. We present in Table~\ref{Table 4families.} the
representation and the family quantum numbers of the left and right handed members of the
lower four families. Since any of the family members ($u^{i}_{L,R}$, $d^{i}_{L,R}$,
$\nu^{i}_{L,R}$, $e^{i}_{L,R}$)  behave equivalently with respect to all the operators concerning
the family groups $\widetilde{SU}(2)_{\widetilde{SO}(1,3)}\times$
$\widetilde{SU}(2)_{\widetilde{SO}(4)}$, the last five columns are the same for all the family 
members.%


We rewrite the interaction, which is in the {\it spin-charge-family} theory responsible for the 
appearance of masses of fermions, presented in  Eq.~(\ref{faction}) in the second line, in a 
slightly different way, expressing $\gamma^7= (\stackrel{78}{(+)} + \stackrel{78}{(-)})$ and
correspondingly $\gamma^8=- i (\stackrel{78}{(+)} - \stackrel{78}{(-)})$.
\begin{eqnarray}
\label{factionmass}
{\mathcal L}_{mass} &=& \frac{1}{2}\, \sum_{+,-} \{ \psi^{\dagger}_{L}
\gamma^0 \, \stackrel{78}{(\pm)} \,( - \sum_{A}  \, \tau^{\alpha}\, A^{\alpha}_{\pm} -
\sum_{\tilde{A}i}\, \tilde{\tau}^{Ai}\,\tilde{A}^{\tilde{A}i}_{\pm}) \psi_{R} \}+ h.c.\,,
\nonumber\\
  \tau^{\alpha}  &= & (Q,Q',Y')\,,    \quad
  \tilde{\tau}^{\tilde{A}i} = (\vec{\tilde{N}}_{L}, \vec{\tilde{\tau}}^{\tilde{1}})\,,\nonumber\\
\gamma^0 \, \stackrel{78}{(\pm)} &=& \gamma^0 \frac{1}{2}\, (\gamma^7 \pm i \,
\gamma^8)\,, \nonumber\\
A^{\alpha}_{\pm}  &=& \sum_{st}\, c^{\alpha}{}_{st}\, \omega^{st}{}_{\pm}\,,\quad
\omega^{st}{}_{\pm} = \omega^{st}{}_{7} \mp\,i \,\omega^{st}{}_{8}\,, \nonumber\\
\vec{\tilde{A}}^{\tilde{A}}_{\pm}  &=& \sum_{ab}\, c^{A}{}_{ab}\,
\tilde{\omega}^{ab}{}_{\pm}\,,
\quad \tilde{\omega}^{ab}{}_{\pm} = \tilde{\omega}^{ab}{}_{7} \mp \,i\,
\tilde{\omega}^{ab}{}_{8}\,.
\end{eqnarray}
In Eq.~(\ref{factionmass}) the term $p_s$ is left out since at low energies its contribution is negligible,
$A$ determines operators, which distinguish among family members ---
($Q,Q',Y'$)~
\footnote{$(Q,Q',Y')$ 
are expressible in terms of $(\tau^{13}, \tau^{23} ,\tau^{4})$ as explained 
in Eq.~(\ref{YQY'Q'andtilde}). The  corresponding superposition of $\omega^{s s'}{}_{\pm}$ 
fields can be found by taking into account Eqs.~(\ref{so42}, \ref{so64}).},  
their eigenvalues on basic states are presented on Table~\ref{Table QYTAU.} ---  ($\tilde{A},i$) 
represent the family operators, determined in Eqs.~(\ref{so1+3tilde}, \ref{so42tilde}, \ref{so64tilde}).
The detailed explanation can be found in Refs.~\cite{JMP2015,norma2014MatterAntimatter,IARD}.

Operators $\tau^{Ai}$ are Hermitian ($(\tau^{Ai})^{\dagger}=$ $\tau^{Ai}$), while
$(\gamma^0 \stackrel{78}{(\pm)})^{\dagger}$ $=\gamma^0 \stackrel{78}{(\mp)}$.  
If the scalar fields $A^{Ai}_s$ are real 
it follows that $(A^{Ai}_{\pm})^{\dagger} = A^{Ai}_{\mp}$.

While the family operators $\tilde{\tau}^{1i}$ and $\tilde{N}^{i}_{L}$ commute with
$\gamma^0 \stackrel{78}{(\pm)}$, $\{S^{ab}, \tilde{S}^{cd}\}_{-}$ =0 for all
$(a,b,c,d)$, the family members operators ($\tau^{13}, \tau^{23}$) 
do not, since  $S^{78} $ does not ($S^{78}\gamma^0 \stackrel{78}{(\mp)} $ $= -
\gamma^0 \stackrel{78}{(\mp)} S^{78}$). However
%
$[\psi^{k\dagger}_{L} \gamma^0 \stackrel{78}{(\mp)} (Q,Q',Y') A^{(Q,Q',Y')}_{\mp}
\psi^{l}_{R}]^{\dagger} =
\psi^{l \dagger}_{R} \, (Q,Q',Y') \,A^{(Q,Q',Y') \dagger}_{\pm} \, \gamma^0
\stackrel{78}{(\pm)} \psi^{k}_{L}\,\delta_{k,l} =$ 
$ \psi^{l \dagger}_{R} \, (Q^{k}_{R},Q'^{k}_{R},Y'^{k}_{R})\,
A^{(Q,Q',Y')}_{\pm} \, \psi^{k}_{R}\,\delta_{k,l}$,
%
where $ (Q^{k}_{R},Q'^{k}_{R},Y'^{k}_{R})$ denote the eigenvalues of the corresponding
operators on the spinor state $\psi^{k}_{R}$. 
This means that we evaluate in both cases quantum numbers of the right handed partners.

But, let us evaluate $\frac{1}{\sqrt{2}}<u^{i}_{L} +u^{i}_{R}|\hat{O}^{\alpha}
|u^{i}_{L} +u^{i}_{R}>\frac{1}{\sqrt{2}}$, with $\hat{O}^{\alpha}=\sum_{+,-} 
\gamma^0 \stackrel{78}{(\pm)} (\tau^{4} A^{4}_{\stackrel{78}{(\pm)}} +
\tau^{23} A^{23}_{\stackrel{78}{(\pm)}}  + \tau^{13} A^{13}_{\stackrel{78}{(\pm)}})$.
One obtains $\frac{1}{\sqrt{2}} \{\frac{1}{6}(A^4_{-} + A^4_{+}) + A^{23}_{-} + 
A^{13}_{+}\}$. Equivalent evaluations for $|d^{i}_{L} + d^{i}_{R}>$ would give
$\frac{1}{\sqrt{2}} \{\frac{1}{6}(A^4_{-} + A^4_{+}) - A^{23}_{-} - 
A^{13}_{+}\}$, while for neutrinos we would obtain $\frac{1}{\sqrt{2}}
 \{-\frac{1}{2}(A^4_{-} + A^4_{+}) + A^{23}_{-} + A^{13}_{+}\}$ and for $e^{i}$ we 
would obtain $\frac{1}{\sqrt{2}} \{- \frac{1}{2}(A^4_{-} + A^4_{+}) - A^{23}_{-} - 
A^{13}_{+}\}$. Let us point out that the fields include also coupling constants, which change 
when the symmetry is broken. This means that we must carefully evaluate expectation values 
of all the operators on each state of broken symmetries. 
We have here much easier work: To see how does the starting symmetry of the 
mass matrices behave under  all possible corrections up to $\infty$ we only have to compare
how do matrix elements, which are equal on the tree level, change in any order of 
corrections.

In Table~\ref{Table 4families.} four families of spinors,  belonging to the group with the nonzero
values of $\vec{\tilde{N}}_{L}$ and $\vec{\tilde{\tau}}^{1}$, are presented.
These are the lower four families, presented also in Table~\ref{Table III.} together with
the upper four families~%
\footnote{The upper four families have the nonzero values of
$\vec{\tilde{N}}_{R}$ and $\vec{\tilde{\tau}}^{2}$. The stable members of the upper four
families offer the explanation for the existence  the dark matter~\cite{gn}.}.
There are indeed the four families of $\psi^{i}_{u_{R}}$ and  $\psi^{i}_{u_{L}}$ presented in
this table. All the $2^{\frac{13+1}{2}-1}$ members of the first family are represented
in Table~\ref{Table so13+1.}.

The three singlet scalar fields ($A^{Q}_{\mp}$, $A^{Q'}_{\mp}$, $A^{Y'}_{\mp}$)
 of Eq.~(\ref{factionmass})  contribute on the tree level the "diagonal" values to the mass 
term  --- 
$ \gamma^0 \stackrel{78}{(\mp)}\,Q A^{Q}_{\mp}$  $+ \gamma^0 \stackrel{78}{(\mp)}\,
Q' A^{Q'}_{\mp}$  $+ \gamma^0 \stackrel{78}{(\mp)}\,Y' A^{Y'}_{\mp}$ --- transforming a
right handed member of one family into the left handed member of the same family, or a left
handed member  of one family into the right handed member of the same family. 
{\it These terms
are different for different family members but the same for all the families.}

 Since  $Q = (\tau^{13} + \tau^{23} +\tau^{4})= (S^{56} + \tau^{4})$,
$Y'= (- \tau^4\, \tan^2 \vartheta_2 + \tau^{23})$  and $Q' = (- (\tau^{4} + \tau^{23})
\tan^2 \vartheta_1 + \tau^{13})$ --- $\vartheta_1$ is the {\it standard model} angle and 
$\vartheta_2$ is the corresponding angle when the second $SU(2)$ symmetry breaks --- 
 we could use instead  of the  operators 
($ \gamma^0 \stackrel{78}{(\mp)}\,Q A^{Q}_{\mp}$  $+ \gamma^0 \stackrel{78}{(\mp)}\,
Q' A^{Q'}_{\mp}$  $+ \gamma^0 \stackrel{78}{(\mp)}\,Y' A^{Y'}_{\mp}$) as well the
operators ($ \gamma^0 \stackrel{78}{(\pm)}\,  \tau^4 \, A^{4}_{\pm} $, $ \gamma^0  
\stackrel{78}{(\pm)} \tau^{23} \, A^{23}_{\pm} $, $ \gamma^0  \stackrel{78}{(\pm)}
\tau^{13} \, A^{13}_{\pm} $), if the fact that the coupling constants of all the fields, 
also of  $\omega_{ab s}$ and $\tilde{\omega}_{ab s}$, change with the 
break of symmetry is taken into account. 

Let us denote by $-a^{\alpha}$ the nonzero vacuum expectation values of the three singlets 
for a family member $\alpha= (u^{i},\nu^{i}, d^{i}, e^{i})$, divided by the energy scale 
(let say TeV), when (if) these scalars have nonzero vacuum expectation values and we use 
the basis $\frac{1}{2} |\psi^{i \alpha}_{L} + \psi^{i \alpha}_{R} >$:
\begin{eqnarray}
\label{a}
&&a^{\alpha}=- 
 \{ \frac{1}{2}<\psi^{i \alpha}_{L}+\psi^{i \alpha}_{R}|
\nonumber\\
&&\sum_{+,-}\gamma^0 \stackrel{78}{(\pm)}
[Q<A^{Q}_{\pm}>+Q'<A^{Q'}_{\pm}>+Y'<A^{Y'}_{\pm}>]
 |\psi^{j \alpha}_{L} +\psi^{j \alpha}_{R}>\frac{1}{2} \} \delta^{ij} + h.c.,  
\end{eqnarray}
%
Each family member has a different value for $a^{\alpha}$.
All the scalar gauge fields $A^{Q}_{\stackrel{78}{(\pm)}}, A^{Q'}_{\stackrel{78}{(\pm)}}, 
A^{Y'}_{\stackrel{78}{(\pm)}}$ have the weak and the hypercharge as higgs scalars:
 ($\pm \frac{1}{2},\mp  \frac{1}{2}$,  respectively). 



%
 \begin{table}
 \begin{center}
 \begin{tabular}{|c|c|c|c|r r r r r|}
 \hline
 &&&&$\tilde{\tau}^{13}$&$\tilde{\tau}^{23}$&$\tilde{N}_{L}^{3}$&$\tilde{N}_{R}^{3}$&
 $\tilde{\tau}^{4}$\\
 \hline
 $\psi^{1}_{u^{ci}_{R}}$&
   $ \stackrel{03}{(+i)}\,\stackrel{12}{[+]}|\stackrel{56}{[+]}\,\stackrel{78}{(+)} ||\cdots$
   &$\psi^{1}_{u^{ci}_{L}}$ &
   $  \stackrel{03}{[-i]}\,\stackrel{12}{[+]}|\stackrel{56}{[+]}\,\stackrel{78}{[-]} ||\cdots$
  &$-\frac{1}{2}$&$0$&$-\frac{1}{2}$&$0$&$-\frac{1}{2}$
 \\
  $\psi^{2}_{u^{ci}_{R}}$ &
   $ \stackrel{03}{[+i]}\,\stackrel{12}{(+)}|\stackrel{56}{[+]}\,\stackrel{78}{(+)} ||\dots$
   &$\psi^{2}_{u^{ci}_{L}}$ &
   $   \stackrel{03}{(-i)}\,\stackrel{12}{(+)}|\stackrel{56}{[+]}\,\stackrel{78}{[-]} ||\cdots$
  &$-\frac{1}{2}$&$0$&$\frac{1}{2}$&$0$&$-\frac{1}{2}$
 \\
 $\psi^{3}_{u^{ci}_{R}}$&
   $ \stackrel{03}{(+i)}\,\stackrel{12}{[+]}|\stackrel{56}{(+)}\,\stackrel{78}{[+]} ||\cdots$
   &$\psi^{3}_{u^{ci}_{L}}$ &
   $   \stackrel{03}{[-i]}\,\stackrel{12}{[+]}|\stackrel{56}{(+)}\,\stackrel{78}{(-)} ||\cdots$
  &$\frac{1}{2}$&$0$&$-\frac{1}{2}$&$0$&$-\frac{1}{2}$
 \\
 $\psi^{4}_{u^{ci}_{R}}$&
  $ \stackrel{03}{[+i]}\,\stackrel{12}{(+)}|\stackrel{56}{(+)}\,\stackrel{78}{[+]} ||\cdots$
   &$\psi^{4}_{u^{ci}_{L}}$ &
  $   \stackrel{03}{(-i)}\,\stackrel{12}{(+)}|\stackrel{56}{(+)}\,\stackrel{78}{(-)} ||\cdots$
  &$\frac{1}{2}$&$0$&$\frac{1}{2}$&$0$&$-\frac{1}{2}$
  \\
  \hline
 \end{tabular}
 \end{center}
\caption{\label{Table 4families.}
Four families of the right handed $u^{c 1}_{R}$ with the weak and the hyper charge 
($\tau^{13}=0, Y=\frac{2}{3}$) and of the left handed $u^{c1}_{L}$ quarks with 
($\tau^{13}=\frac{1}{2}, Y=\frac{1}{6}$), both with spin $\frac{1}{2}$ and colour 
$(\tau^{33}, \tau^{38})=[(1/2,1/(2\sqrt{3}), (-1/2,1/(2\sqrt{3}), (0, -1/(\sqrt{3})]$ 
charges are presented. They represent two of the family members from 
Table~\ref{Table so13+1.} ---  $u^{c_1}_{R}$ and $u^{c_1}_{L}$ --- appearing 
on $1^{st}$ and $7^{th}$ line 
of Table~\ref{Table so13+1.}. 
Spins and charges 
commute with $\tilde{N}^{i}_{L}$, $\tilde{\tau}^{1i}$ and
$\tilde{\tau}^{4}$, and are correspondingly the same for all the families.
}
 \end{table}

Transitions among families for any family member are caused by ($\tilde{N}^{i}_{L} \,
\tilde{A}^{\tilde{N}_{L}\spm}$ and $\tilde{\tau}^{1i}\, \tilde{A}^{\tilde{1}\spm}$),  what
 manifests the symmetry $\widetilde{SU}_{N_{L}}(2)\times $
$\widetilde{SU}_{\tau^{1}}(2)$.  There are corrections in all orders, which make all the matrix 
elements of the mass matrix for any of the family members $\alpha$ dependent on the 
three singlets ($\tau^{4} A^{4}_{\pm}$, $\tau^{23} A^{23}_{\pm}$, $\tau^{13} A^{13}_{\pm}$),
 Eq.~(\ref{a}).

\bottomcaption{\label{Table so13+1.}%
\tiny{
The left handed ($\Gamma^{(13,1)} = -1$,
Eq.~(\ref{hand}))
multiplet of spinors --- the members of the fundamental representation of the $SO(13,1)$ group,
manifesting the subgroup $SO(7,1)$
 of the colour charged quarks and anti-quarks and the colourless
leptons and anti-leptons --- is presented in the massless basis using the technique presented in
App.~\ref{technique}. 
It contains the left handed  ($\Gamma^{(3,1)}=-1$, App.~\ref{technique}) weak ($SU(2)_{I}$)
charged  ($\tau^{13}=\pm \frac{1}{2}$, Eq.~(\ref{so42})),
and $SU(2)_{II}$ chargeless ($\tau^{23}=0$, Eq.~(\ref{so42}))
quarks and leptons and the right handed  ($\Gamma^{(3,1)}=1$, 
 weak  ($SU(2)_{I}$) chargeless and $SU(2)_{II}$
charged ($\tau^{23}=\pm \frac{1}{2}$) quarks and leptons, both with the spin $ S^{12}$  up and
down ($\pm \frac{1}{2}$, respectively). 
Quarks distinguish from leptons only in the $SU(3) \times U(1)$ part: Quarks are triplets
of three colours  ($c^i$ $= (\tau^{33}, \tau^{38})$ $ = [(\frac{1}{2},\frac{1}{2\sqrt{3}}),
(-\frac{1}{2},\frac{1}{2\sqrt{3}}), (0,-\frac{1}{\sqrt{3}}) $], Eq.~(\ref{so64}))
carrying  the "fermion charge" ($\tau^{4}=\frac{1}{6}$, Eq.~(\ref{so64})).
The colourless leptons carry the "fermion charge" ($\tau^{4}=-\frac{1}{2}$).
The same multiplet contains also the left handed weak ($SU(2)_{I}$) chargeless and $SU(2)_{II}$
charged anti-quarks and anti-leptons and the right handed weak ($SU(2)_{I}$) charged and
$SU(2)_{II}$ chargeless anti-quarks and anti-leptons.
Anti-quarks distinguish from anti-leptons again only in the $SU(3) \times U(1)$ part: Anti-quarks are
anti-triplets, 
 carrying  the "fermion charge" ($\tau^{4}=-\frac{1}{6}$).
The anti-colourless anti-leptons carry the "fermion charge" ($\tau^{4}=\frac{1}{2}$).
 $Y=(\tau^{23} + \tau^{4})$ is the hyper charge, the electromagnetic charge
is $Q=(\tau^{13} + Y$).
The states of opposite charges (anti-particle states) are reachable  from the particle states besides by 
$S^{ab}$  also by the application of the discrete symmetry operator
${\cal C}_{{\cal N}}$ ${\cal P}_{{\cal N}}$, presented in Refs.~\cite{discretesym,TDN}.
%
The vacuum state,
on which the nilpotents and projectors operate, is not shown.
The reader can find this  Weyl representation also in
Refs.~\cite{norma2014MatterAntimatter,pikanorma,portoroz03,JMP2015} and in the references
therein. }
}
\tablehead{\hline
i&$$&$|^a\psi_i>$&$\Gamma^{(3,1)}$&$ S^{12}$&
$\tau^{13}$&$\tau^{23}$&$\tau^{33}$&$\tau^{38}$&$\tau^{4}$&$Y$&$Q$\\
\hline
&& ${\rm (Anti)octet},\,\Gamma^{(7,1)} = (-1)\,1\,, \,\Gamma^{(6)} = (1)\,-1$&&&&&&&&& \\
&& ${\rm of \;(anti) quarks \;and \;(anti)leptons}$&&&&&&&&&\\
\hline\hline}
\tabletail{\hline \multicolumn{12}{r}{\emph{Continued on next page}}\\}
\tablelasttail{\hline}
\begin{center}
\tiny{
\begin{supertabular}{|r|c||c||c|c||c|c||c|c|c||r|r|}
1&$ u_{R}^{c1}$&$ \stackrel{03}{(+i)}\,\stackrel{12}{[+]}|
\stackrel{56}{[+]}\,\stackrel{78}{(+)}
||\stackrel{9 \;10}{(+)}\;\;\stackrel{11\;12}{[-]}\;\;\stackrel{13\;14}{[-]} $ &1&$\frac{1}{2}$&0&
$\frac{1}{2}$&$\frac{1}{2}$&$\frac{1}{2\,\sqrt{3}}$&$\frac{1}{6}$&$\frac{2}{3}$&$\frac{2}{3}$\\
\hline
2&$u_{R}^{c1}$&$\stackrel{03}{[-i]}\,\stackrel{12}{(-)}|\stackrel{56}{[+]}\,\stackrel{78}{(+)}
||\stackrel{9 \;10}{(+)}\;\;\stackrel{11\;12}{[-]}\;\;\stackrel{13\;14}{[-]}$&1&$-\frac{1}{2}$&0&
$\frac{1}{2}$&$\frac{1}{2}$&$\frac{1}{2\,\sqrt{3}}$&$\frac{1}{6}$&$\frac{2}{3}$&$\frac{2}{3}$\\
\hline
3&$d_{R}^{c1}$&$\stackrel{03}{(+i)}\,\stackrel{12}{[+]}|\stackrel{56}{(-)}\,\stackrel{78}{[-]}
||\stackrel{9 \;10}{(+)}\;\;\stackrel{11\;12}{[-]}\;\;\stackrel{13\;14}{[-]}$&1&$\frac{1}{2}$&0&
$-\frac{1}{2}$&$\frac{1}{2}$&$\frac{1}{2\,\sqrt{3}}$&$\frac{1}{6}$&$-\frac{1}{3}$&$-\frac{1}{3}$\\
\hline
4&$ d_{R}^{c1} $&$\stackrel{03}{[-i]}\,\stackrel{12}{(-)}|
\stackrel{56}{(-)}\,\stackrel{78}{[-]}
||\stackrel{9 \;10}{(+)}\;\;\stackrel{11\;12}{[-]}\;\;\stackrel{13\;14}{[-]} $&1&$-\frac{1}{2}$&0&
$-\frac{1}{2}$&$\frac{1}{2}$&$\frac{1}{2\,\sqrt{3}}$&$\frac{1}{6}$&$-\frac{1}{3}$&$-\frac{1}{3}$\\
\hline
5&$d_{L}^{c1}$&$\stackrel{03}{[-i]}\,\stackrel{12}{[+]}|\stackrel{56}{(-)}\,\stackrel{78}{(+)}
||\stackrel{9 \;10}{(+)}\;\;\stackrel{11\;12}{[-]}\;\;\stackrel{13\;14}{[-]}$&-1&$\frac{1}{2}$&
$-\frac{1}{2}$&0&$\frac{1}{2}$&$\frac{1}{2\,\sqrt{3}}$&$\frac{1}{6}$&$\frac{1}{6}$&$-\frac{1}{3}$\\
\hline
6&$d_{L}^{c1} $&$  \stackrel{03}{(+i)}\,\stackrel{12}{(-)}|\stackrel{56}{(-)}\,\stackrel{78}{(+)}
||\stackrel{9 \;10}{(+)}\;\;\stackrel{11\;12}{[-]}\;\;\stackrel{13\;14}{[-]} $&-1&$-\frac{1}{2}$&
$-\frac{1}{2}$&0&$\frac{1}{2}$&$\frac{1}{2\,\sqrt{3}}$&$\frac{1}{6}$&$\frac{1}{6}$&$-\frac{1}{3}$\\
\hline
7&$ u_{L}^{c1}$&$  \stackrel{03}{[-i]}\,\stackrel{12}{[+]}|\stackrel{56}{[+]}\,\stackrel{78}{[-]}
||\stackrel{9 \;10}{(+)}\;\;\stackrel{11\;12}{[-]}\;\;\stackrel{13\;14}{[-]}$ &-1&$\frac{1}{2}$&
$\frac{1}{2}$&0 &$\frac{1}{2}$&$\frac{1}{2\,\sqrt{3}}$&$\frac{1}{6}$&$\frac{1}{6}$&$\frac{2}{3}$\\
\hline
8&$u_{L}^{c1}$&$\stackrel{03}{(+i)}\,\stackrel{12}{(-)}|\stackrel{56}{[+]}\,\stackrel{78}{[-]}
||\stackrel{9 \;10}{(+)}\;\;\stackrel{11\;12}{[-]}\;\;\stackrel{13\;14}{[-]}$&-1&$-\frac{1}{2}$&
$\frac{1}{2}$&0&$\frac{1}{2}$&$\frac{1}{2\,\sqrt{3}}$&$\frac{1}{6}$&$\frac{1}{6}$&$\frac{2}{3}$\\
\hline\hline
\shrinkheight{0.3\textheight}
9&$ u_{R}^{c2}$&$ \stackrel{03}{(+i)}\,\stackrel{12}{[+]}|
\stackrel{56}{[+]}\,\stackrel{78}{(+)}
||\stackrel{9 \;10}{[-]}\;\;\stackrel{11\;12}{(+)}\;\;\stackrel{13\;14}{[-]} $ &1&$\frac{1}{2}$&0&
$\frac{1}{2}$&$-\frac{1}{2}$&$\frac{1}{2\,\sqrt{3}}$&$\frac{1}{6}$&$\frac{2}{3}$&$\frac{2}{3}$\\
\hline
10&$u_{R}^{c2}$&$\stackrel{03}{[-i]}\,\stackrel{12}{(-)}|\stackrel{56}{[+]}\,\stackrel{78}{(+)}
||\stackrel{9 \;10}{[-]}\;\;\stackrel{11\;12}{(+)}\;\;\stackrel{13\;14}{[-]}$&1&$-\frac{1}{2}$&0&
$\frac{1}{2}$&$-\frac{1}{2}$&$\frac{1}{2\,\sqrt{3}}$&$\frac{1}{6}$&$\frac{2}{3}$&$\frac{2}{3}$\\
\hline
11&$d_{R}^{c2}$&$\stackrel{03}{(+i)}\,\stackrel{12}{[+]}|\stackrel{56}{(-)}\,\stackrel{78}{[-]}
||\stackrel{9 \;10}{[-]}\;\;\stackrel{11\;12}{(+)}\;\;\stackrel{13\;14}{[-]}$
&1&$\frac{1}{2}$&0&
$-\frac{1}{2}$&$ - \frac{1}{2}$&$\frac{1}{2\,\sqrt{3}}$&$\frac{1}{6}$&$-\frac{1}{3}$&$-\frac{1}{3}$\\
\hline
12&$ d_{R}^{c2} $&$\stackrel{03}{[-i]}\,\stackrel{12}{(-)}|
\stackrel{56}{(-)}\,\stackrel{78}{[-]}
||\stackrel{9 \;10}{[-]}\;\;\stackrel{11\;12}{(+)}\;\;\stackrel{13\;14}{[-]} $
&1&$-\frac{1}{2}$&0&
$-\frac{1}{2}$&$-\frac{1}{2}$&$\frac{1}{2\,\sqrt{3}}$&$\frac{1}{6}$&$-\frac{1}{3}$&$-\frac{1}{3}$\\
\hline
13&$d_{L}^{c2}$&$\stackrel{03}{[-i]}\,\stackrel{12}{[+]}|\stackrel{56}{(-)}\,\stackrel{78}{(+)}
||\stackrel{9 \;10}{[-]}\;\;\stackrel{11\;12}{(+)}\;\;\stackrel{13\;14}{[-]}$
&-1&$\frac{1}{2}$&
$-\frac{1}{2}$&0&$-\frac{1}{2}$&$\frac{1}{2\,\sqrt{3}}$&$\frac{1}{6}$&$\frac{1}{6}$&$-\frac{1}{3}$\\
\hline
14&$d_{L}^{c2} $&$  \stackrel{03}{(+i)}\,\stackrel{12}{(-)}|\stackrel{56}{(-)}\,\stackrel{78}{(+)}
||\stackrel{9 \;10}{[-]}\;\;\stackrel{11\;12}{(+)}\;\;\stackrel{13\;14}{[-]} $&-1&$-\frac{1}{2}$&
$-\frac{1}{2}$&0&$-\frac{1}{2}$&$\frac{1}{2\,\sqrt{3}}$&$\frac{1}{6}$&$\frac{1}{6}$&$-\frac{1}{3}$\\
\hline
15&$ u_{L}^{c2}$&$  \stackrel{03}{[-i]}\,\stackrel{12}{[+]}|\stackrel{56}{[+]}\,\stackrel{78}{[-]}
||\stackrel{9 \;10}{[-]}\;\;\stackrel{11\;12}{(+)}\;\;\stackrel{13\;14}{[-]}$ &-1&$\frac{1}{2}$&
$\frac{1}{2}$&0 &$-\frac{1}{2}$&$\frac{1}{2\,\sqrt{3}}$&$\frac{1}{6}$&$\frac{1}{6}$&$\frac{2}{3}$\\
\hline
16&$u_{L}^{c2}$&$\stackrel{03}{(+i)}\,\stackrel{12}{(-)}|\stackrel{56}{[+]}\,\stackrel{78}{[-]}
||\stackrel{9 \;10}{[-]}\;\;\stackrel{11\;12}{(+)}\;\;\stackrel{13\;14}{[-]}$&-1&$-\frac{1}{2}$&
$\frac{1}{2}$&0&$-\frac{1}{2}$&$\frac{1}{2\,\sqrt{3}}$&$\frac{1}{6}$&$\frac{1}{6}$&$\frac{2}{3}$\\
\hline\hline
17&$ u_{R}^{c3}$&$ \stackrel{03}{(+i)}\,\stackrel{12}{[+]}|
\stackrel{56}{[+]}\,\stackrel{78}{(+)}
||\stackrel{9 \;10}{[-]}\;\;\stackrel{11\;12}{[-]}\;\;\stackrel{13\;14}{(+)} $ &1&$\frac{1}{2}$&0&
$\frac{1}{2}$&$0$&$-\frac{1}{\sqrt{3}}$&$\frac{1}{6}$&$\frac{2}{3}$&$\frac{2}{3}$\\
\hline
18&$u_{R}^{c3}$&$\stackrel{03}{[-i]}\,\stackrel{12}{(-)}|\stackrel{56}{[+]}\,\stackrel{78}{(+)}
||\stackrel{9 \;10}{[-]}\;\;\stackrel{11\;12}{[-]}\;\;\stackrel{13\;14}{(+)}$&1&$-\frac{1}{2}$&0&
$\frac{1}{2}$&$0$&$-\frac{1}{\sqrt{3}}$&$\frac{1}{6}$&$\frac{2}{3}$&$\frac{2}{3}$\\
\hline
19&$d_{R}^{c3}$&$\stackrel{03}{(+i)}\,\stackrel{12}{[+]}|\stackrel{56}{(-)}\,\stackrel{78}{[-]}
||\stackrel{9 \;10}{[-]}\;\;\stackrel{11\;12}{[-]}\;\;\stackrel{13\;14}{(+)}$&1&$\frac{1}{2}$&0&
$-\frac{1}{2}$&$0$&$-\frac{1}{\sqrt{3}}$&$\frac{1}{6}$&$-\frac{1}{3}$&$-\frac{1}{3}$\\
\hline
20&$ d_{R}^{c3} $&$\stackrel{03}{[-i]}\,\stackrel{12}{(-)}|
\stackrel{56}{(-)}\,\stackrel{78}{[-]}
||\stackrel{9 \;10}{[-]}\;\;\stackrel{11\;12}{[-]}\;\;\stackrel{13\;14}{(+)} $&1&$-\frac{1}{2}$&0&
$-\frac{1}{2}$&$0$&$-\frac{1}{\sqrt{3}}$&$\frac{1}{6}$&$-\frac{1}{3}$&$-\frac{1}{3}$\\
\hline
21&$d_{L}^{c3}$&$\stackrel{03}{[-i]}\,\stackrel{12}{[+]}|\stackrel{56}{(-)}\,\stackrel{78}{(+)}
||\stackrel{9 \;10}{[-]}\;\;\stackrel{11\;12}{[-]}\;\;\stackrel{13\;14}{(+)}$&-1&$\frac{1}{2}$&
$-\frac{1}{2}$&0&$0$&$-\frac{1}{\sqrt{3}}$&$\frac{1}{6}$&$\frac{1}{6}$&$-\frac{1}{3}$\\
\hline
22&$d_{L}^{c3} $&$  \stackrel{03}{(+i)}\,\stackrel{12}{(-)}|\stackrel{56}{(-)}\,\stackrel{78}{(+)}
||\stackrel{9 \;10}{[-]}\;\;\stackrel{11\;12}{[-]}\;\;\stackrel{13\;14}{(+)} $&-1&$-\frac{1}{2}$&
$-\frac{1}{2}$&0&$0$&$-\frac{1}{\sqrt{3}}$&$\frac{1}{6}$&$\frac{1}{6}$&$-\frac{1}{3}$\\
\hline
23&$ u_{L}^{c3}$&$  \stackrel{03}{[-i]}\,\stackrel{12}{[+]}|\stackrel{56}{[+]}\,\stackrel{78}{[-]}
||\stackrel{9 \;10}{[-]}\;\;\stackrel{11\;12}{[-]}\;\;\stackrel{13\;14}{(+)}$ &-1&$\frac{1}{2}$&
$\frac{1}{2}$&0 &$0$&$-\frac{1}{\sqrt{3}}$&$\frac{1}{6}$&$\frac{1}{6}$&$\frac{2}{3}$\\
\hline
24&$u_{L}^{c3}$&$\stackrel{03}{(+i)}\,\stackrel{12}{(-)}|\stackrel{56}{[+]}\,\stackrel{78}{[-]}
||\stackrel{9 \;10}{[-]}\;\;\stackrel{11\;12}{[-]}\;\;\stackrel{13\;14}{(+)}$&-1&$-\frac{1}{2}$&
$\frac{1}{2}$&0&$0$&$-\frac{1}{\sqrt{3}}$&$\frac{1}{6}$&$\frac{1}{6}$&$\frac{2}{3}$\\
\hline\hline
25&$ \nu_{R}$&$ \stackrel{03}{(+i)}\,\stackrel{12}{[+]}|
\stackrel{56}{[+]}\,\stackrel{78}{(+)}
||\stackrel{9 \;10}{(+)}\;\;\stackrel{11\;12}{(+)}\;\;\stackrel{13\;14}{(+)} $ &1&$\frac{1}{2}$&0&
$\frac{1}{2}$&$0$&$0$&$-\frac{1}{2}$&$0$&$0$\\
\hline
26&$\nu_{R}$&$\stackrel{03}{[-i]}\,\stackrel{12}{(-)}|\stackrel{56}{[+]}\,\stackrel{78}{(+)}
||\stackrel{9 \;10}{(+)}\;\;\stackrel{11\;12}{(+)}\;\;\stackrel{13\;14}{(+)}$&1&$-\frac{1}{2}$&0&
$\frac{1}{2}$ &$0$&$0$&$-\frac{1}{2}$&$0$&$0$\\
\hline
27&$e_{R}$&$\stackrel{03}{(+i)}\,\stackrel{12}{[+]}|\stackrel{56}{(-)}\,\stackrel{78}{[-]}
||\stackrel{9 \;10}{(+)}\;\;\stackrel{11\;12}{(+)}\;\;\stackrel{13\;14}{(+)}$&1&$\frac{1}{2}$&0&
$-\frac{1}{2}$&$0$&$0$&$-\frac{1}{2}$&$-1$&$-1$\\
\hline
28&$ e_{R} $&$\stackrel{03}{[-i]}\,\stackrel{12}{(-)}|
\stackrel{56}{(-)}\,\stackrel{78}{[-]}
||\stackrel{9 \;10}{(+)}\;\;\stackrel{11\;12}{(+)}\;\;\stackrel{13\;14}{(+)} $&1&$-\frac{1}{2}$&0&
$-\frac{1}{2}$&$0$&$0$&$-\frac{1}{2}$&$-1$&$-1$\\
\hline
29&$e_{L}$&$\stackrel{03}{[-i]}\,\stackrel{12}{[+]}|\stackrel{56}{(-)}\,\stackrel{78}{(+)}
||\stackrel{9 \;10}{(+)}\;\;\stackrel{11\;12}{(+)}\;\;\stackrel{13\;14}{(+)}$&-1&$\frac{1}{2}$&
$-\frac{1}{2}$&0&$0$&$0$&$-\frac{1}{2}$&$-\frac{1}{2}$&$-1$\\
\hline
30&$e_{L} $&$  \stackrel{03}{(+i)}\,\stackrel{12}{(-)}|\stackrel{56}{(-)}\,\stackrel{78}{(+)}
||\stackrel{9 \;10}{(+)}\;\;\stackrel{11\;12}{(+)}\;\;\stackrel{13\;14}{(+)} $&-1&$-\frac{1}{2}$&
$-\frac{1}{2}$&0&$0$&$0$&$-\frac{1}{2}$&$-\frac{1}{2}$&$-1$\\
\hline
31&$ \nu_{L}$&$  \stackrel{03}{[-i]}\,\stackrel{12}{[+]}|\stackrel{56}{[+]}\,\stackrel{78}{[-]}
||\stackrel{9 \;10}{(+)}\;\;\stackrel{11\;12}{(+)}\;\;\stackrel{13\;14}{(+)}$ &-1&$\frac{1}{2}$&
$\frac{1}{2}$&0 &$0$&$0$&$-\frac{1}{2}$&$-\frac{1}{2}$&$0$\\
\hline
32&$\nu_{L}$&$\stackrel{03}{(+i)}\,\stackrel{12}{(-)}|\stackrel{56}{[+]}\,\stackrel{78}{[-]}
||\stackrel{9 \;10}{(+)}\;\;\stackrel{11\;12}{(+)}\;\;\stackrel{13\;14}{(+)}$&-1&$-\frac{1}{2}$&
$\frac{1}{2}$&0&$0$&$0$&$-\frac{1}{2}$&$-\frac{1}{2}$&$0$\\
\hline\hline
33&$ \bar{d}_{L}^{\bar{c1}}$&$ \stackrel{03}{[-i]}\,\stackrel{12}{[+]}|
\stackrel{56}{[+]}\,\stackrel{78}{(+)}
||\stackrel{9 \;10}{[-]}\;\;\stackrel{11\;12}{(+)}\;\;\stackrel{13\;14}{(+)} $ &-1&$\frac{1}{2}$&0&
$\frac{1}{2}$&$-\frac{1}{2}$&$-\frac{1}{2\,\sqrt{3}}$&$-\frac{1}{6}$&$\frac{1}{3}$&$\frac{1}{3}$\\
\hline
34&$\bar{d}_{L}^{\bar{c1}}$&$\stackrel{03}{(+i)}\,\stackrel{12}{(-)}|\stackrel{56}{[+]}\,\stackrel{78}{(+)}
||\stackrel{9 \;10}{[-]}\;\;\stackrel{11\;12}{(+)}\;\;\stackrel{13\;14}{(+)}$&-1&$-\frac{1}{2}$&0&
$\frac{1}{2}$&$-\frac{1}{2}$&$-\frac{1}{2\,\sqrt{3}}$&$-\frac{1}{6}$&$\frac{1}{3}$&$\frac{1}{3}$\\
\hline
35&$\bar{u}_{L}^{\bar{c1}}$&$  \stackrel{03}{[-i]}\,\stackrel{12}{[+]}|\stackrel{56}{(-)}\,\stackrel{78}{[-]}
||\stackrel{9 \;10}{[-]}\;\;\stackrel{11\;12}{(+)}\;\;\stackrel{13\;14}{(+)}$&-1&$\frac{1}{2}$&0&
$-\frac{1}{2}$&$-\frac{1}{2}$&$-\frac{1}{2\,\sqrt{3}}$&$-\frac{1}{6}$&$-\frac{2}{3}$&$-\frac{2}{3}$\\
\hline
36&$ \bar{u}_{L}^{\bar{c1}} $&$  \stackrel{03}{(+i)}\,\stackrel{12}{(-)}|
\stackrel{56}{(-)}\,\stackrel{78}{[-]}
||\stackrel{9 \;10}{[-]}\;\;\stackrel{11\;12}{(+)}\;\;\stackrel{13\;14}{(+)} $&-1&$-\frac{1}{2}$&0&
$-\frac{1}{2}$&$-\frac{1}{2}$&$-\frac{1}{2\,\sqrt{3}}$&$-\frac{1}{6}$&$-\frac{2}{3}$&$-\frac{2}{3}$\\
\hline
37&$\bar{d}_{R}^{\bar{c1}}$&$\stackrel{03}{(+i)}\,\stackrel{12}{[+]}|\stackrel{56}{[+]}\,\stackrel{78}{[-]}
||\stackrel{9 \;10}{[-]}\;\;\stackrel{11\;12}{(+)}\;\;\stackrel{13\;14}{(+)}$&1&$\frac{1}{2}$&
$\frac{1}{2}$&0&$-\frac{1}{2}$&$-\frac{1}{2\,\sqrt{3}}$&$-\frac{1}{6}$&$-\frac{1}{6}$&$\frac{1}{3}$\\
\hline
38&$\bar{d}_{R}^{\bar{c1}} $&$  \stackrel{03}{[-i]}\,\stackrel{12}{(-)}|\stackrel{56}{[+]}\,\stackrel{78}{[-]}
||\stackrel{9 \;10}{[-]}\;\;\stackrel{11\;12}{(+)}\;\;\stackrel{13\;14}{(+)} $&1&$-\frac{1}{2}$&
$\frac{1}{2}$&0&$-\frac{1}{2}$&$-\frac{1}{2\,\sqrt{3}}$&$-\frac{1}{6}$&$-\frac{1}{6}$&$\frac{1}{3}$\\
\hline
39&$ \bar{u}_{R}^{\bar{c1}}$&$\stackrel{03}{(+i)}\,\stackrel{12}{[+]}|\stackrel{56}{(-)}\,\stackrel{78}{(+)}
||\stackrel{9 \;10}{[-]}\;\;\stackrel{11\;12}{(+)}\;\;\stackrel{13\;14}{(+)}$ &1&$\frac{1}{2}$&
$-\frac{1}{2}$&0 &$-\frac{1}{2}$&$-\frac{1}{2\,\sqrt{3}}$&$-\frac{1}{6}$&$-\frac{1}{6}$&$-\frac{2}{3}$\\
\hline
40&$\bar{u}_{R}^{\bar{c1}}$&$\stackrel{03}{[-i]}\,\stackrel{12}{(-)}|\stackrel{56}{(-)}\,\stackrel{78}{(+)}
||\stackrel{9 \;10}{[-]}\;\;\stackrel{11\;12}{(+)}\;\;\stackrel{13\;14}{(+)}$
&1&$-\frac{1}{2}$&
$-\frac{1}{2}$&0&$-\frac{1}{2}$&$-\frac{1}{2\,\sqrt{3}}$&$-\frac{1}{6}$&$-\frac{1}{6}$&$-\frac{2}{3}$\\
\hline\hline
41&$ \bar{d}_{L}^{\bar{c2}}$&$ \stackrel{03}{[-i]}\,\stackrel{12}{[+]}|
\stackrel{56}{[+]}\,\stackrel{78}{(+)}
||\stackrel{9 \;10}{(+)}\;\;\stackrel{11\;12}{[-]}\;\;\stackrel{13\;14}{(+)} $
&-1&$\frac{1}{2}$&0&
$\frac{1}{2}$&$\frac{1}{2}$&$-\frac{1}{2\,\sqrt{3}}$&$-\frac{1}{6}$&$\frac{1}{3}$&$\frac{1}{3}$\\
\hline
42&$\bar{d}_{L}^{\bar{c2}}$&$\stackrel{03}{(+i)}\,\stackrel{12}{(-)}|\stackrel{56}{[+]}\,\stackrel{78}{(+)}
||\stackrel{9 \;10}{(+)}\;\;\stackrel{11\;12}{[-]}\;\;\stackrel{13\;14}{(+)}$
&-1&$-\frac{1}{2}$&0&
$\frac{1}{2}$&$\frac{1}{2}$&$-\frac{1}{2\,\sqrt{3}}$&$-\frac{1}{6}$&$\frac{1}{3}$&$\frac{1}{3}$\\
\hline
43&$\bar{u}_{L}^{\bar{c2}}$&$  \stackrel{03}{[-i]}\,\stackrel{12}{[+]}|\stackrel{56}{(-)}\,\stackrel{78}{[-]}
||\stackrel{9 \;10}{(+)}\;\;\stackrel{11\;12}{[-]}\;\;\stackrel{13\;14}{(+)}$
&-1&$\frac{1}{2}$&0&
$-\frac{1}{2}$&$\frac{1}{2}$&$-\frac{1}{2\,\sqrt{3}}$&$-\frac{1}{6}$&$-\frac{2}{3}$&$-\frac{2}{3}$\\
\hline
44&$ \bar{u}_{L}^{\bar{c2}} $&$  \stackrel{03}{(+i)}\,\stackrel{12}{(-)}|
\stackrel{56}{(-)}\,\stackrel{78}{[-]}
||\stackrel{9 \;10}{(+)}\;\;\stackrel{11\;12}{[-]}\;\;\stackrel{13\;14}{(+)} $
&-1&$-\frac{1}{2}$&0&
$-\frac{1}{2}$&$\frac{1}{2}$&$-\frac{1}{2\,\sqrt{3}}$&$-\frac{1}{6}$&$-\frac{2}{3}$&$-\frac{2}{3}$\\
\hline
45&$\bar{d}_{R}^{\bar{c2}}$&$\stackrel{03}{(+i)}\,\stackrel{12}{[+]}|\stackrel{56}{[+]}\,\stackrel{78}{[-]}
||\stackrel{9 \;10}{(+)}\;\;\stackrel{11\;12}{[-]}\;\;\stackrel{13\;14}{(+)}$
&1&$\frac{1}{2}$&
$\frac{1}{2}$&0&$\frac{1}{2}$&$-\frac{1}{2\,\sqrt{3}}$&$-\frac{1}{6}$&$-\frac{1}{6}$&$\frac{1}{3}$\\
\hline
46&$\bar{d}_{R}^{\bar{c2}} $&$  \stackrel{03}{[-i]}\,\stackrel{12}{(-)}|\stackrel{56}{[+]}\,\stackrel{78}{[-]}
||\stackrel{9 \;10}{(+)}\;\;\stackrel{11\;12}{[-]}\;\;\stackrel{13\;14}{(+)} $
&1&$-\frac{1}{2}$&
$\frac{1}{2}$&0&$\frac{1}{2}$&$-\frac{1}{2\,\sqrt{3}}$&$-\frac{1}{6}$&$-\frac{1}{6}$&$\frac{1}{3}$\\
\hline
47&$ \bar{u}_{R}^{\bar{c2}}$&$\stackrel{03}{(+i)}\,\stackrel{12}{[+]}|\stackrel{56}{(-)}\,\stackrel{78}{(+)}
||\stackrel{9 \;10}{(+)}\;\;\stackrel{11\;12}{[-]}\;\;\stackrel{13\;14}{(+)}$
 &1&$\frac{1}{2}$&
$-\frac{1}{2}$&0 &$\frac{1}{2}$&$-\frac{1}{2\,\sqrt{3}}$&$-\frac{1}{6}$&$-\frac{1}{6}$&$-\frac{2}{3}$\\
\hline
48&$\bar{u}_{R}^{\bar{c2}}$&$\stackrel{03}{[-i]}\,\stackrel{12}{(-)}|\stackrel{56}{(-)}\,\stackrel{78}{(+)}
||\stackrel{9 \;10}{(+)}\;\;\stackrel{11\;12}{[-]}\;\;\stackrel{13\;14}{(+)}$
&1&$-\frac{1}{2}$&
$-\frac{1}{2}$&0&$\frac{1}{2}$&$-\frac{1}{2\,\sqrt{3}}$&$-\frac{1}{6}$&$-\frac{1}{6}$&$-\frac{2}{3}$\\
\hline\hline
49&$ \bar{d}_{L}^{\bar{c3}}$&$ \stackrel{03}{[-i]}\,\stackrel{12}{[+]}|
\stackrel{56}{[+]}\,\stackrel{78}{(+)}
||\stackrel{9 \;10}{(+)}\;\;\stackrel{11\;12}{(+)}\;\;\stackrel{13\;14}{[-]} $ &-1&$\frac{1}{2}$&0&
$\frac{1}{2}$&$0$&$\frac{1}{\sqrt{3}}$&$-\frac{1}{6}$&$\frac{1}{3}$&$\frac{1}{3}$\\
\hline
50&$\bar{d}_{L}^{\bar{c3}}$&$\stackrel{03}{(+i)}\,\stackrel{12}{(-)}|\stackrel{56}{[+]}\,\stackrel{78}{(+)}
||\stackrel{9 \;10}{(+)}\;\;\stackrel{11\;12}{(+)}\;\;\stackrel{13\;14}{[-]} $&-1&$-\frac{1}{2}$&0&
$\frac{1}{2}$&$0$&$\frac{1}{\sqrt{3}}$&$-\frac{1}{6}$&$\frac{1}{3}$&$\frac{1}{3}$\\
\hline
51&$\bar{u}_{L}^{\bar{c3}}$&$  \stackrel{03}{[-i]}\,\stackrel{12}{[+]}|\stackrel{56}{(-)}\,\stackrel{78}{[-]}
||\stackrel{9 \;10}{(+)}\;\;\stackrel{11\;12}{(+)}\;\;\stackrel{13\;14}{[-]} $&-1&$\frac{1}{2}$&0&
$-\frac{1}{2}$&$0$&$\frac{1}{\sqrt{3}}$&$-\frac{1}{6}$&$-\frac{2}{3}$&$-\frac{2}{3}$\\
\hline
52&$ \bar{u}_{L}^{\bar{c3}} $&$  \stackrel{03}{(+i)}\,\stackrel{12}{(-)}|
\stackrel{56}{(-)}\,\stackrel{78}{[-]}
||\stackrel{9 \;10}{(+)}\;\;\stackrel{11\;12}{(+)}\;\;\stackrel{13\;14}{[-]}  $&-1&$-\frac{1}{2}$&0&
$-\frac{1}{2}$&$0$&$\frac{1}{\sqrt{3}}$&$-\frac{1}{6}$&$-\frac{2}{3}$&$-\frac{2}{3}$\\
\hline
53&$\bar{d}_{R}^{\bar{c3}}$&$\stackrel{03}{(+i)}\,\stackrel{12}{[+]}|\stackrel{56}{[+]}\,\stackrel{78}{[-]}
||\stackrel{9 \;10}{(+)}\;\;\stackrel{11\;12}{(+)}\;\;\stackrel{13\;14}{[-]} $&1&$\frac{1}{2}$&
$\frac{1}{2}$&0&$0$&$\frac{1}{\sqrt{3}}$&$-\frac{1}{6}$&$-\frac{1}{6}$&$\frac{1}{3}$\\
\hline
54&$\bar{d}_{R}^{\bar{c3}} $&$  \stackrel{03}{[-i]}\,\stackrel{12}{(-)}|\stackrel{56}{[+]}\,\stackrel{78}{[-]}
||\stackrel{9 \;10}{(+)}\;\;\stackrel{11\;12}{(+)}\;\;\stackrel{13\;14}{[-]} $&1&$-\frac{1}{2}$&
$\frac{1}{2}$&0&$0$&$\frac{1}{\sqrt{3}}$&$-\frac{1}{6}$&$-\frac{1}{6}$&$\frac{1}{3}$\\
\hline
55&$ \bar{u}_{R}^{\bar{c3}}$&$\stackrel{03}{(+i)}\,\stackrel{12}{[+]}|\stackrel{56}{(-)}\,\stackrel{78}{(+)}
||\stackrel{9 \;10}{(+)}\;\;\stackrel{11\;12}{(+)}\;\;\stackrel{13\;14}{[-]} $ &1&$\frac{1}{2}$&
$-\frac{1}{2}$&0 &$0$&$\frac{1}{\sqrt{3}}$&$-\frac{1}{6}$&$-\frac{1}{6}$&$-\frac{2}{3}$\\
\hline
56&$\bar{u}_{R}^{\bar{c3}}$&$\stackrel{03}{[-i]}\,\stackrel{12}{(-)}|\stackrel{56}{(-)}\,\stackrel{78}{(+)}
||\stackrel{9 \;10}{(+)}\;\;\stackrel{11\;12}{(+)}\;\;\stackrel{13\;14}{[-]} $&1&$-\frac{1}{2}$&
$-\frac{1}{2}$&0&$0$&$\frac{1}{\sqrt{3}}$&$-\frac{1}{6}$&$-\frac{1}{6}$&$-\frac{2}{3}$\\
\hline\hline
57&$ \bar{e}_{L}$&$ \stackrel{03}{[-i]}\,\stackrel{12}{[+]}|
\stackrel{56}{[+]}\,\stackrel{78}{(+)}
||\stackrel{9 \;10}{[-]}\;\;\stackrel{11\;12}{[-]}\;\;\stackrel{13\;14}{[-]} $ &-1&$\frac{1}{2}$&0&
$\frac{1}{2}$&$0$&$0$&$\frac{1}{2}$&$1$&$1$\\
\hline
58&$\bar{e}_{L}$&$\stackrel{03}{(+i)}\,\stackrel{12}{(-)}|\stackrel{56}{[+]}\,\stackrel{78}{(+)}
||\stackrel{9 \;10}{[-]}\;\;\stackrel{11\;12}{[-]}\;\;\stackrel{13\;14}{[-]}$&-1&$-\frac{1}{2}$&0&
$\frac{1}{2}$ &$0$&$0$&$\frac{1}{2}$&$1$&$1$\\
\hline
59&$\bar{\nu}_{L}$&$  \stackrel{03}{[-i]}\,\stackrel{12}{[+]}|\stackrel{56}{(-)}\,\stackrel{78}{[-]}
||\stackrel{9 \;10}{[-]}\;\;\stackrel{11\;12}{[-]}\;\;\stackrel{13\;14}{[-]}$&-1&$\frac{1}{2}$&0&
$-\frac{1}{2}$&$0$&$0$&$\frac{1}{2}$&$0$&$0$\\
\hline
60&$ \bar{\nu}_{L} $&$  \stackrel{03}{(+i)}\,\stackrel{12}{(-)}|
\stackrel{56}{(-)}\,\stackrel{78}{[-]}
||\stackrel{9 \;10}{[-]}\;\;\stackrel{11\;12}{[-]}\;\;\stackrel{13\;14}{[-]} $&-1&$-\frac{1}{2}$&0&
$-\frac{1}{2}$&$0$&$0$&$\frac{1}{2}$&$0$&$0$\\
\hline
61&$\bar{\nu}_{R}$&$\stackrel{03}{(+i)}\,\stackrel{12}{[+]}|\stackrel{56}{(-)}\,\stackrel{78}{(+)}
||\stackrel{9 \;10}{[-]}\;\;\stackrel{11\;12}{[-]}\;\;\stackrel{13\;14}{[-]}$&1&$\frac{1}{2}$&
$-\frac{1}{2}$&0&$0$&$0$&$\frac{1}{2}$&$\frac{1}{2}$&$0$\\
\hline
62&$\bar{\nu}_{R} $&$  \stackrel{03}{[-i]}\,\stackrel{12}{(-)}|\stackrel{56}{(-)}\,\stackrel{78}{(+)}
||\stackrel{9 \;10}{[-]}\;\;\stackrel{11\;12}{[-]}\;\;\stackrel{13\;14}{[-]} $&1&$-\frac{1}{2}$&
$-\frac{1}{2}$&0&$0$&$0$&$\frac{1}{2}$&$\frac{1}{2}$&$0$\\
\hline
63&$ \bar{e}_{R}$&$\stackrel{03}{(+i)}\,\stackrel{12}{[+]}|\stackrel{56}{[+]}\,\stackrel{78}{[-]}
||\stackrel{9 \;10}{[-]}\;\;\stackrel{11\;12}{[-]}\;\;\stackrel{13\;14}{[-]}$ &1&$\frac{1}{2}$&
$\frac{1}{2}$&0 &$0$&$0$&$\frac{1}{2}$&$\frac{1}{2}$&$1$\\
\hline
64&$\bar{e}_{R}$&$\stackrel{03}{[-i]}\,\stackrel{12}{(-)}|\stackrel{56}{[+]}\,\stackrel{78}{[-]}
||\stackrel{9 \;10}{[-]}\;\;\stackrel{11\;12}{[-]}\;\;\stackrel{13\;14}{[-]}$&1&$-\frac{1}{2}$&
$\frac{1}{2}$&0&$0$&$0$&$\frac{1}{2}$&$\frac{1}{2}$&$1$\\
\hline
\end{supertabular}
}
\end{center}


Taking into account Table~\ref{Table so13+1.} and Eqs.~(\ref{snmb:gammatildegamma},
\ref{graphbinoms}) one easily finds  what do operators $\gamma^0\,\stackrel{78}{(\pm)}$ do
on the left handed and the right handed members of any family $i =(1,2,3,4)$.
\begin{eqnarray}
\label{leftright0}
\gamma^0\,\stackrel{78}{(-)} \,|\psi^{i}_{u_{R}, \nu_{R}}> &=& |\psi^{i}_{u_{L}, \nu_{L}}>\,,
\nonumber\\
\gamma^0\,\stackrel{78}{(+)} \,|\psi^{i}_{u_{L}, \nu_{L}}> &=&  |\psi^{i}_{u_{R}, \nu_{R}}>\,,
\nonumber\\
\gamma^0\,\stackrel{78}{(+)} \,|\psi^{i}_{d_{R}, e_{R}}> &=&  |\psi^{i}_{d_{L}, e_{L}}>\,,
\nonumber\\
\gamma^0\,\stackrel{78}{(-)} \,|\psi^{i}_{d_{L}, e_{L}}> &=&  |\psi^{i}_{d_{R}, e_{R}}>\,.
\end{eqnarray}

We need to know also what do operators ($\tilde{\tau}^{1 \pm}= \tilde{\tau}^{11} \pm i \,
\tilde{\tau}^{12}$, $ \tilde{\tau}^{13}$) and ($\tilde{N}^{\pm}_{L}= \tilde{N}^{1}_{L} \pm i \,
\tilde{N}^{2}_{L}$, $ \tilde{N}^{3}_{L}$) do when operating on any member ($u_{L,R}$,
$\nu_{L,R}$, $d_{L,R}$, $e_{L,R}$) of  a particular family $\psi^{i}$, $i =(1,2,3,4)$.

Taking into account, Eqs.~(\ref{signature}, \ref{signature1}, \ref{graphbinoms}, \ref{plusminus},
\ref{grapheigen}, \ref{so1+3tilde}, \ref{so42tilde}),
\begin{eqnarray}
\label{tildetau1nl}
\tilde{N}^{\pm}_{L} &=& - \stackrel{03}{\widetilde{(\mp i)}} \stackrel{12}{\widetilde{(\pm )}}\,,
\quad\quad\;\;
\widetilde{\tau}^{1\pm} = (\mp)\, \stackrel{56}{\widetilde{(\pm )}}
\stackrel{78}{\widetilde{(\mp )}}\,,
\nonumber\\
\tilde{N}^{3}_{L} &=&\frac{1}{2}\,(\tilde{S}^{12} + i\, \tilde{S}^{03})\,, \quad
\tilde{\tau}^{1 3}  = \frac{1}{2}\,(\tilde{S}^{56} - \tilde{S}^{78})\,,\nonumber\\
\stackrel{ab}{\widetilde{(-k)}}\,\stackrel{ab}{(k)}&=&  - i \,\eta^{aa}\,\stackrel{ab}{[k]}\,,
 \quad\quad
\stackrel{ab}{\widetilde{(k)}}\,\stackrel{ab}{(k)} =\, 0\,,\nonumber\\
\stackrel{ab}{\widetilde{(k)}}\,\stackrel{ab}{[k]} &=& i\,\stackrel{ab}{(k)}\,,\quad\quad\quad\;\,
\stackrel{ab}{\widetilde{(k)}}\,\stackrel{ab}{[-k]} =\, 0\,, \nonumber\\
\stackrel{ab}{\widetilde{(k)}} &=& \frac{1}{2}\,(\tilde{\gamma}^{a} + \frac{\eta^{aa}}{ik}
\,\tilde{\gamma}^{b})\,,
\quad  \;\, \stackrel{ab}{\widetilde{[k]}} \,= \frac{1}{2}\,(1 + \frac{i}{k}\,\tilde{\gamma}^{a}
 \tilde{\gamma}^{b})\,,
\end{eqnarray}

one finds 
\begin{eqnarray}
\label{tau1nl}
\tilde{N}_{L}^{+} \, |\psi^{1}> &=&  |\psi^{2}>\,, \quad \tilde{N}_{L}^{+} \, |\psi^{2}> =\,0\,,
\nonumber\\
\tilde{N}_{L}^{-} \, |\psi^{2}> &=&  |\psi^{1}>\,, \quad \tilde{N}_{L}^{-} \, |\psi^{1}> =\,0\,,
\nonumber\\
\tilde{N}_{L}^{+} \, |\psi^{3}> &=&  |\psi^{4}>\,, \quad \tilde{N}_{L}^{+} \, |\psi^{4}> =\,0\,,
\nonumber\\
\tilde{N}_{L}^{-} \, |\psi^{4}> &=&  |\psi^{3}>\,, \quad \tilde{N}_{L}^{-} \, |\psi^{3}> =\,0\,,
\nonumber\\
\tilde{\tau}^{1 +} \, |\psi^{1}> &=&  |\psi^{3}>\,, \quad \tilde{\tau}^{1 +} \, |\psi^{3}> =\,0\,,
\nonumber\\
\tilde{\tau}^{1 - } \, |\psi^{3}> &=&  |\psi^{1}>\,, \quad \tilde{\tau}^{1 -} \, |\psi^{1}> =\,0\,,
\nonumber\\
\tilde{\tau}^{1 - } \, |\psi^{4}> &=&  |\psi^{2}>\,, \quad \tilde{\tau}^{1 -} \, |\psi^{2}> =\,0\,,
\nonumber\\
\tilde{\tau}^{1 +} \, |\psi^{2}> &=&  |\psi^{4}>\,, \quad \tilde{\tau}^{1 +} \, |\psi^{4}> =\,0\,,
\nonumber\\
\tilde{N}_{L}^{3} \, |\psi^{1}> &=& - \frac{1}{2} \,|\psi^{1}>\,,\quad
\tilde{N}_{L}^{3} \, |\psi^{2}> = +\frac{1}{2}\, |\psi^{2}>\,,\nonumber\\
\tilde{N}_{L}^{3} \, |\psi^{3}> &=& - \frac{1}{2}\, |\psi^{3}>\,,\quad
\tilde{N}_{L}^{3} \, |\psi^{4}> = +\frac{1}{2} \,|\psi^{4}>\,,\nonumber\\
\tilde{\tau}^{1 3} \, |\psi^{1}> &=&  - \frac{1}{2} \,|\psi^{1}>\,, \quad
\tilde{\tau}^{1 3} \, |\psi^{2}> =  - \frac{1}{2} \,|\psi^{2}>\,, \nonumber\\
\tilde{\tau}^{1 3} \, |\psi^{3}> &=&  + \frac{1}{2} \,|\psi^{3}>\,, \quad
\tilde{\tau}^{1 3} \, |\psi^{4}> =  + \frac{1}{2} \,|\psi^{4}>\,,
\end{eqnarray}
independent of the family member $\alpha=(u,d,\nu,e)$.

The dependence of the mass matrix on the family quantum numbers can easily be understood 
through Table~\ref{Table 4families.}, where we notice that the operator $\tilde{N}_{L}^{\spm}$
transforms the first family into the second (or the second family into the first) and the third family
to the fourth (or the fourth family into the third), while the operator 
$\tilde{\tau}^{\tilde{1} \spm}$ transforms the first family into the third (or the third family into the 
first) and the second family into the fourth (or the fourth family into the second). 
The application of these two operators, $\tilde{N}_{L}^{\spm}$ and  $\tilde{\tau}^{\tilde{1} \spm}$,
is presented in Eq.~(\ref{tau1nl}) and demonstrated in the diagram
\begin{equation}
\label{diagramNtau}
      \stackrel{\stackrel{\tilde{N}_{L}^{\spm}}{\leftrightarrow} }{\begin{pmatrix}
 \psi^{1} & \psi^{2} \\
\psi^{3} & \psi^{4}   \end{pmatrix}} \updownarrow  \tilde{\tau}^{\tilde{1}\spm}\,.
\end{equation}
%



The operators  
$\tilde{N}_{L}^{3}$ and $\tilde{\tau}^{\tilde{1} 3}$ 
are diagonal, with the eigenvalues  presented in Eq.~(\ref{tau1nl}): 
$\tilde{N}_{L}^{3}$ has the eigenvalue $-\frac{1}{2}$ on $|\psi^{1}>$ and  
$|\psi^{3}>$ and $+\frac{1}{2}$ on $|\psi^{2}>$ and  $|\psi^{4}>$,  while 
$\tilde{\tau}^{\tilde{1} 3}$ has the  eigenvalue $ -\frac{1}{2}$ on $|\psi^{1}>$ and  
$|\psi^{2}>$ and $+\frac{1}{2}$ on $|\psi^{3}>$ and  $|\psi^{4}>$.
 If we count $\frac{1}{2}$ as a part 
of these diagonal fields, then the eigenvalues of both operators on families differ only in the sign.  

 The sign and the values of $Q, Q'$ and $Y'$ 
depend on the family members properties and are the same for all the families. 


Let  the scalars ($\tilde{A}^{N_{L} \spm}_{\stackrel{78}{(\pm)}}$, 
$\tilde{A}^{N_{L} 3}_{\stackrel{78}{(\pm)}}$, $\tilde{A}^{1 \spm}_{\stackrel{78}{(\pm)}}$,
 $\tilde{A}^{1 3}_{\stackrel{78}{(\pm)}}$) be scalar gauge fields of the
operators ($\tilde{N}_{L}^{\pm} $, $\tilde{N}_{L}^{3}$, $\tilde{\tau}^{1 \pm}$,
$\tilde{\tau}^{1 3}$), respectively. Here $\tilde{A}_{\stackrel{78}{(\pm)}}=
 \tilde{A}_{7} \mp i \, \tilde{A}_{8}$
for all the scalar gauge fields, while $\tilde{A}^{N_{L} \spm}_{\stackrel{78}{(\pm)}} =$
$\frac{1}{2}\,(\tilde{A}^{N_{L} 1}_{\stackrel{78}{(\pm)}} \mp i 
\,\tilde{A}^{N_{L} 2}_{\stackrel{78}{(\pm)}})$, respectively,
and  $\tilde{A}^{1 \spm}_{\stackrel{78}{(\pm)}}= $ $\frac{1}{2}\,
(\tilde{A}^{1 1}_{\stackrel{78}{(\pm)}} \mp i \,
\tilde{A}^{12 }_{\stackrel{78}{(\pm)}}$), respectively.
All these fields can be expressed by $\tilde{\omega}_{abc}$, as presented  in
Eq.~(\ref{gaugescalarAiomega}), provided that the coupling constants are the same for all the 
spin connection fields of both kinds, that is if no spontaneous symmetry breaking happens up to the 
weak scale.

We shall from now on use the notation $A^{Ai}_{\pm}$ instead of 
$A^{Ai}_{\stackrel{78}{(\pm)}}$ for all the operators with the space index $(7,8)$.


In what follows
{\it we prove that the symmetry of the  mass matrix of any family member
$\alpha$ remains the same in all orders of loop corrections, while the symmetry in all 
orders of corrections (which includes besides the loop corrections also the repetition of
nonzero vacuum expectation values of the scalar fields) remains unchanged only under 
certain conditions. In general case the break of symmetry can still be evaluated for 
small absolute values of $a^{\alpha}$}, Eq.~(\ref{a}). {\it We shall work in the massless basis.}


Let us introduce the notation  $\hat{O}$ for the operator, which in
 Eq.~(\ref{factionmass})
determines the mass matrices of quarks and leptons.
The operator $\hat{O}$ is equal to, Eq.~(\ref{factionmass}),
\begin{eqnarray}
\label{O}
\hat{O}&=& 
 \sum_{+,-}
\gamma^0 \stackrel{78}{(\pm)} \,( - \sum_{\alpha}  \, \tau^{\alpha}\, A^{\alpha}_{\pm} -
\sum_{\tilde{A}i}\, \tilde{\tau}^{\tilde{A}i}\, \tilde{A}^{\tilde{A}i}_{\pm})\,,
\nonumber\\
\tau^{\alpha}\, A^{\alpha}_{\pm} &=& (Q\, A^{Q}_{\pm}, Q'\, A^{Q'}_{\pm}, Y'\, A^{Y'}_{\pm})\,,
\nonumber\\
\tilde{\tau}^{\tilde{A}i}\,\tilde{A}^{\tilde{A}i}_{\pm} &=&
 (\tilde{\tau}^{\tilde{1}i}\,\tilde{A}^{\tilde{1}i}_{\pm},
\tilde{N}_{L}^{i}\, \tilde{A}^{\tilde{N}_{L}i}_{\pm})\,,\nonumber\\
\{\tau^{\alpha}, \tau^{\beta}\}_{-}&=& 0 \,, \quad \{\tilde{\tau}^{\tilde{A}i},
\tilde{\tau}^{\tilde{B}j}\}_{-}= i\, \delta^{\tilde{A}\tilde{B}}\,
 f^{ijk} \,\tilde{\tau}^{\tilde{A}k}\,,\quad \{\tau^{\alpha}, \tilde{\tau}^{\tilde{Bj}}\}_{-}=0\,.
\end{eqnarray}
Each of the fields in Eq.~(\ref{O}) consists in general of the nonzero vacuum expectation value 
and the dynamical part: $\tilde{A}^{\tilde{A}i}_{\pm}=$ $(<\tilde{A}^{\tilde{1}i}_{\pm}>+
\tilde{A}^{\tilde{1}i}_{\pm} (x)$, $<\tilde{A}^{\tilde{N}_{L}i}_{\pm}>+
\tilde{A}^{\tilde{N}_{L}i}_{\pm} (x)$, $<A^{\alpha}_{\pm}>+ A^{\alpha}_{\pm} (x)$), where
a common notation for all three singlets is used, since their eigenvalues depend only 
on the family members $(\alpha= (u,d,\nu,e))$ quantum numbers and are the same for all the 
families.

We further find that
\begin{eqnarray}
\label{Ocom}
\{\gamma^0 \stackrel{78}{(\pm)}, \,\tau^{4}\}_{-}&=& 0\,,\quad
\{\gamma^0 \stackrel{78}{(\pm)},\,\vec{\tilde{\tau}}^{\tilde{1}}\}_{-}= 0\,,\quad
\{\gamma^0 \stackrel{78}{(\pm)},\, \vec{\tilde{N}}_{L}\}_{-}= 0\,,\nonumber\\
\{\gamma^0 \stackrel{78}{(\pm)},\, \tau^{13}\}_{-} &=& - 2\, \gamma^0
\stackrel{78}{(\pm)}\, S^{78}\,, \quad \{\gamma^0  \stackrel{78}{(\pm)}, \,\tau^{23}\}_{-}
= + 2\, \gamma^0 \stackrel{78}{(\pm)}\, S^{78}\,.
\end{eqnarray}
%


%
To calculate the mass matrices of family members $\alpha =(u,d,\nu,e)$ the operator $\hat{O}$
must be taken into account in all orders. Since for our proof the dependence of the operator
$\hat{O}$ on the time and space does not play any role (it is the same for all the operators), we 
introduce the dimensionless operator 
$\mathbf{\hat{O}}$, in which all the degrees of freedom, except the internal ones determined by 
the family and family members quantum numbers, are integrated away~\footnote{$\hat{O}$ 
is measured in $TeV$ units (as all the scalar and vector gauge fields).
If the time evolution is concerned then $\mathbf{\hat{O}}$ $=\hat{O}\cdot (t-t_{0})/ TeV $ is in 
units $\hbar=1 =c$ dimensionless quantity. 
We assume that also the integration over space coordinates is in  
$<\psi^{\alpha \,i}_R|\hat{O}|\psi^{\alpha \,i}_R>$ already taken into account, only the integration  
over the family and family members  is left to be evaluated.}.

\begin{figure}
\centering\includegraphics[width=\textwidth]{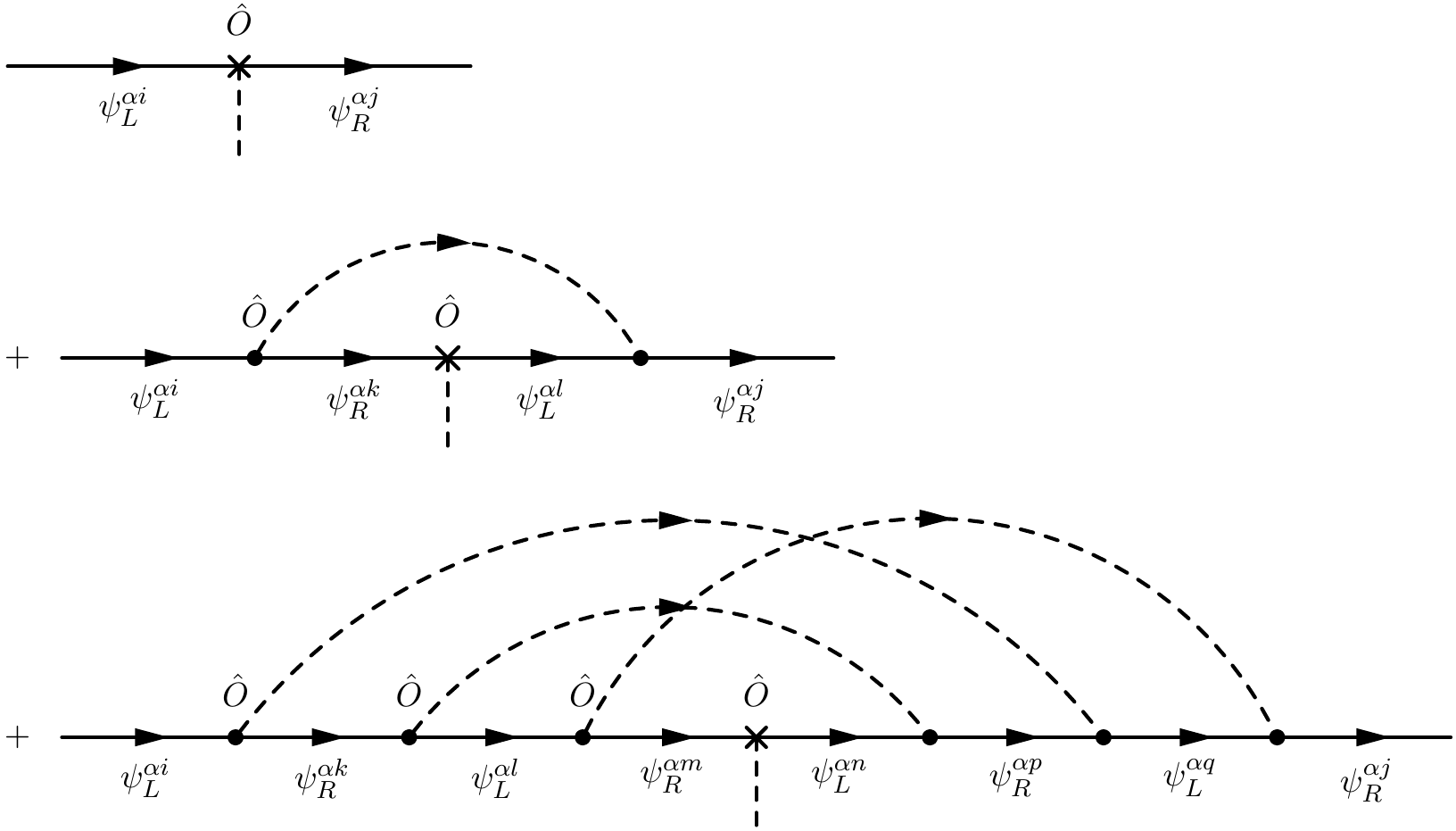}
\caption{\label{Figgeneral} The tree level contributions, one loop contributions
 (not all possibilities  are drawn, the tree level contributions
occurs namely also to the left or to the right of the loop, while to $\mathbf{\hat{O}}$ three singlets 
and two triplets, presented in Eq.~(\ref{O}), contribute) and two loop contributions are drawn
(again not all the possibilities are shown up).
Each $(i,j,k,l,m\dots)$ determines a family quantum number (running within the four families --- 
$(1,2,3,4)$), $\alpha$ denotes one of the family members ($\alpha=(u,\nu,d,e)$) quantum 
numbers, all in the massless basis $\psi^{i \alpha}_{(R,L)}$. Dynamical fields start and end 
with dots ${\bullet}$, while ${\bf x}$ with the  vertical slashed line represents the interaction 
of the fermion fields with the nonzero vacuum expectation values of the scalar fields.
}
\end{figure}

Then the change of the massless state of the $i^{th}$ family of the family member $\alpha$
of the  left or right handedness (${}_{L,R}$), $|\psi^{\alpha \,i}_{L,R}>$, changes in all orders 
of corrections as follows
\begin{eqnarray}
\label{U}
\hat{U} \,|\psi^{\alpha \,i}_{L,R}>&=& i\, \sum_{n=0}^{\infty} \frac{(-1)^n\,
\mathbf{\hat{O}}^{2n+1}}{(2n+1)!}\,|\psi^{\alpha \,i}_{L,R}>\,.
\end {eqnarray}
In Eq.~(\ref{U}) $|\psi^{\alpha \,i}_{(L,R)}>$ represents the internal degrees of freedom of 
the $i^{th}$, $i=(1,2,3,4)$, family state for a particular family member $\alpha$ in the massless
basis. The mass matrix element in all orders of corrections between the left handed $\alpha^{th}$ 
family member of the $i^{th}$ family $<\psi^{\alpha \,i}_L|$ and the right handed $\alpha^{th}$ 
family member of the $j^{th}$ family $|\psi^{\alpha \,j}_R>$, both in the massless basis,
is then equal to  $<\psi^{\alpha \,i}_L|$ $\hat{U} \,|\psi^{\alpha \,i}_R>$.
Only an odd number of operators $\mathbf{\hat{O}}^{2n+1}$ contribute to the mass matrix elements,
transforming $|\psi^{\alpha \,i}_R>$ into $|\psi^{\alpha \,j}_L>$ or opposite. The product of 
an even number of operators  $\mathbf{\hat{O}}^{2n}$ does not change the handedness and 
consequently contributes nothing.
Correspondingly without the nonzero vacuum expectation values of scalar fields all the matrix 
elements would remain zero, since only nonzero vacuum expectation values may appear in an 
odd orders, while the contribution of the loop corrections always contribute to the mass matrix 
elements an even contribution (see Fig.~(\ref{Figgeneral})).

Our purpose is to show how do the matrix elements behave in all orders of corrections
\begin{eqnarray}
\label{Upsi}
<\psi^{\alpha \,j}_L| \hat{U} \,|\psi^{\alpha \,i}_R>&=& i\, \sum_{n=0}^{\infty}
\frac{(-1)^n}{(2n+1)!}\,<\psi^{\alpha \,i}_L| \sum_{k_1=1}^{4} \mathbf{\hat{O}}|
\psi^{\alpha \,k_1}_R><\psi^{\alpha \,k_1}_R| \sum_{k_2=1}^{4} \mathbf{\hat{O}}|
\psi^{\alpha \,k_2}_L>\cdots
\nonumber\\
& &<\psi^{\alpha \,k_n}_L| \sum_{k_i =1}^{4} \mathbf{\hat{O}}|\psi^{\alpha \,k_i}_R>\,.
\end {eqnarray}
%
Let be repeated again that all the matrix elements $<\psi^{\alpha \,k_1}_R| \mathbf{\hat{O}}|$%
$\psi^{\alpha \,k_2}_{L}>$ or $~<~\psi^{\alpha \,k_1}_L| \sum_{k_2=1}^{4} \mathbf{\hat{O}}|$%
$\psi^{\alpha \,k_2}_{R}>$ only evaluate the internal degrees of freedom, that is the family and 
family members ones, while all the rest are assumed to be already evaluated. Since the mass 
matrix is in this notation the dimensionless object, also all the scalar fields are already divided by the
energy unit (let say 1 TeV). We correspondingly introduce the dimensionless scalars 
$ (\mathbf{A}^{Q}_{\pm}, \mathbf{A}^{Q'}_{\pm}, \mathbf{A}^{Y'}_{\pm})$,
$\Vec{\tilde{\mathbf{A}}}^{\tilde{1}}_{\pm},\Vec{\tilde{\mathbf{A}}}^{\tilde{N}_{L}}_{\pm}$.

The only operators $\tau^{\alpha}$, which distinguish among family members, are ($\tau^{4}, 
\tau^{13}, \tau^{23}$), included in $Q=(\tau^{13} +Y) $, $Y=(\tau^{23} + \tau^{4})$,
$Q'= (\tau^{13} - Y \tan^2\vartheta_1)$ and in  
$Y'=(\tau^{23} -\tau^{4} \tan^2\vartheta_2)$.
All the operators contributing to the mass matrices of each family member $\alpha$ have a 
factor $\gamma^0\, \stackrel{78}{(\pm)} $, which transforms the right handed family member
to the corresponding left  handed family member and opposite. 

When taking into account ${\bf \hat{O}}^{2n+1}$
in all orders, the operators  $\tau^{\alpha} \, A^{\alpha}_{\pm}$, $\tau^{\alpha}= (Q, Q', Y')$, 
contribute to all the matrix elements, the diagonal and the off diagonal ones.





To simplify the discussions let us introduce a bit more detailed notation
\begin{eqnarray}
\label{Oi}
\mathbf{\hat{O}} \;\;\;&=&\sum_{i} \mathbf{\hat{O}}^{i} =\mathbf{\hat{O}}^{\alpha} +
\mathbf{\hat{\tilde{O}}}{}^{\tilde{1}3} + \mathbf{\hat{\tilde{O}}}{}^{\tilde{N}_{L}3} + 
\mathbf{\hat{\tilde{O}}}{}^{\tilde{1} \spm} + \mathbf{\hat{\tilde{O}}}{}^{\tilde{N}_{L} \spm}\, 
\nonumber\\
\mathbf{\hat{O}}^{\alpha}\;\; &=& - \sum_{+,-}
\gamma^0 \stackrel{78}{(\pm)} \,( Q \,\mathbf{A}^{Q}_{\pm}, Q' \, \mathbf{A}^{Q'}_{\pm}, 
Y' \, \mathbf{A}^{Y'}_{\pm})\,,
\nonumber\\
\mathbf{\hat{\tilde{O}}}{}^{\tilde{1}3}\;\; &=& - \sum_{+,-} \gamma^0 \stackrel{78}{(\pm)}
\,\tilde{\tau}^{\tilde{1}3}\, \mathbf{\tilde{A}}^{\tilde{1}3}_{\pm} \,,\nonumber\\
\mathbf{\hat{\tilde{O}}}{}^{\tilde{N}_{L}3}&=& - \sum_{+,-} \gamma^0 \stackrel{78}{(\pm)}
\,\tilde{N}_{L}^{3} \, \mathbf{\tilde{A}}^{\tilde{N}_{L}3}_{\pm}\,,\nonumber\\
\mathbf{\hat{\tilde{O}}}{}^{\tilde{1} \spm} \;&=& - \sum_{+,-} \gamma^0 \stackrel{78}{(\pm)}
\,\tilde{\tau}^{\tilde{1}\spm}\, \mathbf{\tilde{A}}^{\tilde{1}\spm}_{\pm} \,,\nonumber\\
\mathbf{\hat{\tilde{O}}}{}^{\tilde{N}_{L} \spm} &=& - \sum_{+,-} \gamma^0 \stackrel{78}{(\pm)}
\,\tilde{N}_{L}^{\spm}\, \mathbf{\tilde{A}}^{\tilde{N}_{L}\spm}_{\pm}\,.
\end{eqnarray}
%
We shall use the notation for the expectation values among the states 
$<\psi^{i}_{L}| =<i|, |\psi^{j}_{R}>= |j>$  for the zero vacuum expectation values and the 
dynamical parts as follows:\\
$\;\;{\bf i.}$ $<i|\mathbf{\hat{O}}^{\alpha}|j> \;\,=<i| \sum_{+,-} \gamma^0 \stackrel{78}{(\pm)} 
\tau^{\alpha} (<\mathbf{A}^{\alpha}_{\pm}> + \mathbf{A}^{\alpha}_{\pm}(x))|j> $. \\
$\;{\bf ii.}$  $<i|\mathbf{\hat{\tilde{O}}}^{\tilde{1}3}|j>\; =<i| - \sum_{+,-} \gamma^0 
\stackrel{78}{(\pm)} \tilde{\tau}^{\tilde{1}3} (<\mathbf{\tilde{A}}^{\tilde{1}3}_{\pm}> + 
\mathbf{\tilde{A}}^{\tilde{1}3}_{\pm}(x))|j>$.\\
${\bf iii.}$ $<i| \mathbf{\hat{\tilde{O}}}^{\tilde{N}_{L}3} |j>= <i|- \sum_{+,-} \gamma^0 
\stackrel{78}{(\pm)} \tilde{N}_{L}^{3} (<\mathbf{\tilde{A}}^{\tilde{N}_{L}3}_{\pm}> + 
\mathbf{\tilde{A}}^{\tilde{N}_{L}3}_{\pm}(x))|j>$.\\
${\bf iv.}$ $<i|\mathbf{\hat{\tilde{O}}}^{\tilde{1} \spm} |j>\;\;= <i| - \sum_{+,-} \gamma^0 
\stackrel{78}{(\pm)} \tilde{\tau}^{\tilde{1}\spm} (<\mathbf{\tilde{A}}^{\tilde{1}\spm}_{\pm}>  + 
\mathbf{\tilde{A}}^{\tilde{1}\spm}_{\pm}(x))|j>$.\\
${\bf v.}$ $<i| \mathbf{\hat{\tilde{O}}}^{\tilde{N}_{L} \spm} |j>= <i|- \sum_{+,-} \gamma^0 
\stackrel{78}{(\pm)} \tilde{N}_{L}^{\spm} (<\mathbf{\tilde{A}}^{\tilde{N}_{L}\spm}_{\pm}> +
\mathbf{\tilde{A}}^{\tilde{N}_{L}\spm}_{\pm}(x))|j>$.\\
${\bf vi.}$ $<i|\mathbf{\hat{\tilde{O}}}^{\alpha}_{\dia}|i>= <i| \sum_{+,-} \gamma^0 
\stackrel{78}{(\pm)} 
\{\tau^{\alpha} (<\mathbf{A}^{\alpha}_{\pm}> + \mathbf{A}^{\alpha}_{\pm}$ $(x)) - $  
$\tilde{\tau}^{\tilde{1}3} (<\tilde{\mathbf{A}}^{\tilde{1}3}_{\pm}>+ 
\tilde{\mathbf{A}}^{\tilde{1}3}_{\pm}$$(x)) -
\tilde{N}_{L}^{3} (<\tilde{\mathbf{A}}^{\tilde{N}_{L}3}_{\pm}>+ 
\tilde{\mathbf{A}}^{\tilde{N}_{L}3}_{\pm}$$(x)) \}|i>$.\\
($<\mathbf{A}^{\alpha}_{\pm}>, <\mathbf{\tilde{A}}^{\tilde{1}3}_{\pm}>, 
<\mathbf{\tilde{A}}^{\tilde{N}_{L}3}_{\pm}>, 
<\mathbf{\tilde{A}}^{\tilde{1}\spm}_{\pm}>, <\mathbf{\tilde{A}}^{\tilde{N}_{L}\spm}_{\pm}>$) 
represent nonzero vacuum expectation values  and ($\mathbf{A}^{\alpha}_{\pm}(x), 
 \mathbf{\tilde{A}}^{\tilde{1}3}_{\pm} $$(x),
\mathbf{\tilde{A}}^{\tilde{N}_{L}3}_{\pm} $$(x), \mathbf{\tilde{A}}^{\tilde{1}\spm}_{\pm} $$(x), 
 \mathbf{\tilde{A}}^{\tilde{N}_{L}\spm}_{\pm}$$ (x)$) the corresponding dynamical fields.

In the case ${\bf i.}$  $<{\mathbf{A}}^{\alpha}_{\pm}>$  represent
the sum of the vacuum expectation values of ($Q^{\alpha} {\mathbf{A}}^{Q}_{(\pm)}$,
 $Q'^{\alpha} {\bf A^{Q'}}_{(\pm)}$, $Y'^{\alpha} {\bf A}^{Y'}_{(\pm)}$)
%
of a particular family member $\alpha$, where ($Q^{\alpha}, Q'^{\alpha}, Y'^{\alpha}$) 
are the corresponding quantum numbers of a family member $\alpha$.
${\bf A}^{\alpha}_{\pm} (x)$ represent the corresponding dynamical fields.

In the case ${\bf vi.}$ we correspondingly have for the four diagonal terms on the tree level,
that is for $n=0$ in Eq.~(\ref{Upsi}) (after taking into account  Eq.~(\ref{tau1nl}):
 $<~1~|\mathbf{\tilde{O}}^{\alpha}_{\dia} |1>$ $= {\bf a}^{\alpha} - 
(\tilde{{\bf a}}_1 + \tilde{{\bf a}}_2)$, 
 $<2|\mathbf{\tilde{O}}^{\alpha}_{\dia} |1> |2>= {\bf a}^{\alpha} - (\tilde{{\bf a}}_1 -
 \tilde{{\bf a}}_2)$,
 $<3| \mathbf{\tilde{O}}^{\alpha}_{\dia} |3>= {\bf a}^{\alpha} + (\tilde{{\bf a}}_1 - 
\tilde{{\bf a}}_2)$ and
 $<4| \mathbf{\tilde{O}}^{\alpha}_{\dia} |4>= {\bf a}^{\alpha} + (\tilde{{\bf a}}_1 +
 \tilde{{\bf a}}_2)$,
where $(\tilde{{\bf a}}_1, \tilde{{\bf a}}_2, {\bf a}^{\alpha})$ 
represent the nonzero vacuum expectation values of  $\frac{1}{2}\, 
\frac{1}{\sqrt{2}}\, (<\mathbf{\tilde{A}}^{\tilde{1}3}_{(+)}> +
<\mathbf{\tilde{A}}^{\tilde{1} 3}_{(-)}>)$, $ \frac{1}{2}\,  \frac{1}{\sqrt{2}}\,
(<\mathbf{\tilde{A}}^{\tilde{N}_{L}3}_{(+)}> + <\mathbf{\tilde{A}}^{\tilde{N}_{L}3}_{(-)}>)$,  
 $\frac{1}{2}\,\frac{1}{\sqrt{2 }}\,(<\mathbf{A}^{\alpha}_{(+)}> + <\mathbf{A}^{\alpha}_{(-)}>)$, all in 
dimensionless units.\\

{\it We are now prepared to show under which conditions the mass matrix elements for any of 
the family members
keep the symmetry $\widetilde{SU}(2) \times \widetilde{SU}(2) \times U(1)$ at each step of 
corrections}, what means that the values of the matrix elements obtained in each correction 
respect the symmetry of  mass matrices on the tree level.



{\it We use the massless basis $|\psi^{i}_{L,R}> $}, 
making for the basis the choice  $\frac{1}{\sqrt{2}}\,(|\psi^{i}_{L}>+\, |\psi^{i}_{R}>) $. 

The diagrams for the tree level, one loop and three loop contributions of the operator 
${\bf \hat{O}}$, 
determining the masses of quarks and leptons, Eqs.~(\ref{O}, \ref{Oi}), are presented in 
Fig.~(\ref{Figgeneral}). 

%


%
\subsection{Mass matrices on the tree level}
\label{treelevel}

Let us first present the mass matrix on the tree level for an $\alpha^{th}$ family member,
 that is for $n=0$ in Eq.~(\ref{Upsi}).

\begin{figure}
\centering\includegraphics[width=\textwidth]{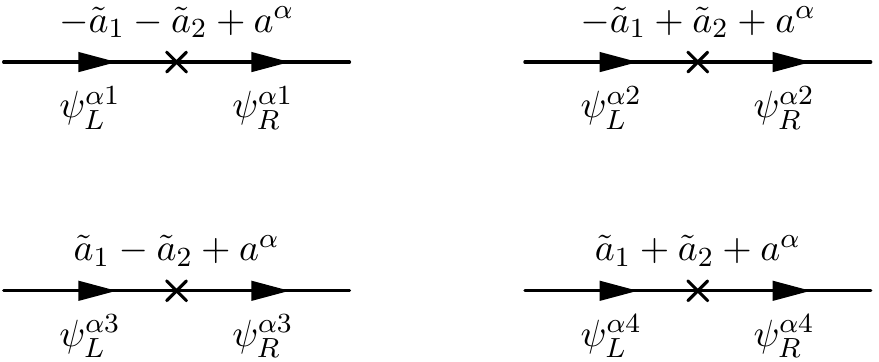}
\caption{\label{Figdiagonaltree} The tree level contributions to the diagonal matrix elements 
of the operator $\hat{O}^{\alpha}_{\dia}$, Eq.~(\ref{Oi}). 
The eigenvalues of the operators $\tilde{N}^{3}_{L}$ and  $\tilde{\tau}^{\tilde{1}3}$ on a family
state $i$ can be read in Eq.~(\ref{tau1nl}).
}
\end{figure}

Taking into account Eq.~(\ref{tau1nl})  one obtains for the diagonal matrix elements on the tree 
level (for $n=0$ in Eq.~(\ref{Upsi}))
 [${\bf a}^{\alpha} - (\tilde{{\bf a}}_1 + \tilde{{\bf a}}_2), {\bf a}^{\alpha} - (\tilde{{\bf a}}_1
 - \tilde{{\bf a}}_2), {\bf a}^{\alpha} +  (\tilde{{\bf a}}_1 - \tilde{{\bf a}}_2), {\bf a}^{\alpha}
 + (\tilde{{\bf a}}_1 + \tilde{{\bf a}}_2)$], respectively.
 The corresponding diagrams are presented 
in Fig.~(\ref{Figdiagonaltree}). 
%


Taking into account Eq.~(\ref{tau1nl}) one finds for the off diagonal elements on the tree level:\\
 $<\psi^1|..|\psi^2>$ $= <\psi^3|..|\psi^4>$ $=<\psi^2|..|\psi^1>^{\dagger} =$ 
$<\psi^4|..|\psi^3>^{\dagger}$ $ =<\mathbf{\tilde{A}}^{\tilde{N}_L\sminus}>$,\\
 $<\psi^1|..|\psi^3>$ $= <\psi^2|..|\psi^4>$ $=<\psi^3|..|\psi^1>^{\dagger} =$ 
$<\psi^4|..|\psi^2>^{\dagger}$ $ =<\mathbf{\tilde{A}}^{\tilde{1} \sminus}>$. 

\begin{figure}
\centering\includegraphics[width=\textwidth]{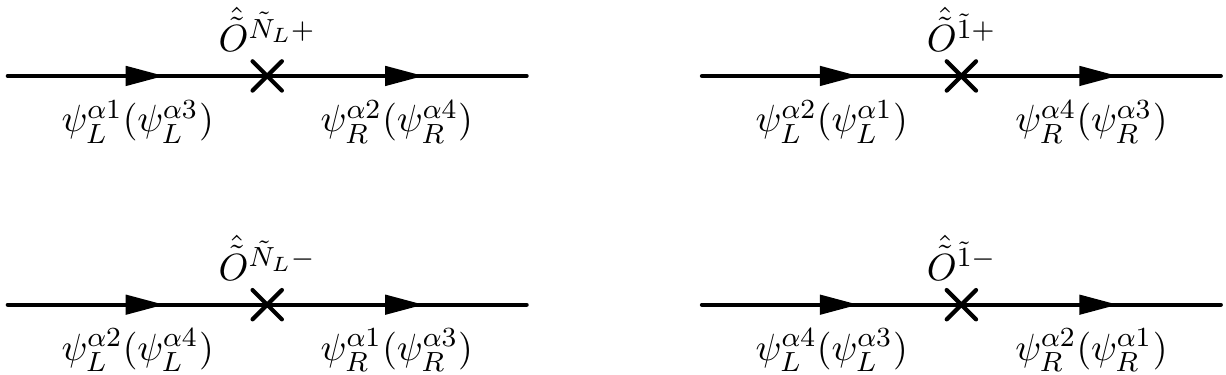}
\caption{\label{Fignodiagonaltree} The tree level contributions to the off diagonal matrix elements 
of the operators  $\hat{\tilde{O}}^{\tilde{1} \spm}$  
and    $\hat{\tilde{O}}^{\tilde{N}_{L} \spm}$, 
 Eq.~(\ref{Oi}) are presented.  
The application of the operators  $\tilde{N}_{L}^{\spm}$ 
and                                       $\tilde{\tau}^{\tilde{1}\spm}$  
on a family
state $i$ can be read in Eq.~(\ref{tau1nl}).
}
\end{figure}

The corresponding diagrams for $<\psi^1|..|\psi^2>$, $<\psi^2|..|\psi^1>$, 
$<\psi^2|..|\psi^3>$ and $<\psi^3|..|\psi^2>$
are presented in Fig.~(\ref{Fignodiagonaltree}). The vacuum expectation values of this matrix 
elements on the tree level are presented in the mass matrix of Eq.(\ref{M0a}). 

\begin{figure}[h]
\centering\includegraphics[width=0.6\textwidth]{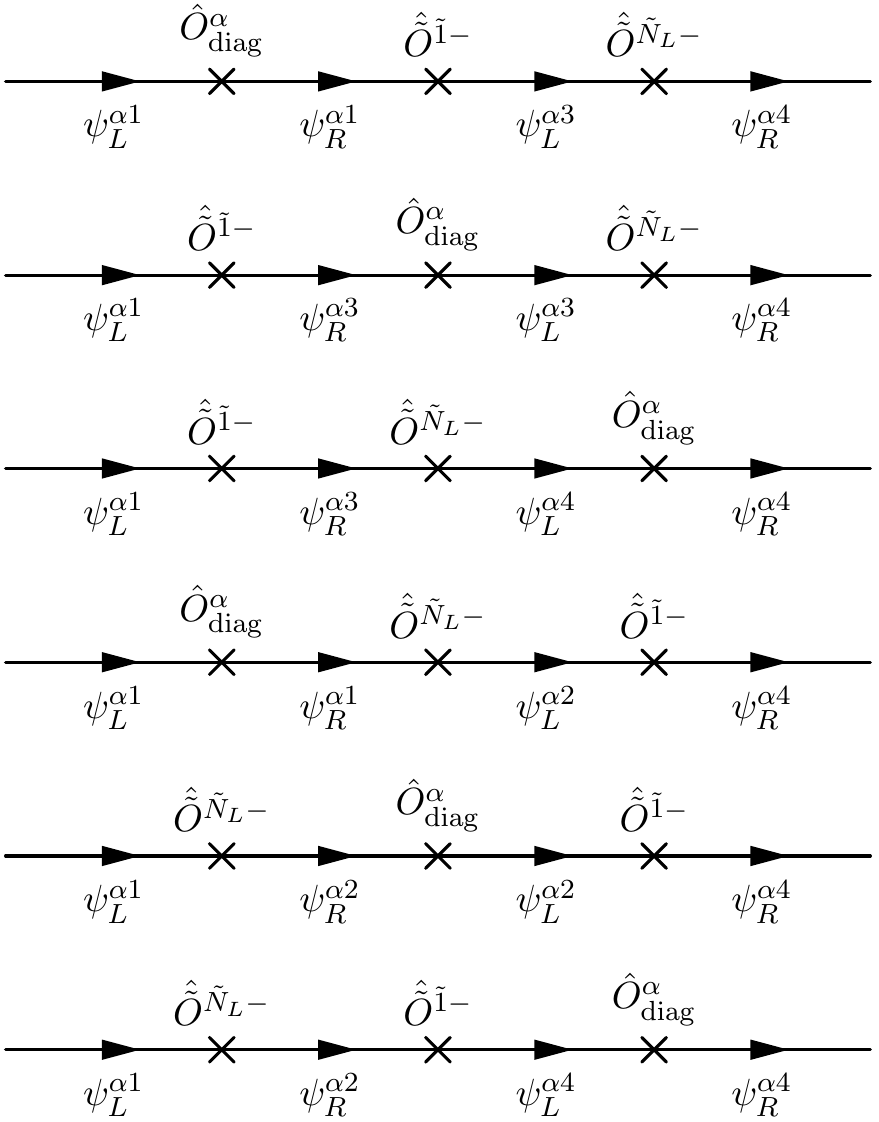}
\caption{\label{Figtreelevel14} The tree level contribution to the matrix element 
$<\psi^1|b|\psi^4>$} is presented. 
One comes from $<\psi^1|$ to $|\psi^4>$ in three steps: $<\psi^1|\sum_{+,-} 
\tilde{\tau}^{\tilde{1}\spm}\,<\tilde{{\bf A}}^{\tilde{1} \spm}>\,$ $\sum_{k}|\psi^k><\psi^k|
 \sum_{+,-} \,\tilde{N}_{L}^{\spm}\,<\tilde{{\bf A}}^{\tilde{N}_{L} \spm}>\,|\psi^4>$ 
$<\psi^4|\,(\tilde{{\bf a}}_1  + \tilde{{\bf a}}_2 + {\bf a}^{\alpha})
|\psi^4>$. 
There are all together six such terms, since the diagonal  term appears also at the beginning 
as $(- \tilde{{\bf a}}_1 - \tilde{{\bf a}}_2 + {\bf a}^{\alpha})$ and in the middle as 
$(\tilde{{\bf a}}_1  - \tilde{{\bf a}}_2 + {\bf a}^{\alpha}) $,
and since the operators $\sum_{+,-}\tilde{\tau}^{\tilde{1}\spm}\, 
<\tilde{{\bf A}}^{\tilde{1} \spm}>\,$ and $\sum_{+,-} \,\tilde{N}_{L}^{\spm} \,
<\tilde{{\bf A}}^{\tilde{N}_{L} \spm}>$ appear in the opposite order as well.
\end{figure}

The contributions to  the off diagonal matrix elements $<\psi^1|..|\psi^4>$, $<\psi^2|..|\psi^3>$, 
$<\psi^3|..|\psi^2>$ and $<\psi^4|..|\psi^1>$ are 
nonzero only, if one makes three steps 
 (not two, due to the left right jumps in each step), that is indeed in the third order of correction.
 For  $<\psi^1|..|\psi^4>$ we have
 (in the basis  $\frac{1}{\sqrt{2}}\,\,(|\psi^{i}_{L}> +\,|\psi^{i}_{R}>)$ and with the 
notation $<\mathbf{\tilde{A}}^{\tilde{N}_L\spm}>=$ $ \frac{1}{\sqrt{2}}\,
(<\mathbf{\tilde{A}}^{\tilde{N}_L \spm}_{(+)}> + <\mathbf{\tilde{A}}^{\tilde{N}_L \spm}_{(-)}>)$ 
after we take intoaccount that $\gamma^0 \stackrel{78}{(\pm)}$ transform the right handed 
family members into the left handed ones and opposite): 
$<\psi^1|\sum_{+,-}  \,\tilde{\tau}^{\tilde{1}\spm}\,
<\mathbf{\tilde{A}}^{\tilde{1} \spm}>\,$ $\sum_{k}|\psi^k><\psi^k| \sum_{+,-} \,
\tilde{N}_{L}^{\spm}<\mathbf{\tilde{A}}^{\tilde{N}_{L} \spm}>\,
|\psi^4>$ $<\psi^4|\,(\tilde{{\bf a}}_1 + \tilde{{\bf a}}_2 + {\bf a}^{\alpha})|\psi^4>$. 
There are all together six such terms, presented in Fig.~(\ref{Figtreelevel14}), 
since the diagonal  term appears also at the beginning as $(- \tilde{{\bf a}}_1 -
\tilde{{\bf a}}_2 + {\bf a}^{\alpha})$ and in the middle as $(\tilde{{\bf a}}_1  - \tilde{{\bf a}}_2 
+ {\bf a}^{\alpha}) $, and since the operators $\sum_{+,-}\tilde{\tau}^{\tilde{1}\spm}\,
<\mathbf{\tilde{A}}^{\tilde{1} \spm}>\,$ and $\sum_{+,-} \,\tilde{N}_{L}^{\spm} 
<\mathbf{\tilde{A}}^{\tilde{N}_{L} \spm}>$ appear in the opposite
 order as well. 
We simplify the notation from $|\psi^k>$ to $|k>$. Summing all these six terms for each of 
four matrix elements ($<1|..|4>$, $<2|..|3>$, $<3|..|2>$, $<4|..|1>$) one gets 
(taking into account Eqs.~(\ref{Upsi}, \ref{tau1nl})):
%
\begin{eqnarray}
\label{b}
<1|..|4>&=&  {\bf a}^{\alpha}\, <\mathbf{\tilde{A}}^{\tilde{1}\sminus}>\,
<\mathbf{\tilde{A}}^{\tilde{N}_L\sminus}>\,,\nonumber\\
<2|..|3>&=&  {\bf a}^{\alpha}\, <\mathbf{\tilde{A}}^{\tilde{1}\sminus}>\,
<\mathbf{\tilde{A}}^{\tilde{N}_L\splus}>\,,\nonumber\\
<3|..|2>&=&  {\bf a}^{\alpha}\, <\mathbf{\tilde{A}}^{\tilde{1}\splus}>\,
<\mathbf{\tilde{A}}^{\tilde{N}_L\sminus}>\,,\nonumber\\
<4|..|1>&=&  {\bf a}^{\alpha}\, <\mathbf{\tilde{A}}^{\tilde{1}\splus}>\,
<\mathbf{\tilde{A}}^{\tilde{N}_L\splus}>\,.
\end{eqnarray}
%
 Each matrix element is in Eq.~(\ref{b}) divided by $3!$, since it is the contribution in the third
order!
One notices that $<4|..|1>^{\dagger}= <1|..|4>$ and $<3|..|2>^{\dagger}= <2|..|3>$. 
These matrix elements are included into the mass matrix, Eq.~(\ref{M0a}).


%
To show up the symmetry of the mass matrix on the lowest level  we put all  the matrix elements
 in Eq.~(\ref{M0a}).

%
\begin{eqnarray}
\label{M0a}
&& \qquad \qquad \qquad  \qquad \qquad \qquad \qquad \qquad \qquad
{}^{\alpha}{\cal M}_{(o)} =\nonumber\\
&& \begin{pmatrix} - \tilde{{\bf a}}_1  - \tilde{{\bf a}}_2 + {\bf a}^{\alpha} &
<\mathbf{\tilde{A}}^{\tilde{N}_L\sminus}>& <\mathbf{\tilde{A}}^{\tilde{1}\sminus}> &
 {\bf a}^{\alpha}<\mathbf{\tilde{A}}^{\tilde{1}\sminus}> <\mathbf{\tilde{A}}^{\tilde{N}_L\sminus}>\\
<\mathbf{\tilde{A}}^{\tilde{N}_L\splus}>  & - \tilde{{\bf a}}_1 +   \tilde{{\bf a}}_2 +
 {\bf a}^{\alpha} &
{\bf a}^{\alpha}<\mathbf{\tilde{A}}^{\tilde{1}\sminus}> <\mathbf{\tilde{A}}^{\tilde{N}_L\splus}> &
<\mathbf{\tilde{A}}^{\tilde{1}\sminus}>\\
<\mathbf{\tilde{A}}^{\tilde{1}\splus}> &
{\bf a}^{\alpha}<\mathbf{\tilde{A}}^{\tilde{1}\splus}> <\mathbf{\tilde{A}}^{\tilde{N}_L\sminus}> &
 \tilde{{\bf a}}_1  - \tilde{{\bf a}}_2 + {\bf a}^{\alpha} & 
<\mathbf{\tilde{A}}^{\tilde{N}_L\sminus}>\\
 {\bf a}^{\alpha}<\mathbf{\tilde{A}}^{\tilde{1}\splus}> <\mathbf{\tilde{A}}^{\tilde{N}_L\splus}> &
<\mathbf{\tilde{A}}^{\tilde{1}\splus}>  & <\mathbf{\tilde{A}}^{\tilde{N}_L\splus}> &  \tilde{{\bf a}}_1  +
\tilde{{\bf a}}_2 + {\bf a}^{\alpha}
\end{pmatrix}
\nonumber\\
\end{eqnarray}
%
Mass matrix is dimensionless.
One notices that the diagonal terms have on the tree level the symmetry $<\psi^1|..|\psi^1>+
~<~\psi^4|..|\psi^4>=2\, a^{\alpha}= $ $<\psi^2|..|\psi^2>+<\psi^3|..|\psi^3> $,
and that in the off diagonal elements with "three steps 
needed" the contribution of the fields, which depend on particular
family member $\alpha=(u,d,\nu,e)$, enters.

We also notice that $<\psi^i|..|\psi^j>^{\dagger}=<\psi^j|..|\psi^i>$.
We see that $<1|..|3>=~<~2|..|4>= <3|..|1>^{\dagger} =<4|..|2>^{\dagger}$,  that
$<1|..|2>=<3|..|4>= <2|..|1>^{\dagger} =<4|..|3>^{\dagger}$ and that
$~<~4|..|1>^{\dagger}= <1|..|4>$ and $<3|..|2>^{\dagger}= <2|..|3>$, what is already written
below Eq.~(\ref{b}), $<i|..|j>$ denotes $<\psi^i|..|\psi^j>$.

In the case that $a=<\tilde{{\bf A}}^{\tilde{1}\sminus}>=<\tilde{{\bf A}}^{\tilde{1}\splus}>= e$ 
and
$<\tilde{{\bf A}}^{\tilde{N}_L\sminus}>=<\tilde{{\bf A}}^{\tilde{N}_L\splus}>=d$, which would 
mean that all the matrix elements are real, the mass matrix simplifies to
\begin{equation}
\label{M0a1}
{\cal M}_{(o)}^{\alpha} = \begin{pmatrix} - \tilde{a}_1  - \tilde{a}_2 + a^{\alpha} &
d& e & b\\
d  & - \tilde{a}_1 +   \tilde{a}_2 + a^{\alpha} &
b & e\\
e & b &  \tilde{a}_1  - \tilde{a}_2 + a^{\alpha} & d\\
b &  e  & d &  \tilde{a}_1  +\tilde{a}_2 + a^{\alpha}
\end{pmatrix}\,,
\end{equation}
with $b={\bf a}^{\alpha} ed$.


%
\subsection{Mass matrices beyond the tree level}
\label{beyond}
%

We   discuss in this subsection  the  matrix elements of the mass matrix in all orders 
of corrections, Eq.~(\ref{Upsi}), the tree level, $n=0$, of which is 
presented in Eq.~(\ref{M0a}). The tree level mass matrix manifests the $\widetilde{SU}(2)
\times \widetilde{SU}(2) \times U(1)$ symmetry as seen in Eq.~(\ref{M0a}), with
 $(<1|x|1>+ <4|x|4>) - (<2|x|2>+<3|x|3>) =0$ and $<1|x|3>= <2|x|4>=
<3|x|1>^{\dagger}= <4|x|1>^{\dagger}$ and with $(<1|xxx|4>, <2|xxx|3>, <3|xxx|2>,
<4|xxx|1>)$ related so that all are equal if $<\mathbf{\tilde{A}}^{\tilde{1}\spm}>$ and 
$<\mathbf{\tilde{A}}^{\tilde{N}_{L}\spm}>$  are real. 

Let us repeat that the generators of the two groups which operate among families commute:
$\{\tilde{\tau}^{\tilde{1}i}, \tilde{N}^{j}_{L}\}_{-} =0$\,,
and that these generators commute also with generators which distinguish among family members:
 $\{\tilde{\tau}^{\tilde{1}i}, \tau^{\alpha}\}_{-} =0$\,,
 $\{\tau^{\alpha}, \tilde{N}^{j}_{L}\}_{-} =0$\,,
where $\tau^{\alpha}$ represents $(Q, Q', Y')$ (or $\tau^4, \tau^{23}, \tau^{13}$).
 
To study the symmetry $\widetilde{SU}(2) \times$
$ \widetilde{SU}(2) \times U(1)$ of the mass matrix, Eq.~(\ref{M0a}), in all orders of loop 
corrections, of repetition of nonzero vacuum expectation values and of both together
--- loop corrections and nonzero vacuum expectation values --- we just have to calculate at each 
order of corrections the difference between each pair of the matrix elements which are equal 
on the three level, as well as the Hermitian conjugated difference  of such a pair.

Since the dependence of all the scalar fields on ordinary coordinates are in all cases the same, we 
only have to evaluate the application of the operators to the internal space of basic state, that is
 on the space of family and family members degrees of freedom.  Correspondingly we pay 
attention only on this internal part --- on the interaction of scalar fields with the space index $(7,8)$ 
with any family member of any of four families separately with respect to their internal space. 
The dependence of the mass matrix elements 
on the family member quantum numbers appears through the nonzero vacuum expectation 
value ${\bf a}^{\alpha}$, Eq.~(\ref{M0a}), as well as through the dynamical part of 
$\hat{{\bf O}}^{\alpha}$, Eq.~(\ref{Oi}).

We demonstrate in this subsection how does the repetition of the nonzero vacuum expectation 
values of the scalar fields and  loop corrections in all orders influence matrix elements, presented
on the tree level  in Eq.~(\ref{M0a}). 

In the case that ${\bf a}^{\alpha}=0$ (that is for $<{\bf A}^{Q}>=0$, $<{\bf A}^{Q'}>=0$ 
and $<{\bf A}^{Y'}>=0$)  the symmetry in all  corrections, that is in all loop corrections
and all the repetition of nonzero vacuum expectation values of the scalar fields, and of both ---
the loop corrections and the repetitions of nonzero vacuum expectation values nonzero  of 
all the scalar fields except ${\bf a}^{\alpha}$ --- keep the symmetry of the tree level, 
presented in Eq.~(\ref{M0a}).


We prove in this subsection that in the case that $<{\bf A}^{Q}>=0$, $<{\bf A}^{Q'}>
=0$ and $<{\bf A}^{Y'}>=0$, that is for ${\bf a}^{\alpha}$ $=0$,  the symmetry of mass 
matrices remains unchanged in all orders of corrections:  the loop ones of dynamical fields  ---
${\bf A}^{Q}$, ${\bf A}^{Q'}$, ${\bf A}^{Y'}$, $\vec{\tilde{{\bf A}}}^{\tilde{N}_L}$, 
$\vec{\tilde{{\bf A}}}^{\tilde{1}}$ ---  in the repetition of nonzero 
vacuum expectation values of the scalar fields carrying the family quantum numbers ---
$<\vec{\tilde{{\bf A}}}^{\tilde{N}_L}>$ and $<\vec{\tilde{{\bf A}}}^{\tilde{1}}>$ --- and of all 
together. The symmetry of mass matrices remains in all orders of corrections the one of the tree
level also if  ${\bf a}^{\alpha}\ne 0$ while ${\bf \tilde{a}}_1=0$ and ${\bf \tilde{a}}_2=0$.
The symmetry changes if the nonzero vacuum expectation values of all the scalar fields 
are nonzero.

In the case, however, that ${\bf a}^{\alpha}=0$, the matrix elements,
which are in the lowest order proportional to ${\bf a}^{\alpha}$ in Eq.~(\ref{M0a}), remain 
zero in all orders of corrections, while the nonzero matrix elements become dependent on family 
members quantum numbers due to the participations in loop corrections in all orders of the 
dynamical  fields ${\bf A}^{Q}$, ${\bf A}^{Q'}$ and ${\bf A}^{Y'}$. 

 We study in what follows first the symmetry of mass matrices in all orders of  corrections  in the 
case that ${\bf a}^{\alpha}=0$,  and then  the symmetry of the mass matrices,
again in all orders of corrections, when ${\bf a}^{\alpha}\ne 0$. We also comment that the symmetry 
of the tree level remain the same in all orders of corrections, if ${\bf a}^{\alpha}\ne 0$, while 
${\bf \tilde{a}_1}=0={\bf \tilde{a}_2}$.

%

%


%
\subsubsection{Mass matrices beyond the tree level, if ${\bf a}^{\alpha}=0$}
\label{aiszero}
%

We study corrections  to which the scalar fields which distinguish among families, contribute --- 
with their nonzero vacuum expectation values 
$<\vec{\tilde{{\bf A}}}^{\tilde{N}_L}>$ and $<\vec{\tilde{{\bf A}}}^{\tilde{1}}>$  and
their dynamical parts $\vec{\tilde{{\bf A}}}^{\tilde{N}_L}$ and 
$\vec{\tilde{{\bf A}}}^{\tilde{1}}$ ---
while we assume ${\bf a}^{\alpha}=0$ (${\bf a}^{\alpha}$ denotes the vacuum expectation
values to which the tree singlet fields, distinguishing among family members, contribute, 
that is ($<{\bf A}^{Q}>$, $<{\bf A}^{Q'}>$, $<{\bf A}^{Y'}>$),  taking into account the 
loop corrections of the corresponding dynamical parts  (${\bf A}^{Q}$, ${\bf A}^{Q'}$, 
${\bf A}^{Y'}$) in all orders.

We show that in such a case --- that is in the case that ${\bf a}^{\alpha}=0$ while all the other 
scalar fields determining mass matrices have nonzero vacuum expectation values 
($\tilde{{\bf a}}_{1}\ne 0$, $\tilde{{\bf a}}_{2} \ne 0$, $<\tilde{{\bf A}}^{\tilde{N}_L\spm}>
\ne 0$, $<\tilde{{\bf A}}^{\tilde{1}\spm}>\ne 0$)  --- the matrix elements, evaluated in all 
orders of corrections, keep the symmetry of the tree level.

We also show, that in this case the off diagonal matrix elements, represented 
in Eq.~(\ref{M0a}) as  (${\bf a}^{\alpha}<\tilde{{\bf A}}^{\tilde{1}\sminus}>
<\tilde{{\bf A}}^{\tilde{N}_{L}\sminus}>$, ${\bf {\bf a}}^{\alpha}
<\tilde{{\bf A}}^{\tilde{1}\sminus}>
<\tilde{{\bf A}}^{\tilde{N}_{L}\splus}>$, ${\bf a}^{\alpha}<\tilde{{\bf A}}^{\tilde{1}\splus}>
<\tilde{{\bf A}}^{\tilde{N}_{L}\sminus}>$, ${\bf a}^{\alpha}<\tilde{{\bf A}}^{\tilde{1}\splus}>
<\tilde{{\bf A}}^{\tilde{N}_{L}\splus}>$), remain zero in all orders of corrections.


Let us look how the corrections in all orders manifest for each matrix element separately.\\

{\bf i.} $\;\;$ We start with diagonal terms: $<\psi^{i}|.....|\psi^{i}>, i =(1,2,3,4)$.\\
On the tree level the symmetry is:\\
 $\{<\psi^{1}|<\hat{\mathbf{O}}^{\alpha}_{\dia}>|\psi^{1}> + 
<\psi^{4}|<\hat{\mathbf{O}}^{\alpha}_{\dia}>|\psi^{4}>\} - 
\{<\psi^{2}|<\hat{\mathbf{O}}^{\alpha}_{\dia}>
|\psi^{2}> + \{<\psi^{3}|<\hat{\mathbf{O}}^{\alpha}_{\dia}>|\psi^{3}> \}=0$. \\
{\bf i.a.} $\;$
It is easy to see that the tree level symmetry, 
$\{<\psi^{1}|<\hat{\mathbf{O}}^{\alpha}_{\dia}>|\psi^{1}> +
<\psi^{4}|<\hat{\mathbf{O}}^{\alpha}_{\dia}>|\psi^{4}>\} - 
\{<\psi^{2}|<\hat{\mathbf{O}}^{\alpha}_{\dia}>|\psi^{2}> + 
<\psi^{3}|<\hat{O}^{\alpha}_{\dia}>|\psi^{3}> \}=0$, remains in all orders of corrections, if
only the nonzero vacuum expectation values of  $<\tilde{{\bf A}}^{\tilde{1}3}>=
\tilde{{\bf a}}_{1}$ and $<\tilde{{\bf A}}^{\tilde{N}_{L}3}>=\tilde{{\bf a}}_{2}$ contribute
 in operators 
$\gamma^0 \stackrel{78}{(\pm)} \,\tilde{\tau}^{\tilde{1}3}\,
<\tilde{{\bf A}}^{\tilde{1}3}>$ and $\gamma^0 \stackrel{78}{(\pm)} \,\tilde{N}_{L}^{3}\,
<\tilde{{\bf A}}^{\tilde{N}_{L}3}> $. 
At, let say, $(2k+1)^{st}$  order of corrections we namely have $\{ (- (\tilde{{\bf a}}_{1}+ 
\tilde{{\bf a}}_{2}))^{(2k+1)} +  (\tilde{{\bf a}}_{1}+ \tilde{{\bf a}}_{2})^{(2k+1)}\} - 
\{ (- (\tilde{{\bf a}}_{1} - \tilde{{\bf a}}_{2}))^{(2k+1)} +  (\tilde{{\bf a}}_{1} - 
\tilde{{\bf a}}_{2})^{(2k+1)}\}=0.$ \\
%
{\bf i.b.} $\;$
The contributions of the dynamical terms, either ($ A^{Q}$, $ A^{Q'}$, $ A^{Y'}$) 
or ($\tilde{A}^{\tilde{1}3}, \tilde{A}^{\tilde{N}_{L}3}$) do not break the three level symmetry.
Each of them namely always appears in an even power, Fig.~(\ref{Figgeneral}), changing the 
order of corrections by a factor of two or $2n$
($| {\bf A}^{\alpha}|^{2(n-k-l)}, |\tilde{{\bf A}}^{\tilde{1}3}|^{2k}, 
|\tilde{{\bf A}}^{\tilde{N}_{L}3}|^{2l}$), where $(n-k-l, k, l)$ are nonnegative integers, while 
$\tau^{A \alpha}$ represents $(Q^{\alpha}, Q'^{\alpha}, Y'^{\alpha})$. The contribution to
 $|{\bf A}^{\alpha}|^{2m}, m=(n-k-l)$, origins in the product of $|{\bf A}^{Q}|^{2(m-p-r)}\cdot 
|{\bf A}^{Q'}|^{2p}\cdot |{\bf A}^{Y'}|^{2r}$. Again $(m-p-r, p, r)$ are nonnegative integers. \\
{\bf i.c.} $\;$
%
There are also other contributions, either those with only nonzero vacuum expectation values or 
with dynamical fields in addition to nonzero vacuum expectation values of scalars, in which  
$\hat{\tilde{{\bf O}}}^{\tilde{1}\spm}$ and $\hat{\tilde{{\bf O}}}^{\tilde{N}_{L}\spm}$
 together with all kinds of diagonal terms contribute.
Let us repeat again what do the operators $\hat{\tilde{O}}^{\tilde{1}\spm}$ and
$\hat{\tilde{O}}^{\tilde{N}_{L}\spm}$, Eq.~(\ref{Oi}), do when they apply on $\psi^{i}$. 
 The operators $\hat{\tilde{O}}^{\tilde{1}\splus}$ transforms $\psi^{1}$ into $\psi^{3}$ and 
$\psi^{2}$ into $\psi^{4}$. Correspondingly the states $\psi^{1}$ and $\psi^{4}$ take  under 
the application of $\hat{\tilde{O}}^{\tilde{1}\splus}$ the role of $\psi^{2}$ 
and $\psi^{3}$, while $\psi^{2}$ and $\psi^{3}$ take the role of $\psi^{1}$ and $\psi^{4}$, 
 all carrying the correspondingly changed eigenvalues of  $\tilde{\tau}^{\tilde{1} 3}$.
The operator $\hat{\tilde{O}}^{\tilde{N}_{L}\splus}$ transforms
$\psi^{1}$ into $\psi^{2}$ and $\psi^{3}$ into $\psi^{4}$.
Correspondingly the states $\psi^{1}$ and $\psi^{2}$ take  under the application 
of $\hat{\tilde{O}}^{\tilde{N}_{L}\splus}$ the role of $\psi^{3}$ and $\psi^{4}$,
while $\psi^{3}$ and $\psi^{4}$ take the role of $\psi^{1}$ and $\psi^{2}$, carrying 
the correspondingly changed eigenvalues of  $\tilde{N_{L}}^{3}$.
 Either the dynamical fields or the nonzero vacuum expectation values 
of these scalar fields, $\hat{\tilde{O}}^{\tilde{1}\spm}$ and 
$\hat{\tilde{O}}^{\tilde{N}_{L}\spm}$, must in diagonal terms appear in the second power or 
in $n\times$ the second power. We easily see that 
also in such cases the tree level symmetry remains in all orders.\\
{\bf i.c.1.} $\;$
To better understand the contributions in all orders to the diagonal terms, discussing here, 
let us calculate the contribution of the third order 
corrections either from the loop or from the nonzero vacuum expectation values to the diagonal 
matrix elements $<\psi^{i}|...|\psi^{i}>$ under the assumption that ${\bf a}^{\alpha}=0$. 
Let us evaluate the contributions of the  operators  
$<\hat{\tilde{{\bf O}}}^{\tilde{1}3}>$, $\hat{\tilde{{\bf O}}}^{\tilde{N}_{L}3}$, 
$\hat{\tilde{{\bf O}}}^{\tilde{1}\spm}$ and $\hat{\tilde{{\bf O}}}^{\tilde{N}_{L}\spm}$ in the 
third order. 
We see that $\tilde{\tau}^{\tilde{1} \sminus}$ transforms $\psi^{3}$ into $\psi^{1}$ and 
$\psi^{4}$ into $\psi^{2}$, while $\tilde{\tau}^{\tilde{1} \splus}$ transforms $\psi^{2}$ into 
$\psi^{4}$ and $\psi^{1}$ into $\psi^{3}$. We see that  $\tilde{N}_{L}^{\sminus}$ transforms
$\psi^{2}$ into $\psi^{1}$ and $\psi^{4}$ into $\psi^{3}$, while $\tilde{N}_{L}^{\splus}$
transforms $\psi^{1}$ into $\psi^{2}$ and $\psi^{3}$ into $\psi^{4}$.
It then follows that $\{<\psi^{1}|xxx|\psi^{1}> + <\psi^{4}|xxx|\psi^{4}>\} -
\{<\psi^{2}|xxx|\psi^{2}> + <\psi^{3}|xxx|\psi^{3}>\}=0$, where $xxx$ represent all possible
acceptable combination of $<\hat{\tilde{{\bf O}}}^{\tilde{1}\spm}>$,  
$<\hat{\tilde{{\bf O}}}^{\tilde{N}_{L}\spm}>$ and the diagonal terms 
$<\hat{\tilde{{\bf O}}}^{\tilde{1}3}>$ and $<\hat{\tilde{{\bf O}}}^{\tilde{N}_{L}3}>$. 
One namely obtains that the contribution of $\{<\psi^{1}|xxx|\psi^{1}> + <\psi^{4}|xxx|\psi^{4}>\}=$
\{$|<\tilde{{\bf A}}^{\tilde{1} \sminus}>|^2 [-2(\tilde{{\bf a}}_1 + \tilde{{\bf a}}_2) + 
(\tilde{{\bf a}}_1 - \tilde{{\bf a}}_2)]
+ |<\tilde{{\bf A}}^{\tilde{N}_{L}\sminus}>~|^2 [- 2(\tilde{{\bf a}}_1 + \tilde{{\bf a}}_2) - 
(\tilde{{\bf a}}_1 - \tilde{{\bf a}}_2)]+ (-(\tilde{{\bf a}}_1 + \tilde{{\bf a}}_2)^{3}) +
|<\tilde{{\bf A}}^{\tilde{1}\sminus}>|^2 [ +2 (\tilde{{\bf a}}_1 + \tilde{{\bf a}}_2) - 
(\tilde{{\bf a}}_1 - \tilde{{\bf a}}_2)]+ |<\tilde{{\bf A}}^{\tilde{N}_{L}-}>|^2 [ +
2 (\tilde{{\bf a}}_1 + \tilde{{\bf a}}_2) + (\tilde{{\bf a}}_1 - \tilde{{\bf a}}_2)]
+ (\tilde{{\bf a}}_1 + \tilde{{\bf a}}_2)^{3}\} =0$, and for 
$\{<\psi^{2}|xxx|\psi^{2}> + <\psi^{3}|xxx|\psi^{3}>\}$ one obtains 
 $=\{|<\tilde{{\bf A}}^{\tilde{1}\sminus}>~|^2 [-2 (\tilde{{\bf a}}_1 - \tilde{{\bf a}}_2) + 
(\tilde{{\bf a}}_1 + \tilde{{\bf a}}_2)]
+ |<\tilde{{\bf A}}^{\tilde{N}_{L}\sminus}>~|^2 [ -2 (\tilde{{\bf a}}_1 - \tilde{{\bf a}}_2) - 
(\tilde{{\bf a}}_1 + \tilde{{\bf a}}_2)]+ (-(\tilde{{\bf a}}_1 - \tilde{{\bf a}}_2)^{3}) +
|<\tilde{{\bf A}}^{\tilde{1}\sminus}>~|^2 [ +2(\tilde{{\bf a}}_1 - \tilde{{\bf a}}_2) - 
(\tilde{{\bf a}}_1 + \tilde{{\bf a}}_2)]
+ |<\tilde{{\bf A}}^{\tilde{N}_{L}-}>|^2 [+2(\tilde{{\bf a}}_1 - \tilde{{\bf a}}_2) + 
(\tilde{{\bf a}}_1 + \tilde{{\bf a}}_2)] + (\tilde{{\bf a}}_1 - \tilde{{\bf a}}_2)^{3}\} =0$.
Also the dynamical fields keep the tree level symmetry of mass matrices. To prove one only must
replace in the above calculation $|<\tilde{{\bf A}}^{\tilde{1}\sminus}>|^2$ by
 $|\tilde{{\bf A}}^{\tilde{1}\sminus}|^2$ and  $|<\tilde{{\bf A}}^{\tilde{N}_{L}\sminus}>|^2$
 by $|\tilde{{\bf A}}^{\tilde{N}_{L}\sminus}|^2$.

To the diagonal terms the three singlets contribute
 in absolute squared values ($|{\bf A}^{Q}|^2$, $|{\bf A}^{Q'}|^2$, $|{\bf A}^{Y'}|^2$, 
each on a power, which depend on the order of corrections. This makes  all the diagonal matrix 
elements,  
$<\psi^{1}|.....|\psi^{1}>$, $<\psi^{2}|.....|\psi^{2}>$,  $<\psi^{3}|.....|\psi^{3}>$ and 
$<\psi^{4}|.....|\psi^{4}>$, dependent on the family member quantum numbers.

Such behaviour of matrix elements remains unchanged in all orders of corrections, either due to loops 
of dynamical fields or due to repetitions of nonzero vacuum expectation values. The reason
 is in the fact that the operators  $<\hat{\tilde{{\bf O}}}^{\tilde{1}\spm}>$ and 
$<\hat{\tilde{{\bf O}}}^{\tilde{N}_{L}\spm}>$ exchange the role of the states in the way that the odd 
power of diagonal contributions to the diagonal matrix elements always keep the symmetry 
$\{<\psi^{1}|\hat{U}|\psi^{1}> + <\psi^{4}|\hat{U}|\psi^{4}>\} -
\{<\psi^{2}|\hat{U}|\psi^{2}> + <\psi^{3}|\hat{U}|\psi^{3}>\}=0$.  

These proves the statement that {\it corrections in  all orders keep the symmetry of the tree level
diagonal terms in the case that} $a^{\alpha}=0$.\\  



{\bf ii.} $\;\;$ Let us look at matrix element $<\psi^{1}|.....|\psi^{3}>$ and
 $<\psi^{2}|.....|\psi^{4}>$ in Eq.~(\ref{M0a}),
where we have on the tree level $<1|x|3>=<2|x|4>$ and $<3|x|1>=<4|x|2>= 
<1|x|3>^{\dagger}$. We again simplify the notation $<\psi^{i}|.....|\psi^{j}>$ into $<i|...|j>$.
The two matrix elements --- $<1|x|3>,<2|x|4>$ --- are  on the tree level denoted by 
$<\tilde{{\bf A}}^{\tilde{1}\sminus}>$, while $<3|x|1>$ and $<4|x|2>$ are denoted by 
$<\tilde{{\bf A}}^{\tilde{1}\splus}>$.\\
We have to prove that corrections, either of the loops kind or of the repetitions of the nonzero
vacuum expectation values or of both kinds in any order keeps the symmetry of the tree level. \\
%
%
%
%
{\bf ii.a.} $\;$ Let us start with the corrections in which besides 
$<\tilde{{\bf A}}^{\tilde{1}\sminus}>$ in the first power only $<\tilde{{\bf A}}^{\tilde{1}3}>=
\tilde{{\bf a}}_1$ and $<\tilde{{\bf A}}^{\tilde{N}_{L}3}>= \tilde{{\bf a}}_2$ contribute, 
 the last two together appear in an even power so that all three together contribute in an odd power.  \\ 
The contribution of $(<1|x|1>)^{2k+1} = (-(\tilde{{\bf a}}_1 +\tilde{{\bf a}}_2))^{2k+1}$ in
the $(2k+1)^{th}$ order is up to a sign equal to $(<4|x|4>)^{2k+1} = (\tilde{{\bf a}}_1 +
\tilde{{\bf a}}_2)^{2k+1}$, where $k$ is a 
nonnegative integer, while the contribution of $(<2|x|2>)^{2k+1} = (-(\tilde{{\bf a}}_1 - 
\tilde{{\bf a}}_2))^{2k+1}$ is up to a sign equal to $(<3|x|3>)^{2k+1} =
 (\tilde{{\bf a}}_1 - \tilde{{\bf a}}_2)^{2k+1}$. 
In each of the matrix elements, either $<1|.....|3>$ or $<2|.....|4>$, 
both factors together,  $(-(\tilde{{\bf a}}_1 +\tilde{{\bf a}}_2))^{m}$ $(\tilde{{\bf a}}_1 -
\tilde{{\bf a}}_2)^{n}$ in the case $<1|.....|3>$ and $(-(\tilde{{\bf a}}_1 -
\tilde{{\bf a}}_2))^{m}$ $(\tilde{{\bf a}}_1 +\tilde{{\bf a}}_2)^{n}$
in the case $<2|.....|4>$, with ($m+n$) an even nonnegative integer (since together with 
$<\tilde{{\bf A}}^{\tilde{1}\sminus}>$ must be of an odd integer corrections to take care of 
the left/right nature of matrix elements) one must make the sum over all the terms contributing to 
corrections of the order ($m+n +1$). It is not difficult to see that the contribution  to  
$<1|.....|3>$ is  in any order of corrections equal to the contributions to the same order of 
corrections to $<2|.....|4>$.\\
{\bf ii.a.1.} $\;$ To illustrate the same contribution in each order of corrections to $<1|.....|3>$ and
to $<2|.....|4>$ let us calculate, let say, the third order corrections. 
The contribution of the third order to $<1|xxx|3>$ is 
$-\frac{1}{3!}<\tilde{{\bf A}}^{\tilde{1}\sminus}> \{ (\tilde{{\bf a}}_1 +\tilde{{\bf a}}_2)^{2} +
(\tilde{{\bf a}}_1 - \tilde{{\bf a}}_2)^{2}  - (\tilde{{\bf a}}_1 -\tilde{{\bf a}}_2)(\tilde{{\bf a}}_1
 +\tilde{{\bf a}}_2)\}$ 
 and the contribution of the third order to $<2|xxx|4>$ is 
$-\frac{1}{3!}<\tilde{{\bf A}}^{\tilde{1}\sminus}> 
\{ (\tilde{{\bf a}}_1 - \tilde{{\bf a}}_2)^{2} +(\tilde{a}_1 + \tilde{{\bf a}}_2)^{2}  - 
(\tilde{{\bf a}}_1 +\tilde{{\bf a}}_2)(\tilde{{\bf a}}_1 -\tilde{{\bf a}}_2)\}$, 
that is the contributions in the third order of $<1|xxx|3>$ and $<2|xxx|4>$ are  the same.\\
{\bf ii.b.} $\;$  One can repeat the calculations with  $<\tilde{{\bf A}}^{\tilde{1}\sminus}>$ and 
the dynamical fields $\tilde{{\bf A}}^{\tilde{1}\sminus}$ and $\tilde{{\bf A}}^{\tilde{1}\splus}$, 
with or without the diagonal nonzero vacuum expectation values. In all cases all the contributions 
keep the symmetry on the tree level due to the above discussed properties of the diagonal terms. 
All the dynamical terms must namely appear in absolute values squared in order to contribute to the 
mass matrices, as shown in Fig.~\ref{Figgeneral}. To the diagonal terms the three singlets 
contribute  in absolute squared values ($|{\bf A}^{Q}|^2$, $|{\bf A}^{Q'}|^2$, 
$|{\bf A}^{Y'}|^2$), each on some power, depending on the order of corrections. This makes 
the matrix element $<1|.....|3>$ 
and $<2|.....|4>$,  $<3|.....|1>$ and $<4|.....|2>$, dependent on the family members 
quantum numbers.

 In all cases all the contributions keep the symmetry on the 
 tree level.

{\bf ii.c.} $\;$ The Hermitian conjugate values $<1|.....|3>^{\dagger}=<2|.....|4>^{\dagger}$ 
have the transformed value of $<\tilde{{\bf A}}^{\tilde{1}\sminus}>$, that means that the value
is $<\tilde{{\bf A}}^{\tilde{1}\splus}>$, provided that the diagonal matrix elements of the mass
matrix are real,
keeping the symmetry of the matrix elements $<1|.....|3>^{\dagger}=<2|.....|4>^{\dagger}$ 
in all orders of corrections. 

These proves the statement that {\it corrections in  all orders keep the symmetry of the tree level
of the off-diagonal terms} $<1|.....|3>$ and $<2|.....|4>$ and of their Hermitian conjugated 
matrix elements in the case that ${\bf a}^{\alpha}=0$.\\

{\bf iii.} $\;\;$ Let us look at matrix element $<1|.....|2>$ and $<3|.....|4>$ in Eq.~(\ref{M0a}),
where we have on the tree level $<1|x|2>=<3|x|4>$.
These two matrix elements are  on the tree level denoted by
 $<\tilde{{\bf A}}^{\tilde{N}_{L}\sminus}>$.
We have to prove that corrections, either the loop corrections or the 
repetitions of the nonzero vacuum expectation values or  both kinds of corrections, in any order, 
keep the $\widetilde{SU}(2)$ $\times \widetilde{SU}(2)\times U(1)$ symmetry of the
tree level. \\
The proof for the symmetry of these matrix elements is carried out in equivalent way to the proof
under  {\bf ii.} $\;\;$.\\
{\bf iii.a.} $\;$ Let us start with the corrections in which besides 
 $<\tilde{{\bf A}}^{\tilde{N}_{L}\sminus}>$ in the first power also only 
$<\tilde{{\bf A}}^{\tilde{1}3}>=\tilde{{\bf a}}_1$ and $<\tilde{{\bf A}}^{\tilde{N}_{L}3}>= 
\tilde{{\bf a}}_2$ contribute.  The sum of powers of the last two {\bf a} must be even, so that a 
correction would be of an odd power due to the left/right transitions.  \\ 
Again the contributions of both diagonal terms,  $<1|x|1>$ and $<4|x|4>$, in any power --- 
$(<1|x|1>)^{2k+1} = (-(\tilde{{\bf a}}_1 +\tilde{{\bf a}}_2))^{2k+1}$ and 
$(<4|x|4>)^{2k+1} =(\tilde{{\bf a}}_1 +\tilde{{\bf a}}_2)^{2k+1}$, where $k$ is a
 nonnegative integer --- differ  only up to a sign when they appear in an odd power and are 
equal when they appear in an even power. 
These is true also for the contributions of $<2|x|2>$ and $<3|x|3>$ since $(<2|x|2>)^{2k+1} =
 (-(\tilde{{\bf a}}_1 - \tilde{{\bf a}}_2))^{2k+1}$ is up to a sign equal to $(<3|x|3>)^{2k+1} = 
(\tilde{{\bf a}}_1 - \tilde{{\bf a}}_2)^{2k+1}$. If they appear with an even power, they are equal. 
In each of the ($m+n +1$)$^{th}$ order corrections to the matrix elements, either $<1|.....|2>$ or 
$<3|.....|4>$,  where  $(-(\tilde{{\bf a}}_1 +\tilde{{\bf a}}_2))^{m}$ 
$(-(\tilde{{\bf a}}_1 -\tilde{{\bf a}}_2))^{n}$ contribute to  $<1|.....|2>$ and 
$(\tilde{{\bf a}}_1 -\tilde{{\bf a}}_2)^{m}$ $(\tilde{{\bf a}}_1 +\tilde{{\bf a}}_2)^{n}$
contribute to $<3|.....|4>$, the two contributions are again equal, since both $m$ and $n$ are 
even nonnegative integers.\\
{\bf iii.a.1.} $\;$ Let us, as an example, calculate the fifth order 
corrections to the tree level contributions of $<1|x|2>$ 
$=<\tilde{{\bf A}}^{\tilde{N}_{L}\sminus}>$. 
The contribution of the fifth order $<1|xxxxx|2>$ to
 $<1|x|2>$ is $ \frac{1}{5!} <\tilde{{\bf A}}^{\tilde{N}_{L}\sminus}> 
\{(- (\tilde{{\bf a}}_1 - \tilde{{\bf a}}_2))^{4} +
(-(\tilde{{\bf a}}_1 + \tilde{{\bf a}}_2))^{4} +
3 (- (\tilde{{\bf a}}_1 +\tilde{{\bf a}}_2))$$ (-(\tilde{{\bf a}}_1 - \tilde{{\bf a}}_2))^3+ 
6 (- (\tilde{{\bf a}}_1 +\tilde{{\bf a}}_2))^2 (-(\tilde{{\bf a}}_1 - \tilde{{\bf a}}_2))^2 +
3 (- (\tilde{{\bf a}}_1 +\tilde{{\bf a}}_2))^3
(-(\tilde{{\bf a}}_1 - \tilde{{\bf a}}_2))\}$,
and the contribution of the fifth order $<3|xxxxx|4>$ to $<3|x|4>$ is 
$ \frac{1}{5!} <\tilde{{\bf A}}^{\tilde{N}_{L}\sminus}> \{ (\tilde{{\bf a}}_1 + 
\tilde{{\bf a}}_2)^{4} +(\tilde{{\bf a}}_1 - \tilde{{\bf a}}_2)^{4}  +
3 (\tilde{{\bf a}}_1 -\tilde{{\bf a}}_2)(\tilde{{\bf a}}_1 + \tilde{{\bf a}}_2)^3+
6 (\tilde{{\bf a}}_1 -\tilde{{\bf a}}_2)^2 (\tilde{{\bf a}}_1 +\tilde{{\bf a}}_2)^2 + 
3 (\tilde{{\bf a}}_1 -\tilde{{\bf a}}_2)^3 (\tilde{{\bf a}}_1 +\tilde{{\bf a}}_2)\}$, 
 which is  equal to the contribution of 
the fifth order in the case of $<1|xxxxx|2>$.\\
{\bf iii.b.} $\;$  One can repeat the calculations with dynamical fields 
($\tilde{{\bf A}}^{\tilde{N}_{L}\sminus}$, $\tilde{{\bf A}}^{\tilde{N}_{L}\splus}$) in all orders 
and with $<\tilde{{\bf A}}^{\tilde{1}\sminus}>$  and with the diagonal nonzero vacuum 
expectation values and with the diagonal dynamical terms, paying attention that  the dynamical 
fields  contribute to masses of any of the family members only if they appear in pairs. 

To the diagonal terms the three singlets  (${\bf A}^{Q}$, ${\bf A}^{Q'}$, ${\bf A}^{Y'}$) 
contribute in the absolute squared values  ($|{\bf A}^{Q}|^2$, $|{\bf A}^{Q'}|^2$, 
$|{\bf A}^{Y'}|^2$), each on  a power, which depends on the order of  corrections. 

 In all cases all the contributions keep the symmetry on the 
 tree level.\\
{\bf iii.c.} $\;$ The proof is valid also for $<2|.....|1>= (<1|.....|2>)^{\dagger}$  
 and $<4|.....|3>=(<3|.....|4>)^{\dagger}$ in any order of corrections. Namely, if diagonal mass
 matrix elements are real then in the matrix elements  $<2|.....|1>$ only
 $<\tilde{A}^{\tilde{N}_{L}\sminus}>$ of the matrix element $<1|.....|2>$ must be replaced 
by $<\tilde{A}^{\tilde{N}_{L}\splus}>$.

 These proves the statement that {\it corrections in  all orders keep the symmetry of the tree level
off-diagonal terms} $<1|.....|2>$ and $<3|.....|4>$ in the case that ${\bf a}^{\alpha}=0$.\\  

{\bf iv.} $\;\;$ It remains to check the matrix elements $<1|.....|4>$, $<2|.....|3>$,
$<3|.....|2>$ and $<4|.....|1>$ in all orders of corrections. The matrix elements on the 
third power, ($<1|xxx|4>$, $<2|xxx|3>$, $<3|xxx|2>$, $<4|xxx|1>$), appearing in 
Eqs.~(\ref{b}, \ref{M0a}), are for ${\bf a}^{\alpha}=0$ all equal to zero. 
It is not difficult to prove that these four matrix elements  remain zero in all order of loop 
corrections. The reason is the same as in the above three cases, {\bf i.}, {\bf ii.}, {\bf iii.}.

{\it The proof that the symmetry $\tilde{SU}(2) \times \tilde{SU}(2)\times U(1)$ of the tree level  
remains unchanged in all orders of corrections, provided that ${\bf a}^{\alpha}=0$, is 
completed.}

 There are in all these cases the dynamical singlets contributing in the absolute squared values
 ($|{\bf A}^{Q}|^2$, $|{\bf A}^{Q'}|^2$, $|{\bf A}^{Y'}|^2$ --- each on  a power, which 
depend on the order of  corrections --- which make that all the matrix elements of a mass matrix, 
except the  ($<1|.....|4>$, $<2|.....|3>$, $<3|.....|2>$, $<4|.....|1>$) which remain zero in 
all orders of corrections,  depend on a particular family member.

\subsubsection{Mass matrices beyond the tree level if ${\bf a}^{\alpha} \ne 0$  
}
\label{aisnonzero}

We demonstrated that for ${\bf a}^{\alpha} = 0$ the symmetry of the tree level remains 
in all orders of corrections, the loops corrections and the repetitions of nonzero vacuum expectation 
values of all the scalar fields contributing to mass terms, the same as on the tree level, that is 
$\widetilde{SU}(2)\times\widetilde{SU}(2)\times U(1)$. 

Let us denote all corrections to the diagonal terms in all orders, in which the nonzero vacuum 
expectation values in all orders as well as their dynamical fields in all orders contribute when 
${\bf a}^{\alpha} = 0$ as:\\
$- ({\underline {\tilde{\bf a}}}_{1}+{\underline {\tilde{\bf a}}_{2}} ):= <\psi^{\alpha 1}_{L}|....
|\psi^{\alpha 1}_{R}>$,  
$- ({\underline {\tilde{\bf a}}}_{1}-{\underline {\tilde{\bf a}}}_{2} ):= <\psi^{\alpha 2}_{L}|....
|\psi^{\alpha 2}_{R}>$,\\  
$\;\;\,  ({\underline {\tilde{\bf a}}}_{1}-{\underline {\tilde{\bf a}}}_{2} ):= <\psi^{\alpha 3}_{L}|....
|\psi^{\alpha 3}_{R}>$,  
$ \;\;\, ({\underline {\tilde{\bf a}}}_{1}+{\underline {\tilde{\bf a}}}_{2} ):= <\psi^{\alpha 4}_{L}|....
|\psi^{\alpha 4}_{R}>$. \\ 

We study for  ${\bf a}^{\alpha} \ne 0$ how does the symmetry of the diagonal and the off
diagonal matrix elements  of the family members mass matrices change with respect to the 
symmetry on the tree level, presented in Eq.~(\ref{M0a}), in particular for small values of 
 $|{\bf a}^{\alpha}|$ in comparison with the contributions of all the rest of nonzero vacuum 
expectation values or of dynamical fields.

We discuss diagonal and off diagonal matrix elements separately. The symmetry of all 
depends on ${\bf a}^{\alpha} $.

%

{\bf i.} $\;\;$ Let us start with diagonal terms: $<\psi^{i}|.....|\psi^{i}>$.\\
On the tree level the symmetry is for ${\bf a}^{\alpha}\ne 0$: 
$\{<\psi^{1}|<\hat{{\bf O}}^{\alpha}_{\dia}>|\psi^{1}> +
<\psi^{4}|<\hat{{\bf O}}^{\alpha}_{\dia}>|\psi^{4}>~\} - 
\{~<~\psi^{2}|<\hat{{\bf O}}^{\alpha}_{\dia}>
|\psi^{2}> + \{<\psi^{3}|<\hat{{\bf O}}^{\alpha}_{\dia}>|\psi^{3}> \}=0$. 
 \\



{\bf i.a.} $\;\;$ 
Let us evaluate the matrix elements  $<\psi^{\alpha i}_{L}|....|\psi^{\alpha i}_{R}>$.
To simplify the derivations, we introduce, for a while, the notation: $ n_1=
{\bf a}^{\alpha} - ({\underline {\tilde {\bf a}}}_{1} + {\underline {\tilde {\bf a}}}_{2})$\,, 
$ n_2= {\bf a}^{\alpha} - ({\underline {\tilde {\bf a}}}_{1} - {\underline {\tilde {\bf a}}}_{2})$\,,
$ n_3={\bf a}^{\alpha} + ({\underline {\tilde {\bf a}}}_{1} - {\underline {\tilde {\bf a}}}_{2})$\,,
$ n_4={\bf a}^{\alpha} + ({\underline {\tilde {\bf a}}}_{1} + {\underline {\tilde {\bf a}}}_{2})$\,.
 Then it follows 
\begin{eqnarray}
\label{ianot0}
<\psi^{\alpha i}_{L}|....|\psi^{\alpha i}_{R}>&=& n_i-\frac{1}{3!} (n_i)^3 + \frac{1}{5!}
 (n_1)^5 - \frac{1}{7!} (n_1)^7 + ....\,\,.
\end{eqnarray}
 Taking into account the expressions for $ n_i$  one finds
\begin{eqnarray}
\label{1anot0}
&&(<\psi^{\alpha 1}_{L}|....|\psi^{\alpha 1}_{R}> + <\psi^{\alpha 4}_{L}|....|\psi^{\alpha 4}_{R}>) - 
(<\psi^{\alpha 2}_{L}|....|\psi^{\alpha 2}_{R}> + <\psi^{\alpha 3}_{L}|....|\psi^{\alpha 3}_{R}>) =
\nonumber\\
&& - 4 \,a^{\alpha}\, {\underline {\tilde {\bf a}}}_{1}  {\underline {\tilde {\bf a}}}_{2} \{1- \frac{1}{6}
[( {\underline {\tilde {\bf a}}}_{1})^2 + ( {\underline {\tilde {\bf a}}}_{2})^2 + (a^{\alpha})^2] +.....\}
\end{eqnarray}

For  ${\bf a}^{\alpha}\ne 0$ the relation among the diagonal  matrix elements obviously
changes with respect to the tree level symmetry of mass matrices (which is 
$(<\psi^{\alpha 1}_{L}|\dots |\psi^{\alpha 1}_{R}>+
<\psi^{\alpha 4}_{L}|\dots |\psi^{\alpha 4}_{R}>)-$ 
$(<\psi^{\alpha 2}_{L}|\dots |\psi^{\alpha 2}_{R}>+
<\psi^{\alpha 3}_{L}|\dots |\psi^{\alpha 3}_{R}>)=0$). 
But one sees as well that the contributions of higher terms to this change are for 
 ($|a^{\alpha}|,|{\underline {\tilde {\bf a}}}_{1}|,$ $|{\underline {\tilde {\bf a}}}_{2}|)< 1 $
smaller  the higher is the order of corrections and for ($|a^{\alpha}|,|{\underline {\tilde {\bf a}}}_{1}|,$ 
$|{\underline {\tilde {\bf a}}}_{2}|)<< 1 $, the symmetry of the tree level is almost reproduced. \\



{\bf ii.} $\;\;$ Let us look for the matrix elements $<1|.....|3>$ and $<2|.....|4>$ when 
${\bf a}^{\alpha}\ne 0$. (On the tree level, Eq.~(\ref{M0a}), we find $<1|x|3>$
 $=<2|x|4>=<3|x|1>^{\dagger}=<4|x|2>^{\dagger}$). Let 
$<{\underline {\tilde {\bf A}}}^{\tilde{1}\sminus}>$ represent 
the matrix elements $<1|.....|3>$  and $<2|.....|4>$ in all orders of corrections in the case that 
${\bf a}^{\alpha}=0$, as it is shown in~\ref{aiszero} of Subsect.~\ref{beyond}. We namely showed 
that  in the case that ${\bf a}^{\alpha}=0$ the matrix element $<1|.....|3>$ is  
equal to $<2|.....|4>=$ $<{\underline {\tilde {\bf A}}}^{\tilde{1}\sminus}>$.\\

Taking into account that in the case that ${\bf a}^{\alpha}$ is zero 
$<{\underline {\tilde {\bf A}}}^{\tilde{1}\sminus}>$ includes all the corrections in all 
orders and that also ${\underline {\tilde {\bf a}}}_2$ includes the corrections in all orders, we find
\begin{eqnarray}
\label{sumanot01324}
 (<\psi^{\alpha 1}_{L}|\dots |\psi^{\alpha 3}_{R}> &=& 
<{\underline {\tilde {\bf A}}}^{\tilde{1}\sminus}> [1 - \frac{1}{3!} (n_1^2 +n_1 n_3 + n_3^2) 
+\nonumber\\
&& \frac{1}{5!} (n_1^4 +n_1^3 n_3 +n_1^2 n_3^2 + n_1 n_3^3 + n_3^4) - .... ]\,,\nonumber\\
 (<\psi^{\alpha 2}_{L}|\dots |\psi^{\alpha 4}_{R}> &=& 
<{\underline {\tilde {\bf A}}}^{\tilde{1}\sminus}> [1 - \frac{1}{3!} (n_2^2 +n_2 n_4 + n_4^2) 
+\nonumber\\
&& \frac{1}{5!} (n_2^4 +n_2^3 n_4 +n_2^2 n_4^2 + n_2 n_4^3 + n_4^4)- .... ]\,,\nonumber\\
 (<\psi^{\alpha 1}_{L}|\dots |\psi^{\alpha 3}_{R}> &-&
<\psi^{\alpha 2}_{L}|\dots |\psi^{\alpha 4}_{R}>) = \nonumber\\
 2 a^{\alpha} {\underline {\tilde {\bf a}}}_2 <{\underline {\tilde {\bf A}}}^{\tilde{1}\sminus}>
&& \{1-\frac{1}{6} [({\bf a}^{\alpha})^2 +  ({\underline {\tilde {\bf a}}}_{1})^2 +
({\underline {\tilde {\bf a}}}_{2} )^2] + ...\}  
\end{eqnarray}
For  ${\bf a}^{\alpha}\ne 0$ the relation among these non diagonal  matrix elements obviously
changes with respect to the tree level symmetry of mass matrices (which is 
$<\psi^{\alpha 1}_{L}|\dots |\psi^{\alpha 3}_{R}> - 
<\psi^{\alpha 2}_{L}|\dots |\psi^{\alpha 4}_{R}>) = 0$), but one sees as well that the 
changes are smaller and smaller  the higher is the order of corrections,
and are all for ($|a^{\alpha}|,|{\underline {\tilde {\bf a}}}_{1}|,$ 
$|{\underline {\tilde {\bf a}}}_{2}|)<< 1 $, close to zero. \\

{\bf iii.} $\;\;$ Let us look at the matrix element $<1|.....|2>$ and $<3|.....|4>$. In the case that  
 corrections in all orders manifest the symmetry of  the tree level (which is
 $<1|x|2>$ $=<3|x|4>=<2|x|1>^{\dagger}=<4|x|3>^{\dagger}$, Eq.~(\ref{M0a}).  
Let  $<{\underline {\tilde {\bf A}}}^{\tilde{N}_{L}\sminus}>$ represent  the matrix elements  
$<1|.....|2>$ and $<3|.....|4>$ in all orders of corrections for ${\bf a}^{\alpha}=0$, with 
${\underline {\tilde {\bf a}}}_2$ including in this case corrections in all orders as well, as shown
 in~\ref{aiszero} of Subsect.~\ref{beyond}.

Allowing ${\bf a}^{\alpha}\ne  0$ one obtains
\begin{eqnarray}
\label{sumanot01234}
 <\psi^{\alpha 1}_{L}|\dots |\psi^{\alpha 2}_{R}> &=& 
<{\underline {\tilde {\bf A}}}^{\tilde{N}_{L}\sminus}> [1 - \frac{1}{3!} (n_1^2 +n_1 n_2 + n_2^2) 
+\nonumber\\
&& \frac{1}{5!} (n_1^4 +n_1^3 n_2 +n_1^2 n_2^2 + n_1 n_2^3 + n_2^4) - .... ]\,,\nonumber\\
 (<\psi^{\alpha 3}_{L}|\dots |\psi^{\alpha 4}_{R}> &=& 
<{\underline {\tilde {\bf A}}}^{\tilde{N}_{L}\sminus}> [1 - \frac{1}{3!} (n_3^2 +n_3 n_4 + n_4^2) 
+\nonumber\\
&& \frac{1}{5!} (n_3^4 +n_3^3 n_4 +n_3^2 n_4^2 + n_3 n_4^3 + n_4^4) - .... ]\,,\nonumber\\
 (<\psi^{\alpha 1}_{L}|\dots |\psi^{\alpha 2}_{R}> & - &
<\psi^{\alpha 3}_{L}|\dots |\psi^{\alpha 4}_{R}>)  = \nonumber\\
 2 a^{\alpha} {\underline {\tilde {\bf a}}}_1 <{\underline {\tilde {\bf A}}}^{\tilde{1}\sminus}>
&& \{1-\frac{1}{6} [({\bf a}^{\alpha})^2 +  ({\underline {\tilde {\bf a}}}_{1})^2 +
({\underline {\tilde {\bf a}}}_{2} )^2] + ...\}.  
\end{eqnarray}
The observation is similar as in the above cases: 
For  ${\bf a}^{\alpha}\ne 0$ the relation among the non diagonal  matrix elements
$<\psi^{\alpha 1}_{L}|\dots |\psi^{\alpha 2}_{R}> $ and 
$<\psi^{\alpha 3}_{L}|\dots |\psi^{\alpha 4}_{R}>$ changes with respect to the tree level
symmetry of mass matrices (which is 
$<\psi^{\alpha 1}_{L}|\dots |\psi^{\alpha 2}_{R}> - 
<\psi^{\alpha 3}_{L}|\dots |\psi^{\alpha 4}_{R}>) = 0$). 
But one sees as well that the changes are smaller and smaller  the higher is the order of corrections,
and are all for ($|a^{\alpha}|,|{\underline {\tilde {\bf a}}}_{1}|,$ 
$|{\underline {\tilde {\bf a}}}_{2}|)<< 1 $ close to zero. \\

{\bf iv.} $\;\;$ It remains to calculate for  ${\bf a}^{\alpha}\ne 0$ the symmetry of matrix elements 
$<1|.....|4>$, $<2|.....|3>$, $<3|.....|2>$ and $<4|.....|1>$ in higher order corrections. 
The matrix elements are for $a^{\alpha}\ne 0$ nonzero only in the third order of corrections  
($<1|x|4>=0=<2|x|3>=0=<3|x|2>=<4|x|1>$, the first nonzero terms are
 --- $<1|xxx|4>$, $<2|xxx|3>$, $<3|xxx|2>$, $<4|xxx|1>$ --- appearing in 
Eqs.~(\ref{b}, \ref{M0a}), and are for ${\bf a}^{\alpha}=0$ all equal to zero in all orders of 
corrections. 

We again take into account that for ${\bf a}^{\alpha} =0$ the matrix element 
$<{\underline {\tilde {\bf A}}}^{\tilde{1} \spm}>$ and 
$<{\underline {\tilde {\bf A}}}^{\tilde{N}_{L}\spm}>$ include the corrections in all orders and 
that also ${\underline {\tilde {\bf a}}}_1$ and ${\underline {\tilde {\bf a}}}_2$ include 
the corrections  in all orders. We find when ${\bf a}^{\alpha} \ne 0$
\begin{eqnarray}
\label{sumanot01234}
&&<\psi^{\alpha 1}_{L}|\dots |\psi^{\alpha 4}_{R}>= 
<{\underline {\tilde {\bf A}}}^{\tilde{1} \sminus}>
<{\underline {\tilde {\bf A}}}^{\tilde{N}_{L}\sminus}>\{- \frac{1}{3!}
[2 (n_{4} + n_{1}) + (n_2 +n_3)] + \nonumber\\
&& \frac{1}{5!} [2 (n_{4}^3 + n_{1}^3) +
(n_{2}^3 + n_{3}^3) + 2 n_{4} n_{1}(n_{4} + n_{1}) + \nonumber\\ 
&&  (n_{4} n_{1} + 
n_{4}^2 + n_{1}^2) (n_{3} + n_{2}) + (n_{4} + n_{1}) (n_{3}^2 + n_{2}^2)] +
...\}\nonumber\\
&&<\psi^{\alpha 2}_{L}|\dots |\psi^{\alpha 3}_{R}>= 
<{\underline {\tilde {\bf A}}}^{\tilde{1} \sminus}>
<{\underline {\tilde {\bf A}}}^{\tilde{N}_{L}\splus}>\{- \frac{1}{3!}
[2 (n_{3} + n_{2}) + (n_1 +n_4)] + \nonumber\\
&& \frac{1}{5!} [2 (n_{3}^3 + n_{2}^3) +
(n_{4}^3 + n_{1}^3) + 2 n_{2} n_{3}(n_{3} + n_{2}) + \nonumber\\ 
&&  (n_{2} n_{3} + 
n_{3}^2 + n_{2}^2) (n_{4} + n_{1}) + (n_{3} + n_{2}) (n_{4}^2 + n_{1}^2)] +
...\}
\end{eqnarray}
It then follow 
\begin{eqnarray}
\label{sumanot01234F}
\nonumber\\
&&\frac{<\psi^{\alpha 1}_{L}|\dots |\psi^{\alpha 4}_{R}>}{<{\underline {\tilde {\bf A}}}^{\tilde{1} 
\sminus}><{\underline {\tilde {\bf A}}}^{\tilde{N}_{L}\sminus}>} =
\frac{<\psi^{\alpha 2}_{L}|\dots |\psi^{\alpha 3}_{R}>}{<{\underline {\tilde {\bf A}}}^{\tilde{1} 
\sminus}><{\underline {\tilde {\bf A}}}^{\tilde{N}_{L}\splus}>}  = \nonumber\\ 
&&\frac{<\psi^{\alpha 4}_{L}|\dots |\psi^{\alpha 1}_{R}>}{<{\underline {\tilde {\bf A}}}^{\tilde{1} 
\splus}><{\underline {\tilde {\bf A}}}^{\tilde{N}_{L}\splus}>} =
\frac{<\psi^{\alpha 3}_{L}|\dots |\psi^{\alpha 2}_{R}>}{<{\underline {\tilde {\bf A}}}^{\tilde{1} 
\splus}><{\underline {\tilde {\bf A}}}^{\tilde{N}_{L}\sminus}>}  = \nonumber\\
&& - {\bf a}^{\alpha} \{1- \frac{1}{6} [ (a^{\alpha})^2 +{\underline {\tilde {\bf a}}}_{1})^2 +
 ({\underline {\tilde {\bf a}}}_{2})^2] +\cdots\}
\,\,.
\end{eqnarray}
One sees that these off diagonal matrix elements keep the relations from Eq.~(\ref{M0a}).
 Again is true that the higher order corrections are smaller and smaller  the higher is the order of 
corrections, and are all for ($|a^{\alpha}|,|{\underline {\tilde {\bf a}}}_{1}|,$ 
$|{\underline {\tilde {\bf a}}}_{2}|)<< 1 $ closer and closer to zero.
%

We demonstrated in this subsection that the matrix elements of the mass matrix of Eq.~(\ref{M0a}), 
although do not keep the symmetry of the tree level in all orders of corrections if 
${\bf a}^{\alpha} \ne 0$,  the changes are for ($|{\bf a}^{\alpha}|$, $|{\underline {\tilde {\bf a}}}_{1}|$, 
$|{\underline {\tilde {\bf a}}}_{2}|$) $<< 1$  small, approaching very quickly zero, while almost
reproducing the tree level symmetry of Eq.~(\ref{M0a}). We also show that  the change of the 
tree level symmetry can easily be evaluated.

The  calculations of Subsect.~\ref{aisnonzero} demonstrate, that the symmetry of the  mass
matrices keep the symmetry of the mass matrices on the tree level in all orders of corrections not 
only in the case that ${\bf a}^{\alpha}=0$ but also  in the case that  ${\bf a}^{\alpha}\ne 0$ while 
$\tilde{a}_1=0= \tilde{a}_2$. This choice makes all the matrix elements nonzero, on 
the tree level and correspondingly on all orders of corrections.


\section{Conclusions}
\label{conclusion}

In the {\it spin-charge-family} theory to the mass matrix of any family member the two scalar 
triplets ($\vec{\tilde{A}}^{\tilde{1}}_{s}$,
$\vec{\tilde{A}}^{\tilde{N}_{L}}_{s}$)  and the three scalar singlets ($A^{Q}_{s}$, $A^{Q'}_{s}$,
$A^{Y'}_{s}$), $s=(7,8)$,  contribute, all with the weak and the hyper charge of the {\it standard
model} higgs ($\pm \frac{1}{2}, \mp \frac{1}{2}$, respectively). 
The first two triplets carry the family quantum numbers, while  the last three singlets carry  
the family members quantum numbers. 

The only dependence of the mass matrix on the family members quantum numbers 
 ($\alpha=(u,d,\nu,e)$) is due to the operators $\gamma^0 \stackrel{78}{(\pm)}\,
Q A_{\pm}^Q$, $\gamma^0 \stackrel{78}{(\pm)}\,Q' A_{\pm}^{Q'}$ and
$\gamma^0 \stackrel{78}{(\pm)}\,Y' A_{\pm}^{Y'}$. The operator  
$\gamma^0 \stackrel{78}{(\pm)}$, appearing with the two triplet scalar fields
--- carrying the family quantum numbers --- as well as with the three singlet scalar fields --- carrying
the family members quantum numbers, transforms the right handed members into the left
handed ones, or opposite, while the family operators transform a family member of one family into 
the same family member of another family.

We demonstrate in this paper that the matrix elements of mass matrices $4 \times 4$, predicted 
by the {\it spin-charge-family} theory  for each family member $\alpha=(u,d,\nu,e)$, keep the 
symmetry $\widetilde{SU}(2)_{\widetilde{SO}(4)_{1+3}} \times 
\widetilde{SU}(2)_{\widetilde{SO}(4)_{"weak"}} \times U(1)$ in all orders of corrections under the 
assumption that either the nonzero vacuum expectation values of three singlets $<A^{\alpha}>
= a^{\alpha}$ are equal to zero, Subsect.~\ref{aiszero},  $a^{\alpha}= 0$,  while all the other
scalar fields --- $\vec{\tilde{A}}^{\tilde{1}} $, $\vec{\tilde{A}}^{\tilde{N}_L} $ --- (can) have for all 
the  components nonzero vacuum expectation values, or that $a^{\alpha}$ does not need to be 
zero, $a^{\alpha}\ne 0$, but then the two third components of the two scalar triplets, 
$<\tilde{A}^{\tilde{1}3}>= \tilde{a}_{1}, <\tilde{A}^{\tilde{N}_{L}3}>= \tilde{a}_{2}$,
are zero, $\tilde{a}_{1}=0, \tilde{a}_{2}=0$.

We present the relations among  mass matrix elements of the mass matrix of any family member 
also in the case  that
we take into account higher order corrections when all the scalar fields have nonzero vacuum 
expectation values, Subsect.~\ref{aisnonzero}.

In the first case when $a^{\alpha}= 0$ in any order of corrections all the components of the two
triplet scalar fields --- carrying the family quantum numbers --- contribute, either  with the nonzero 
vacuum expectation values (on the tree level and in all orders of corrections) or as dynamical fields
(in all orders of corrections) or as both (in all orders of corrections), while the three singlet scalar fields
contribute only as dynamical fields.  
The contributions of the dynamical fields of the three singlets --- carrying the family members 
quantum numbers --- in all orders of  loop corrections 
(together with the contributions of the two triplets which interact with spinors through the family 
quantum numbers either with the nonzero vacuum expectation values or as dynamical fields)
 make all the matrix elements dependent on the particular family member quantum numbers.
Correspondingly all the mass matrices  bring different masses to any of the family members 
and also different  mixing matrices to quarks and leptons. However, the choice $a^{\alpha}= 0$ 
makes the four off diagonal terms, which are proportional to $a^{\alpha}$ in Eq.(\ref{M0a}), 
equal to zero in all orders of corrections. 

In the second case when $ \tilde{a}_{1}=0$, $\tilde{a}_{2}=0$, in any order of corrections the
three singlet scalar fields contribute either with nonzero vacuum expectation values or as dynamical 
fields, while the two triplets scalar fields contribute with the nonzero vacuum expectation values and 
the dynamical fields, except the two of the triplets components --- $\tilde{A}^{\tilde{1} 3}$ and 
$\tilde{A}^{\tilde{N}_{L}3}$  --- which contribute only as dynamical fields. This choice makes 
 all the diagonal terms to remain equal in all orders of corrections. 

Both choices, either $a^{\alpha}= 0$ or $ \tilde{a}_{1}=0$ and $\tilde{a}_{2}=0$, keep the symmetry 
of the mass matrices of~Eq.(\ref{M0a}) in all orders of corrections to which either the dynamical fields 
or the nonzero vacuum expectation values contribute.

In  Subsect.~\ref{aisnonzero} we evaluate the symmetry of mass matrices in higher order corrections 
also in the general case, when all the singlet and triplet scalar fields, contributing to mass matrices,
have for all the components nonzero vacuum expectation values ($a^{\alpha}\ne 0$, 
$ \tilde{a}_{1}\ne 0$, $\tilde{a}_{2}\ne 0$, $<\tilde{A}^{\tilde{N}_{L} \spm}\ne 0>$, 
$<\tilde{A}^{\tilde{1}\spm}>\ne 0$). The higher orders 
corrections of all the matrix elements, diagonal  and off diagonal, start to depend on a common 
factor (Eqs.~(\ref{ianot0} - \ref{sumanot01234F}), manifesting that the higher orders corrections
are  smaller and smaller the higher is the order of corrections, and approach zero very quickly
if $|{\bf a}^{\alpha}|$, $|{\underline {\tilde {\bf a}}}_{1}|$ and $|{\underline {\tilde {\bf a}}}_{2}|$ 
$<< 1$.  

We demonstrated in the subsection, Subsect.~\ref{aisnonzero}, that the matrix elements of the mass 
matrix of Eq.~(\ref{M0a}), although do not keep the symmetry of the tree level in all orders of 
corrections if ${\bf a}^{\alpha} \ne 0$,  the changes  can be very well estimated for 
$|{\bf a}^{\alpha}|$, $|{\underline {\tilde {\bf a}}}_{1}|$, 
$|{\underline {\tilde {\bf a}}}_{2}|$) $<< 1$.

 When fitting free parameters of the mass matrices $4 \times 4$ of quarks and leptons, as suggested 
by the {\it spin-charge-family} theory --- $({\underline {\tilde{\bf a}}}_{1},
 {\underline {\tilde{\bf a}}}_{2}, a^{\alpha}$, $<{\underline {\tilde {\bf A}}}^{\tilde{1}\spm}>$,  
$<{\underline {\tilde {\bf A}}}^{\tilde{N}_{L}\spm}>)$ --- (in which the higher order corrections in 
all orders are for $a^{\alpha}= 0$ already taken into account, while corrections with $ a^{\alpha}\ne 0$
are presented in Eqs.~(\ref{ianot0} - \ref{sumanot01234F})) to the measured 
 $3 \times3 $ submatrices of the predicted $4 \times 4$ mass matrices, we are able to predict the
 masses of the fourth family
members as well as the matrix elements of the fourth components to the observed free families,  
provided that the mixing $3 \times3 $ submatrices of the predicted $4 \times 4$ mass 
matrices of quarks and leptons would be measured accurately enough --- since the (accurate) 
$3 \times 3$ submatrix of a $4\times 4$ matrix determines $4\times 4$ matrix 
uniquely~\cite{gn2013,gn2015}. 

This means that although we are so far only in principle able to calculate directly the mass matrix 
elements of the $4 \times 4$ mass matrices, predicted by the {\it spin-charge-family}, yet the 
symmetry of mass matrices, discussed in this paper, enables us --- due to the limited number of 
free parameters --- to predict properties of the four families of quarks and leptons to the observed 
three (the masses of the fourth families  and the corresponding mixing matrices)%
~\cite{gn2013,gn2015}.  
{\it We only have to wait for accurate enough data for the $3 \times 3$ mixing (sub)matrices of 
quarks and leptons.} 

Let us add that the right handed neutrino, which is a regular member of the four families,
Table~\ref{Table so13+1.}, has the nonzero value of the operator $Y' A_{s}^{Y'}$
only.





\appendix

\section{Short presentation of the {\it spin-charge-family} theory}
\label{SCFT}

This section follows similar sections in Refs.~\cite{IARD,JMP2015,norma2014MatterAntimatter,%
JMP,NBled2013}.

The {\it spin-charge-family} theory~\cite{IARD,ND2017,NH2017,JMP2015,%
norma2014MatterAntimatter,JMP,NBled2013,NBled2012,norma92,norma93,norma94,%
pikanorma,portoroz03,norma95,gmdn07,gn,gn2013,gn2015,NPLB,N2014scalarprop}
 assumes:

\noindent
 {\bf a.} $\;\;$  A simple action
(Eq.~(\ref{wholeaction})) in an even dimensional space ($d=2n$, $d>5$), $d$ is chosen to be
$(13+1)$. This choice makes that  the action manifests in $d=(3+1)$ in the low energy regime
all the observed degrees of freedom, explaining all
the assumptions of the {\it standard model}, as well as other observed phenomena.

There are two kinds of the Clifford algebra objects,
$\gamma^a$'s and $\tilde{\gamma}^a$'s in this theory with the properties.
\begin{eqnarray}
\label{gammatildegamma}
&& \{ \gamma^a, \gamma^b\}_{+} = 2\eta^{ab}\,, \quad\quad
\{ \tilde{\gamma}^a, \tilde{\gamma}^b\}_{+}= 2\eta^{ab}\,, \quad \quad
\{ \gamma^a, \tilde{\gamma}^b\}_{+} = 0\,.
\end{eqnarray}
Fermions interact with the vielbeins $f^{\alpha}{}_{a}$ and the two kinds of the spin-connection
fields --- $\omega_{ab \alpha}$ and $\tilde{\omega}_{ab \alpha}$ ---
the  gauge fields of $S^{ab} = \,\frac{i}{4} (\gamma^a\, \gamma^b
- \gamma^b\, \gamma^a)\,$ and $\tilde{S}^{ab} = \,\frac{i}{4} (\tilde{\gamma}^a\,
\tilde{\gamma}^b - \tilde{\gamma}^b\, \tilde{\gamma}^a)$, respectively.

The action
\begin{eqnarray}
{\cal A}\,  &=& \int \; d^dx \; E\;\frac{1}{2}\, (\bar{\psi} \, \gamma^a p_{0a} \psi) + h.c. +
\nonumber\\
               & & \int \; d^dx \; E\; (\alpha \,R + \tilde{\alpha} \, \tilde{R})\,,
%
\label{wholeaction}
\end{eqnarray}
in which $p_{0a } = f^{\alpha}{}_a\, p_{0\alpha} + \frac{1}{2E}\, \{ p_{\alpha},
E f^{\alpha}{}_a\}_- $,
$ p_{0\alpha} =  p_{\alpha}  - \frac{1}{2} \, S^{ab}\, \omega_{ab \alpha} -
                    \frac{1}{2} \,  \tilde{S}^{ab} \,  \tilde{\omega}_{ab \alpha} $,   and
$$R =  \frac{1}{2} \, \{ f^{\alpha [ a} f^{\beta b ]} \;(\omega_{a b \alpha, \beta}
- \omega_{c a \alpha}\,\omega^{c}{}_{b \beta}) \} + h.c., $$
$$\tilde{R}  =  \frac{1}{2} \, \{ f^{\alpha [ a} f^{\beta b ]} \;(\tilde{\omega}_{a b \alpha,\beta} -
\tilde{\omega}_{c a \alpha} \,\tilde{\omega}^{c}{}_{b \beta})\} + h.c.$$~\footnote{Whenever
two indexes are equal the summation over these two is meant.},
 introduces two kinds of the Clifford algebra objects, $\gamma^a$ and
 $\tilde{\gamma}^a$,
%
$\{\gamma^a, \gamma^b\}_{+}= 2 \eta^{ab} =
\{\tilde{\gamma}^a, \tilde{\gamma}^b\}_{+}$.
%
$f^{\alpha}{}_{a}$ are vielbeins inverted to $e^{a}{}_{\alpha}$, Latin letters ($a,b,..$) denote
flat indices, Greek letters ($\alpha,\beta,..$) are Einstein indices,  $(m,n,..)$ and $(\mu,\nu,..)$
denote the corresponding indices in ($0,1,2,3$), while $(s,t,..)$ and $(\sigma,\tau,..)$  denote the
corresponding indices in $d\ge5$:
\begin{eqnarray}
\label{vielfe}
e^{a}{}_{\alpha}f^{\beta}{}_{a} &=&\delta^{\beta}_{\alpha}\,, \quad
e^{a}{}_{\alpha}f^{\alpha}{}_{b}= \delta^{a}_{b}\,,
\end{eqnarray}
$E =\det(e^{a}{}_{\alpha})$.

\noindent
{\bf b.} $\;\;$  The {\it spin-charge-family} theory assumes in addition that the manifold $M^{(13+1)}$
breaks first into $M^{(7+1)}$ $\times$ $M^{(6)}$ (which manifests
as $SO(7,1)$ $\times SU(3)$ $\times U(1)$), affecting both internal degrees of freedom --- the
one represented by $\gamma^a$ and the one represented by $\tilde{\gamma}^a$. Since the
left handed (with respect to $M^{(7+1)}$) spinors couple differently to scalar (with respect to
$M^{(7+1)}$) fields than the right handed ones, the break can leave massless and mass
protected $2^{((7+1)/2-1)}$ families~\cite{NHD}. The rest of families get heavy
masses~\footnote{A toy model~\cite{NHD,DN012} was studied in $d=(5+1)$ with the same
action as in Eq.~(\ref{wholeaction}). The break from $d=(5+1)$ to $d=(3+1) \times$ an almost
$S^{2}$ was studied. For a particular choice of vielbeins and for a class of spin connection fields
the manifold $M^{(5+1)}$ breaks into $M^{(3+1)}$ times an almost $S^2$, while
$2^{((3+1)/2-1)}$ families remain massless and mass protected.
Equivalent assumption, although not yet proved how does it really work, is made in the
$d=(13+1)$ case.
 This study is in progress.}.

\noindent
{\bf c.} $\;\;$
There is additional breaking of symmetry: The manifold $M^{(7+1)}$ breaks further into
$M^{(3+1)} \times$  $M^{(4)}$.

\noindent
{\bf d.} $\;\;$
There is a scalar condensate (Table~\ref{Table con.}) of two right handed neutrinos with the
family quantum numbers of the upper four families, bringing masses of the scale $\propto 10^{16}$
GeV or higher to all the vector and scalar gauge fields, which
interact with the condensate~\cite{norma2014MatterAntimatter}.

\noindent
{\bf e.} $\;\;$
There are  the scalar fields with the space index $(7,8)$ carrying the weak ($\tau^{1i}$) and the
hyper charges ($Y=\tau^{23} + \tau^{4}$, $\tau^{1i}$ and $\tau^{2i}$ are generators of the
 subgroups of
$SO(4)$, $\tau^{4}$ and $\tau^{3i}$  are the generators of $U(1)_{II}$ and $SU(3)$, respectively,
which are subgroups of $SO(6)$), which with their nonzero vacuum expectation values change the
properties of the vacuum and break the weak charge and the hyper charge. Interacting with fermions
and with the weak and hyper bosons, they bring masses to heavy bosons and to twice four groups of
families. Carrying no  electromagnetic ($Q=\tau^{13} + Y$) and colour ($\tau^{3i}$) charges and
no $SO(3,1)$ spin, the scalar fields  leave the electromagnetic, colour and gravity fields in $d=(3+1)$
massless.

The assumed action ${\cal A} $ and the assumptions offer: \\
{\bf o.} the explanation for the origin and all the
 properties  of the observed fermions:\\
{\bf o.i.} of the family members, on Table~\ref{Table so13+1.} the family members 
belonging to one Weyl (fundamental) representation of massless spinors of the group $SO(13,1)$
are presented in the "technique"~\cite{norma92,norma93,norma94,pikanorma,portoroz03,%
norma95,hn02,hn03} and  analyzed with respect to the subgroups $SO(3,1)$, $SU(2)_{I}$,
$SU(2)_{II}$, $SU(3)$, $U(1)_{II})$, Eqs.~(\ref{so1+3}, \ref{so42}, \ref{so64}) with the
generators $\tau^{Ai}=\sum_{s,t} c^{Ai}{}_{st}\,S^{st}$,\\
{\bf o.ii.} of  the families analyzed with respect to the subgroups $(\widetilde{SO}(3,1)$,
$\widetilde{SU}(2)_{I}$, $\widetilde{SU}(2)_{II}$, $\widetilde{U}(1)_{{II}})$
with the generators $\tilde{\tau}^{Ai} =\sum_{ab} c^{Ai}{}_{ab}\,\tilde{S}^{st}$,
 Eqs.~(\ref{so1+3tilde}, \ref{so42tilde}, \ref{so64tilde}) --- they are presented  on 
Table~\ref{Table III.} --- all the families are singlets with respect to  $\widetilde{SU}(3)$, \\
{\bf oo.i.} of the observed vector gauge fields of the charges $(SU(2)_{I}, SU(2)_{II},
SU(3), U(1)_{II})$ discussed in Refs.~(\cite{IARD,JMP2015,ND2017}, and the references
therein), all the vector gauge fields are the superposition of $\omega_{st m} $, $A^{Ai}_{m}=
\sum_{s,t} c^{Ai}{}_{st}\,\omega_{st m}$, Eq.~(\ref{gaugevectorAiomega}),\\
{\bf oo.ii.} of the Higgs's scalar and of the Yukawa couplings,  explainable with the scalar
fields with the space index $(7,8)$, there are two groups of two triplets, which are scalar
gauge fields of the charges $\tilde{\tau}^{Ai}$, expressible with the superposition of 
$\tilde{\omega}_{ab s} $, $A^{Ai}_{s}=
\sum_{a,b} c^{Ai}{}_{ab}\,\omega_{ab s}$, Eq.~(\ref{gaugescalarAiomega}),
 and three singlets, the gauge fields of $Q,Q',Y'$,
Eqs.~(\ref{YQY'Q'andtilde}, \ref{gaugescalarAiomega}), all with the weak and
the hyper charges  as assumed by the {\it standard model} for the Higgs's scalars,\\
{\bf oo.iii.} of the scalar fields explaining the origin of the matter-antimatter asymmetry,
 Ref.~\cite{norma2014MatterAntimatter}, \\
{\bf oo.iv.} of the appearance of the dark matter, there are two decoupled groups of four families,
carrying family charges ($\vec{\tilde{N}}_{L}$, $\vec{\tilde{\tau}}^{1}$) and ($\vec{\tilde{N}}_{R}$,
$\vec{\tilde{\tau}}^{2}$), Eqs.~(\ref{so1+3tilde}, \ref{so42tilde}), both groups carry also the family
members charges ($Q,Q',Y'$), Eq.~(\ref{YQY'Q'andtilde}).

The {\it standard model} groups of spins and charges are the subgroups of the $SO(13,1)$ group
with the generator of the infinitesimal transformations expressible with $S^{ab}$ ($=\frac{i}{2}
 (\gamma^a \gamma^b - \gamma^b \gamma^a)$, $\{S^{ab},S^{cd}\}_{-} $
$=- i (\eta^{ad} S^{bc}+ \eta^{bc} S^{ad} - \eta^{ac} S^{bd} - \eta^{bd} S^{ac})$)
 for the spin
\begin{eqnarray}
\label{so1+3}
\vec{N}_{\pm}(= \vec{N}_{(L,R)}): &=& \,\frac{1}{2} (S^{23}\pm i S^{01},S^{31}\pm i S^{02},
S^{12}\pm i S^{03} )\,,
\end{eqnarray}
for the weak charge, $SU(2)_{I}$, and the second $SU(2)_{II}$, these two groups are the
invariant subgroups of $SO(4)$,
 \begin{eqnarray}
 \label{so42}
 \vec{\tau}^{1}:&=&\frac{1}{2} (S^{58}-  S^{67}, \,S^{57} + S^{68}, \,S^{56}-  S^{78} )\,,
\nonumber\\
 \vec{\tau}^{2}:&=& \frac{1}{2} (S^{58}+  S^{67}, \,S^{57} - S^{68}, \,S^{56}+  S^{78} )\,,
 \end{eqnarray}
for the colour charge  $SU(3)$ and for the "fermion charge" $U(1)_{II}$, these two groups are
subgroups of $SO(6)$,
 \begin{eqnarray}
 \label{so64}
 \vec{\tau}^{3}: = &&\frac{1}{2} \,\{  S^{9\;12} - S^{10\;11} \,,
  S^{9\;11} + S^{10\;12} ,\, S^{9\;10} - S^{11\;12} ,\nonumber\\
 && S^{9\;14} -  S^{10\;13} ,\,  S^{9\;13} + S^{10\;14} \,,
  S^{11\;14} -  S^{12\;13}\,,\nonumber\\
 && S^{11\;13} +  S^{12\;14} ,\,
 \frac{1}{\sqrt{3}} ( S^{9\;10} + S^{11\;12} -
 2 S^{13\;14})\}\,,\nonumber\\
 \tau^{4}: = &&-\frac{1}{3}(S^{9\;10} + S^{11\;12} + S^{13\;14})\,,
 \end{eqnarray}
$  \tau^{4}$ is the "fermion charge", while the hyper charge $Y=\tau^{23} + \tau^{4}$.

The generators of the family quantum numbers are the superposition of the generators
$\tilde{S}^{ab}$ ($\tilde{S}^{ab}= \frac{i}{4}\, \{\tilde{\gamma}^a, \tilde{\gamma}^b\}_{-}$,
 $\{\tilde{S}^{ab},\tilde{S}^{cd}\}_{-}=- i (\eta^{ad} \tilde{S}^{bc}+
 \eta^{bc} \tilde{S}^{ad} - \eta^{ac} \tilde{S}^{bd}- \eta^{bd} \tilde{S}^{ac})$,
$\{\tilde{S}^{ab},S^{cd}\}_{-} =0$).
One correspondingly finds the generators of the subgroups of $\widetilde{SO}(7,1)$,
\begin{eqnarray}
\label{so1+3tilde}
\vec{\tilde{N}}_{L,R}:&=& \,\frac{1}{2} (\tilde{S}^{23}\pm i \tilde{S}^{01},
\tilde{S}^{31}\pm i \tilde{S}^{02}, \tilde{S}^{12}\pm i \tilde{S}^{03} )\,,
\end{eqnarray}
which determine representations of the two $\widetilde{SU}(2)$ invariant subgroups of
$\widetilde{SO}(3,1)$,
while
 \begin{eqnarray}
 \label{so42tilde}
 \vec{\tilde{\tau}}^{1}:&=&\frac{1}{2} (\tilde{S}^{58}-  \tilde{S}^{67}, \,\tilde{S}^{57} +
 \tilde{S}^{68}, \,\tilde{S}^{56}-  \tilde{S}^{78} )\,,\;\;\nonumber\\
 \vec{\tilde{\tau}}^{2}:&=&\frac{1}{2} (\tilde{S}^{58}+  \tilde{S}^{67}, \,\tilde{S}^{57} -
 \tilde{S}^{68}, \,\tilde{S}^{56}+  \tilde{S}^{78} )\,,
 \end{eqnarray}
 determine representations of $\widetilde{SU}(2)_{I}\times $ $\widetilde{SU}(2)_{II}$ of $\widetilde{SO}(4)$.
 Both, $\widetilde{SO}(3,1)$ and $\widetilde{SO}(4)$,
 are the subgroups of $\widetilde{SO}(7,1)$. One finds for the infinitesimal generator $\tilde{\tau}^{4}$
 of $\widetilde{U}(1)$, originating in $\widetilde{SO}(6)$, the expression
 \begin{eqnarray}
 \label{so64tilde}
 \tilde{\tau}^{4}: = &&-\frac{1}{3}(\tilde{S}^{9\;10} + \tilde{S}^{11\;12} + \tilde{S}^{13\;14})\,.
 \end{eqnarray}

The operators for the charges $Y$ and $Q$ of the {\it standard model},
 together with  $Q'$ and $Y'$,  and the  corresponding operators of the family charges
$\tilde{Y}$, $\tilde{Y'}$,  $\tilde{Q}$, $\tilde{Q'}$, are defined as follows:
 \begin{eqnarray}
  \label{YQY'Q'andtilde}
  Y= \tau^{4} + \tau^{23}\,,\;\; Y'= -\tau^{4}\tan^2\vartheta_2 + \tau^{23}\,,\;\;
  Q =  \tau^{13} + Y\,,\;\; Q'= -Y \tan^2\vartheta_1 + \tau^{13}&&,\nonumber\\
  \tilde{Y}= \tilde{\tau}^{4} + \tilde{\tau}^{23}\,,\;\; \tilde{Y'}= -\tilde{\tau}^{4}
  \tan^2 \vartheta_2 + \tilde{\tau}^{23}\,,\;\;
  \tilde{Q}= \tilde{Y} + \tilde{\tau}^{13}\;\; \tilde{Q'}= -\tilde{Y} \tan^2 \vartheta_1
  + \tilde{\tau}^{13}&&.
  \end{eqnarray}
 The families split into two groups of four families, each manifesting the
$\widetilde{SU}(2) \times$$\widetilde{SU}(2) \times U(1)$,  with the generators  of the
infinitesimal transformations ($\vec{\tilde{N}}_{L}, \vec{\tilde{\tau}}^{1}, Q,Q',Y'$) and
 ($\vec{\tilde{N}}_{R}, \vec{\tilde{\tau}}^{2}, Q,Q',Y'$), respectively. The generators
of $U(1)$ group ($Q,Q',Y'$), Eq.~\ref{YQY'Q'andtilde}, distinguish  among family members
and are the  same for both groups of four families, presented on Table~\ref{Table III.}, taken
from Ref.~\cite{JMP2015}.
 \begin{table}
 \begin{center}
 \begin{tabular}{|c|c|c|c|c|r r r r r|}
 \hline
 &&&&&$\tilde{\tau}^{13}$&$\tilde{\tau}^{23}$&$\tilde{N}_{L}^{3}$&$\tilde{N}_{R}^{3}$&$\tilde{\tau}^{4}$\\
 \hline
 $I$&$u^{c1}_{R\,1}$&
   $ \stackrel{03}{(+i)}\,\stackrel{12}{[+]}|\stackrel{56}{[+]}\,\stackrel{78}{(+)} ||
   \stackrel{9 \;10}{(+)}\;\;\stackrel{11\;12}{[-]}\;\;\stackrel{13\;14}{[-]}$ &
   $\nu_{R\,1}$&
   $ \stackrel{03}{(+i)}\,\stackrel{12}{[+]}|\stackrel{56}{[+]}\,\stackrel{78}{(+)} ||
   \stackrel{9 \;10}{(+)}\;\;\stackrel{11\;12}{(+)}\;\;\stackrel{13\;14}{(+)}$
  &$-\frac{1}{2}$&$0$&$-\frac{1}{2}$&$0$&$-\frac{1}{2}$
 \\
  $I$&$u^{c1}_{R\,2}$&
   $ \stackrel{03}{[+i]}\,\stackrel{12}{(+)}|\stackrel{56}{[+]}\,\stackrel{78}{(+)} ||
   \stackrel{9 \;10}{(+)}\;\;\stackrel{11\;12}{[-]}\;\;\stackrel{13\;14}{[-]}$ &
   $\nu_{R\,2}$&
   $ \stackrel{03}{[+i]}\,\stackrel{12}{(+)}|\stackrel{56}{[+]}\,\stackrel{78}{(+)} ||
   \stackrel{9 \;10}{(+)}\;\;\stackrel{11\;12}{(+)}\;\;\stackrel{13\;14}{(+)}$
  &$-\frac{1}{2}$&$0$&$\frac{1}{2}$&$0$&$-\frac{1}{2}$
 \\
  $I$&$u^{c1}_{R\,3}$&
   $ \stackrel{03}{(+i)}\,\stackrel{12}{[+]}|\stackrel{56}{(+)}\,\stackrel{78}{[+]} ||
   \stackrel{9 \;10}{(+)}\;\;\stackrel{11\;12}{[-]}\;\;\stackrel{13\;14}{[-]}$ &
   $\nu_{R\,3}$&
   $ \stackrel{03}{(+i)}\,\stackrel{12}{[+]}|\stackrel{56}{(+)}\,\stackrel{78}{[+]} ||
   \stackrel{9 \;10}{(+)}\;\;\stackrel{11\;12}{(+)}\;\;\stackrel{13\;14}{(+)}$
  &$\frac{1}{2}$&$0$&$-\frac{1}{2}$&$0$&$-\frac{1}{2}$
 \\
 $I$&$u^{c1}_{R\,4}$&
  $ \stackrel{03}{[+i]}\,\stackrel{12}{(+)}|\stackrel{56}{(+)}\,\stackrel{78}{[+]} ||
  \stackrel{9 \;10}{(+)}\;\;\stackrel{11\;12}{[-]}\;\;\stackrel{13\;14}{[-]}$ &
  $\nu_{R\,4}$&
  $ \stackrel{03}{[+i]}\,\stackrel{12}{(+)}|\stackrel{56}{(+)}\,\stackrel{78}{[+]} ||
  \stackrel{9 \;10}{(+)}\;\;\stackrel{11\;12}{(+)}\;\;\stackrel{13\;14}{(+)}$
  &$\frac{1}{2}$&$0$&$\frac{1}{2}$&$0$&$-\frac{1}{2}$
  \\
  \hline
  $II$& $u^{c1}_{R\,5}$&
        $ \stackrel{03}{[+i]}\,\stackrel{12}{[+]}|\stackrel{56}{[+]}\,\stackrel{78}{[+]}||
        \stackrel{9 \;10}{(+)}\;\;\stackrel{11\;12}{[-]}\;\;\stackrel{13\;14}{[-]}$ &
        $\nu_{R\,5}$&
        $ \stackrel{03}{[+i]}\,\stackrel{12}{[+]}|\stackrel{56}{[+]}\,\stackrel{78}{[+]}||
        \stackrel{9 \;10}{(+)}\;\;\stackrel{11\;12}{(+)}\;\;\stackrel{13\;14}{(+)}$
        &$0$&$-\frac{1}{2}$&$0$&$-\frac{1}{2}$&$-\frac{1}{2}$
 \\
  $II$& $u^{c1}_{R\,6}$&
      $ \stackrel{03}{(+i)}\,\stackrel{12}{(+)}|\stackrel{56}{[+]}\,\stackrel{78}{[+]}||
      \stackrel{9 \;10}{(+)}\;\;\stackrel{11\;12}{[-]}\;\;\stackrel{13\;14}{[-]}$ &
      $\nu_{R\,6}$&
      $ \stackrel{03}{(+i)}\,\stackrel{12}{(+)}|\stackrel{56}{[+]}\,\stackrel{78}{[+]}||
      \stackrel{9 \;10}{(+)}\;\;\stackrel{11\;12}{(+)}\;\;\stackrel{13\;14}{(+)}$
      &$0$&$-\frac{1}{2}$&$0$&$\frac{1}{2}$&$-\frac{1}{2}$
 \\
 $II$& $u^{c1}_{R\,7}$&
 $ \stackrel{03}{[+i]}\,\stackrel{12}{[+]}|\stackrel{56}{(+)}\,\stackrel{78}{(+)}||
 \stackrel{9 \;10}{(+)}\;\;\stackrel{11\;12}{[-]}\;\;\stackrel{13\;14}{[-]}$ &
      $\nu_{R\,7}$&
      $ \stackrel{03}{[+i]}\,\stackrel{12}{[+]}|\stackrel{56}{(+)}\,\stackrel{78}{(+)}||
      \stackrel{9 \;10}{(+)}\;\;\stackrel{11\;12}{(+)}\;\;\stackrel{13\;14}{(+)}$
    &$0$&$\frac{1}{2}$&$0$&$-\frac{1}{2}$&$-\frac{1}{2}$
  \\
   $II$& $u^{c1}_{R\,8}$&
    $ \stackrel{03}{(+i)}\,\stackrel{12}{(+)}|\stackrel{56}{(+)}\,\stackrel{78}{(+)}||
    \stackrel{9 \;10}{(+)}\;\;\stackrel{11\;12}{[-]}\;\;\stackrel{13\;14}{[-]}$ &
    $\nu_{R\,8}$&
    $ \stackrel{03}{(+i)}\,\stackrel{12}{(+)}|\stackrel{56}{(+)}\,\stackrel{78}{(+)}||
    \stackrel{9 \;10}{(+)}\;\;\stackrel{11\;12}{(+)}\;\;\stackrel{13\;14}{(+)}$
    &$0$&$\frac{1}{2}$&$0$&$\frac{1}{2}$&$-\frac{1}{2}$
 \\
 \hline
 \end{tabular}
 \end{center}
\caption{\label{Table III.}
Eight families of the right handed $u^{c1}_{R}$ (\ref{Table so13+1.})
quark with spin $\frac{1}{2}$, the colour charge $(\tau^{33}=1/2$, $\tau^{38}=1/(2\sqrt{3})$
[the definition of the operators is presented in Eqs.~(\ref{so42},\ref{so64}), 
the definition of the operators, expressible with $\tilde{S}^{ab}$ is presented: 
 $\vec{\tilde{N}}_{L,R}$  (Eq.~(\ref{so1+3tilde})), 
$\vec{\tilde{\tau}}^{1}$   (Eq.~(\ref{so42tilde})),
 $\vec{\tilde{\tau}}^{2}$   (Eq.~(\ref{so42tilde})), 
$\tilde{\tau}^{4} $   (Eq.~(\ref{so64tilde}))] 
and of  the colourless right handed neutrino $\nu_{R}$ of spin $\frac{1}{2}$ 
are presented in the  left and in the right column, respectively.
They belong to two groups of four families, one ($I$) is a doublet with respect to
($\vec{\tilde{N}}_{L}$ and  $\vec{\tilde{\tau}}^{1}$) and  a singlet with respect to
($\vec{\tilde{N}}_{R}$ and  $\vec{\tilde{\tau}}^{2}$), the other ($II$) is a singlet with respect to
($\vec{\tilde{N}}_{L}$ and  $\vec{\tilde{\tau}}^{1}$) and  a doublet with with respect to
($\vec{\tilde{N}}_{R}$ and  $\vec{\tilde{\tau}}^{2}$).
All the families follow from the starting one by the application of the operators
($\tilde{N}^{\pm}_{R,L}$, $\tilde{\tau}^{(2,1)\pm}$), Eq.~(\ref{plusminus}).  The generators
($N^{\pm}_{R,L} $, $\tau^{(2,1)\pm}$) (Eq.~(\ref{plusminus}))
transform $u_{Ri}, i=(1,\cdots,8),$ to all the members of the same colour of the $i^{th}$ family.
The same generators transform equivalently the right handed   neutrino $\nu_{Ri},  i=(1,\cdots,8)$,
 to all the colourless members of the  $i^{th}$  family.
}
 \end{table}

The vector gauge fields of the charges $\vec{\tau}^{1}$, $\vec{\tau}^{2}$, $\vec{\tau}^{3}$
and $\tau^{4}$ follow from the requirement $\sum_{Ai} \tau^{Ai} A^{Ai}_{m}=\sum_{s,t}
\frac{1}{2}\,S^{st}\, \omega_{st m}$ and the requirement that
%
$\tau^{Ai} = \sum_{a,b} \;c^{Ai}{ }_{ab} \; S^{ab}$, Eq.~(\ref{tau}),
%
fulfilling the commutation relations
%
$\{\tau^{Ai}, \tau^{Bj}\}_- = i \delta^{AB} f^{Aijk} \tau^{Ak}$, Eq.~(\ref{taucom}).
%
Correspondingly we find
%
$A^{Ai}_{m} = \sum_{s,t} \;c^{Ai}{ }_{st} \; \omega^{st}{}_{m}$, Eq.~(\ref{AAiomega}),
%
with $(s,t)$  either in $ (5,6,7,8)$ or in
$ (9,\dots,14)$.

The explicit expressions for these vector gauge fields in terms of $\omega_{stm}$~%
[\cite{JMP2015}, Eq. (22)],~ \cite{norma2014MatterAntimatter}]
are presented in the case that the electroweak $\vartheta_{1}=\vartheta_{W}$ is zero and so is 
$\vartheta_{2}$ and in the case that the two angles,  $(\vartheta_{1}, \vartheta_{2})$, are not zero.
\begin{eqnarray}
\vec{A}^{1}_{m} &=& (\omega_{58 m} - \omega_{67 m}, \omega_{57 m} +
\omega_{68 m}, \omega_{56 m} - \omega_{78 m})\,,   \nonumber\\
\vec{A}^{2}_{m} &=& (\omega_{58 m} + \omega_{67 m}, \omega_{57 m} -
\omega_{68 m}, \omega_{56 m} +  \omega_{78 m})\,,   \nonumber\\
A^{Q}_{m} &=& \omega_{56 m} - (\omega_{9\,10 m} + \omega_{11\,12 m}
+ \omega_{13\,14 m})\,, \nonumber\\
A^{Y}_{m} &=& (\omega_{56 m} + \omega_{78 m}) - (\omega_{9\,10 m} +
\omega_{11\,12 m} + \omega_{13\,14 m})\,, \nonumber\\
\vec{A}^{3}_{m} &=& (\omega_{9\,12 m} - \omega_{10\,11 m}, \omega_{9\,11 m}
+ \omega_{10\,12 m}, \omega_{9\,10 m} - \omega_{11\,12 m}, \nonumber\\
&& \omega_{9\,14 m} - \omega_{10\,13 m}, \omega_{9\,13 m} +
\omega_{10\,14 m}, \omega_{11\,14 m} - \omega_{12\,13 m},\nonumber\\
&& \omega_{11\,13 m} + \omega_{12\,14 m}, \frac{1}{\sqrt{3}}\,(\omega_{9\,10 m} +
\omega_{11\,12 m} - 2 \omega_{13\,14 m}))\,,   \nonumber\\
A^{4}_{m} &=& (\omega_{9\,10 m} + \omega_{11\,12 m} + \omega_{13\,14 m})\,,
\nonumber\\
A^{Q}_{m} &=& \sin \vartheta_{1} \,A^{13}_{m} + \cos \vartheta_{1} \,A^{Y}_{m}\,,
\nonumber\\
A^{Q'}_{m}  &=& \cos \vartheta_{1} \,A^{13}_{m} - \sin \vartheta_{1} \,A^{Y}_{m}\,,
\nonumber\\
A^{Y'}_{m}&=&\cos \vartheta_{2} \,A^{23}_{m} - \sin \vartheta_{2} \,A^{4}_{m}\,,
\nonumber\\
          & &(m\in (0,1,2,3))\,. 
\label{gaugevectorAiomega}
\end{eqnarray}
All $\omega_{st m}$ vector gauge fields are real fields.
Here the fields contain in general the coupling constants which are not necessarily the same for 
all of them.  The angle $\vartheta_1$  is the angle of the electroweak break, while
$\vartheta_2$ is the angle of breaking the $SU(2)_{II}$ and $U(1)_{II}$ at much higher
scale~[\cite{norma2014MatterAntimatter,JMP2015} and references therein].

One obtains in a similar way the scalar gauge fields, which determine mass matrices of family
members. They carry the space index $s=(7,8)$. The scalar fields contain in general the coupling 
constants.  Before the electroweak break the electroweak angle $\vartheta_{1}=\vartheta_{W}$
 is zero,  while $\vartheta_{2}$ is the angle determined by the break of symmetry at much higher scale.%
\begin{eqnarray}
\vec{\tilde{A}}^{1}_{s} &=& (\tilde{\omega}_{58 s} - \tilde{\omega}_{67 s},
\tilde{\omega}_{57 s} + \tilde{\omega}_{68 s}, \tilde{\omega}_{56 s} -
\tilde{\omega}_{78 s})\,,   \nonumber\\
\vec{\tilde{A}}^{2}_{s} &=& (\tilde{\omega}_{58 s} + \tilde{\omega}_{67 s},
\tilde{\omega}_{57 s} - \tilde{\omega}_{68 s}, \tilde{\omega}_{56 s} +
\tilde{\omega}_{78 s})\,,   \nonumber\\
\vec{\tilde{A}}^{N_{L}}_{s} &=& (\tilde{\omega}_{23 s} + i \tilde{\omega}_{01 s},
\tilde{\omega}_{31 s} + i \tilde{\omega}_{02s}, \tilde{\omega}_{12 s} +
i\tilde{\omega}_{03 s})\,,   \nonumber\\
\vec{\tilde{A}}^{N_{R}}_{s} &=& (\tilde{\omega}_{23 s} - i \tilde{\omega}_{01 s},
\tilde{\omega}_{31 s} - i \tilde{\omega}_{02s}, \tilde{\omega}_{12 s} - i
\tilde{\omega}_{03 s})\,,   \nonumber\\
A^{Q}_{s} &=& \omega_{56 s} - (\omega_{9\,10 s} + \omega_{11\,12 s}
+ \omega_{13\,14 s})\,, \nonumber\\
A^{Y}_{s} &=& (\omega_{56 s} + \omega_{78 s}) - (\omega_{9\,10 s} +
\omega_{11\,12 s} + \omega_{13\,14 s})\, \nonumber\\
A^{4}_{s} &=&- (\omega_{9\,10 s} + \omega_{11\,12 s}
+ \omega_{13\,14 s})\,,\nonumber\\
A^{Q}_{s} &=& \sin \vartheta_{1} \,A^{13}_{s} + \cos \vartheta_{1} \,A^{Y}_{s}\,,\quad
A^{Q'}_{s}  = \cos \vartheta_{1} \,A^{13}_{s} - \sin \vartheta_{1} \,A^{Y}_{s}\,,\nonumber\\
A^{Y'}_{s}&=&\cos \vartheta_{2} \,A^{23}_{s} - \sin \vartheta_{2} \,A^{4}_{s}\,,\nonumber\\
          & &(s\in (7,8))\,. 
\label{gaugescalarAiomega}
\end{eqnarray}
All $\omega_{st s'}$, $\tilde{\omega}_{st s'} $, $(s,t,s')=(5,\dots,14)$, $\tilde{\omega}_{i,j, s'}$
and  $i\, \tilde{\omega}_{0, s'}$, $(i,j) =(1,2,3)$ scalar gauge fields are real fields.

The theory predicts, due to commutation relations of generators of the infinitesimal transformations
of the family groups, $\widetilde{SU}(2)_{I}$ $\times \widetilde{SU}(2)_{I}$ 
and  $\widetilde{SU}(2)_{II}$ $\times \widetilde{SU}(2)_{II}$, 
the first one with the generators $\vec{\tilde{N}}_{L}$ and $\vec{\tilde{\tau}}^{1}$, and the
second one with the generators  $\vec{\tilde{N}}_{R}$ and $\vec{\tilde{\tau}}^{2}$,
Eqs.~(\ref{so1+3tilde},\ref{so42tilde}), two groups of four families.

 The theory offers (so far) several predictions:\\
 {\bf i.} several new scalars, those coupled to the lower group of four families --- two triplets and
three singlets, the superposition of ($\vec{\tilde{A}}^{1}_{s}$, $\vec{\tilde{A}}^{N}_{Ls}$ and
$A^{Q}_{s}, A^{Y}_{s}, A^{4}_{s}$, Eq.~(\ref{gaugescalarAiomega})) --- some of them
to be observed at the LHC~(\cite{IARD,norma2014MatterAntimatter,JMP2015}),\\
{\bf ii.} the fourth family  to the observed three to be observed at the
LHC~(\cite{IARD,norma2014MatterAntimatter,JMP2015} and the references therein),\\
{\bf iii.} new nuclear force among nucleons among quarks of the upper four families.

The theory offers also the explanation for several phenomena, like it is the "miraculous"
cancellation of the {\it standard model}  triangle anomalies~\cite{NH2017}.


The breaks of the symmetries, manifesting
in Eqs.~(\ref{so1+3}, \ref{so1+3tilde}, \ref{so42},  \ref{so42tilde}, \ref{so64}, \ref{so64tilde}),
are in the {\it spin-charge-family} theory caused by
the scalar condensate of the two right handed neutrinos belonging to one group of four families,
Table~\ref{Table con.}, and by the nonzero vacuum expectation values of the scalar fields
carrying the space index
$(7,8)$ (Refs.~\cite{JMP2015,IARD} and the references therein). The space breaks first to
$SO(7,1)$ $\times SU(3) \times U(1)_{II}$ and then further to $SO(3,1)\times SU(2)_{I} $
$\times U(1)_{I}$ $\times SU(3) \times U(1)_{II}$, what explains the connections between
 the weak and the hyper charges and the handedness of spinors~\cite{NH2017}.


%
 \begin{table}
 \begin{center}
 \begin{tabular}{c|c c c r c r r |c c c c c c c}
 \hline
 state & $S^{03}$& $ S^{12}$ & $\tau^{13}$& $\tau^{23}$ &$\tau^{4}$& $Y$&$Q$&$\tilde{\tau}^{13}$&
 $\tilde{\tau}^{23}$&$\tilde{\tau}^{4}$&$\tilde{Y} $& $\tilde{Q}$&$\tilde{N}_{L}^{3}$& $\tilde{N}_{R}^{3}$
 \\
 \hline
 ${\bf (|\nu_{1 R}^{VIII}>_{1}\,|\nu_{2 R}^{VIII}>_{2})}$
 & $0$& $0$& $0$& $1$ & $-1$ & $0$ & $0$&$0$&$1$&$-1$& $0$& $0$& $0$& $1$\\
 \hline
 $ (|\nu_{1 R}^{VIII}>_{1}|e_{2 R}^{VIII}>_{2})$
 & $0$& $0$& $0$& $0$ & $-1$ & $-1$& $-1$ & $0$ &$1$&$-1$& $0$& $0$& $0$& $1$\\
 $ (|e_{1 R}^{VIII}>_{1}|e_{2 R}^{VIII}>_{2})$
 & $0$& $0$& $0$& $-1$& $-1$ & $-2$& $-2$ & $0$ &$1$&$-1$& $0$& $0$& $0$& $1$\\
 \hline
 \end{tabular}
 \end{center}
\caption{\label{Table con.} This table is taken from~\cite{norma2014MatterAntimatter}.
The condensate of the two right handed neutrinos $\nu_{R}$,  with the $VIII^{th}$
family quantum numbers, coupled to spin zero and belonging to a triplet with
respect to the generators $\tau^{2i}$, is presented together with its two partners.
The right handed neutrino has $Q=0=Y$. The triplet carries $\tau^4=-1$, $\tilde{\tau}^{23}=1$,
$\tilde{\tau}^{4} =-1$, $\tilde{N}_{R}^3 = 1$, $\tilde{N}_{L}^3 = 0$, $\tilde{Y}=0 $,
$\tilde{Q}=0$, $\tilde{\tau}^{31}=0$.
The family quantum numbers are presented in Table~\ref{Table III.}.
}
 \end{table}
The stable of the upper four families is the candidate for the dark matter, the fourth of the lower
four families is predicted to be measured at the LHC.

\section{Short presentation of spinor technique~\cite{IARD,JMP2015,norma93,hn02,hn03}}
\label{technique}

This appendix is a short review (taken from~\cite{JMP2015}) of the technique~\cite{norma93,DKhn,%
hn02,hn03}, initiated and developed in Ref.~\cite{norma93} by one of the authors (N.S.M.B.), while
proposing the {\it spin-charge-family} theory~\cite{ND2017,JMP2015,%
norma2014MatterAntimatter,NBled2013,NBled2012,IARD,pikanorma,portoroz03,norma92,%
norma93,norma94,norma95,gmdn07,gn,gn2013,gn2015,NPLB,N2014scalarprop}.
All the internal degrees of freedom of spinors, with family quantum numbers included, are describable
with two kinds of the Clifford algebra objects, besides with $\gamma^a$'s, used in this theory to describe
spins and all the charges of fermions, also with  $\tilde{\gamma}^a$'s, used in this theory to describe
families of spinors:
\begin{eqnarray}
\label{gammatildegamma}
&& \{ \gamma^a, \gamma^b\}_{+} = 2\eta^{ab}\,, \quad\quad
\{ \tilde{\gamma}^a, \tilde{\gamma}^b\}_{+}= 2\eta^{ab}\,, \quad\quad
\{ \gamma^a, \tilde{\gamma}^b\}_{+} = 0\,.
\end{eqnarray}
We assume  the ``Hermiticity'' property for $\gamma^a$'s  (and $\tilde{\gamma}^a$'s)
$\gamma^{a\dagger} = \eta^{aa} \gamma^a$ (and $\tilde{\gamma}^{a\dagger} =
\eta^{aa} \tilde{\gamma}^a$),
%
in order that $\gamma^a$ (and $\tilde{\gamma}^a$) are compatible with (\ref{gammatildegamma})
and formally unitary, i.e. $\gamma^{a \,\dagger} \,\gamma^a=I$
 (and $\tilde{\gamma}^{a\,\dagger} \tilde{\gamma}^a=I$).
One correspondingly finds  that $(S^{ab})^{\dagger} = \eta^{aa} \eta^{bb}S^{ab}$ (and
 $(\tilde{S}^{ab})^{\dagger} = \eta^{aa} \eta^{bb} \tilde{S}^{ab}$).

Spinor states are represented as products of nilpotents and projectors, formed as odd and even
objects of $\gamma^a$'s, respectively, chosen to be the eigenstates of a Cartan subalgebra of
the Lorentz  groups defined by $\gamma^a$'s
\begin{eqnarray}
\stackrel{ab}{(k)}:&=&
\frac{1}{2}(\gamma^a + \frac{\eta^{aa}}{ik} \gamma^b)\,,\quad \quad
\stackrel{ab}{[k]}:=
\frac{1}{2}(1+ \frac{i}{k} \gamma^a \gamma^b)\,,
\label{signature}
\end{eqnarray}
where $k^2 = \eta^{aa} \eta^{bb}$.
We further have~\cite{JMP2015}
\begin{eqnarray}
\gamma^{a}\,\stackrel{ab}{(k)}:&=&
\frac{1}{2}(\gamma^a  \gamma^a + \frac{\eta^{aa}}{ik} \gamma^a \gamma^b)=
\eta^{aa}\,\stackrel{ab}{[-k]}\,,\quad \gamma^{a}\,\stackrel{ab}{[k]}:=
\frac{1}{2}(\gamma^a+ \frac{i}{k} \gamma^a \gamma^a \gamma^b)=\stackrel{ab}{(-k)}\,,\nonumber\\
\tilde{\gamma}^{a}\,\stackrel{ab}{(k)}:&=&
-i \frac{1}{2}(\gamma^a + \frac{\eta^{aa}}{ik} \gamma^b) \gamma^a= -i \eta^{aa}
 \stackrel{ab}{[k]}\,,\quad \tilde{\gamma}^{a}\,\stackrel{ab}{[k]}:=
i \frac{1}{2}(1+ \frac{ i}{k}\gamma^a \gamma^b) \gamma^a=- i \stackrel{ab}{(k)}\,,
\label{signature1}
\end{eqnarray}
where we assume that all the operators apply on the vacuum state  $|\psi_0\rangle$.
We define a vacuum state $|\psi_0>$ so that one finds
%
$< \;\stackrel{ab}{(k)}^{\dagger}  \stackrel{ab}{(k)}\; > = 1\,, \quad
< \;\stackrel{ab}{[k]}^{\dagger}
 \stackrel{ab}{[k]}\; > = 1$.
%

We recognize 
that $\gamma^a$
transform  $\stackrel{ab}{(k)}$ into  $\stackrel{ab}{[-k]}$, never to $\stackrel{ab}{[k]}$,
while $\tilde{\gamma}^a$ transform  $\stackrel{ab}{(k)}$ into $\stackrel{ab}{[k]}$, never to
$\stackrel{ab}{[-k]}$
\begin{eqnarray}
&&\gamma^a \stackrel{ab}{(k)}= \eta^{aa}\stackrel{ab}{[-k]},\;
\gamma^b \stackrel{ab}{(k)}= -ik \stackrel{ab}{[-k]}, \;
\gamma^a \stackrel{ab}{[k]}= \stackrel{ab}{(-k)},\;
\gamma^b \stackrel{ab}{[k]}= -ik \eta^{aa} \stackrel{ab}{(-k)}\,,\nonumber\\
&&\tilde{\gamma^a} \stackrel{ab}{(k)} = - i\eta^{aa}\stackrel{ab}{[k]},\;
\tilde{\gamma^b} \stackrel{ab}{(k)} =  - k \stackrel{ab}{[k]}, \;
\tilde{\gamma^a} \stackrel{ab}{[k]} =  \;\;i\stackrel{ab}{(k)},\;
\tilde{\gamma^b} \stackrel{ab}{[k]} =  -k \eta^{aa} \stackrel{ab}{(k)}\,.
\label{snmb:gammatildegamma}
\end{eqnarray}

The Clifford algebra objects $S^{ab}$ and $\tilde{S}^{ab}$ close the algebra of the Lorentz
group
\begin{eqnarray}
\label{sabtildesab}
S^{ab}: &=& (i/4) (\gamma^a \gamma^b - \gamma^b \gamma^a)\,, \nonumber\\
\tilde{S}^{ab}: &=& (i/4) (\tilde{\gamma}^a \tilde{\gamma}^b
- \tilde{\gamma}^b \tilde{\gamma}^a)\,,
\end{eqnarray}
$ \{S^{ab}, \tilde{S}^{cd}\}_{-}= 0\,$, 
$\{S^{ab},S^{cd}\}_{-} = $ $ i(\eta^{ad} S^{bc} + \eta^{bc} S^{ad} - \eta^{ac} S^{bd} - \eta^{bd} S^{ac})\,$,
$\{\tilde{S}^{ab},\tilde{S}^{cd}\}_{-} $ $= i(\eta^{ad} \tilde{S}^{bc} + \eta^{bc} \tilde{S}^{ad}
- \eta^{ac} \tilde{S}^{bd} - \eta^{bd} \tilde{S}^{ac})\,$.

One can easily check that the nilpotent $\stackrel{ab}{(k)}$ and the projector $\stackrel{ab}{[k]}$
are "eigenstates" of $S^{ab}$ and $\tilde{S}^{ab}$
\begin{eqnarray}
        &&S^{ab}\, \stackrel{ab}{(k)}= \frac{1}{2}\,k\, \stackrel{ab}{(k)}\,,\quad \quad
        S^{ab}\, \stackrel{ab}{[k]}= \;\;\frac{1}{2}\,k \,\stackrel{ab}{[k]}\,,\nonumber\\
&&\tilde{S}^{ab}\, \stackrel{ab}{(k)}= \frac{1}{2}\,k \,\stackrel{ab}{(k)}\,,\quad \quad
\tilde{S}^{ab}\, \stackrel{ab}{[k]}=-\frac{1}{2}\,k \,\stackrel{ab}{[k]}\,,
\label{grapheigen}
\end{eqnarray}
where the vacuum state $|\psi_0\rangle$ is meant to stay on the right hand sides of projectors
and nilpotents. This means that multiplication of nilpotents $\stackrel{ab}{(k)}$
and projectors $\stackrel{ab}{[k]}$ by $S^{ab}$  get the same objects back multiplied
by the constant $\frac{1}{2}k$, while $\tilde{S}^{ab}$ multiply $\stackrel{ab}{(k)}$ by
$\frac{k}{2}$ and $\stackrel{ab}{[k]}$ by $(-\frac{k}{2})$ (rather than by $\frac{k}{2}$).
This also means that when
$\stackrel{ab}{(k)}$ and $\stackrel{ab}{[k]}$ act from the left hand side on  a
vacuum state $|\psi_0\rangle$ the obtained states are the eigenvectors of $S^{ab}$.

The technique can be used to construct a spinor basis for any dimension $d$ and any signature in an
easy and transparent way. Equipped with nilpotents and projectors of Eq.~(\ref{signature}),
the technique offers an elegant way to see all the quantum numbers of states with respect to the
two Lorentz groups, as well as transformation properties of the states under the application of any
Clifford algebra object.

Recognizing from Eq.(\ref{sabtildesab})  that the two Clifford algebra objects ($S^{ab}, S^{cd}$)
with all indexes different commute (and equivalently for ($\tilde{S}^{ab},\tilde{S}^{cd}$)), we
select  the  Cartan subalgebra of the algebra of the
two groups, which  form  equivalent representations with respect to one another
\begin{eqnarray}
S^{03}, S^{12}, S^{56}, \cdots, S^{d-1\; d}, \quad {\rm if } \quad d &=& 2n\ge 4,
\nonumber\\
\tilde{S}^{03}, \tilde{S}^{12}, \tilde{S}^{56}, \cdots, \tilde{S}^{d-1\; d},
\quad {\rm if } \quad d &=& 2n\ge 4\,.
\label{choicecartan}
\end{eqnarray}

The choice of  the Cartan subalgebra in $d <4$ is straightforward.
It is  useful  to define one of the Casimirs of the Lorentz group ---
the  handedness $\Gamma$ ($\{\Gamma, S^{ab}\}_- =0$) (as well as $\tilde{\Gamma}$)
in any $d=2n$
\begin{eqnarray}
\Gamma^{(d)} :&=&(i)^{d/2}\; \;\;\;\;\;\prod_a \quad (\sqrt{\eta^{aa}} \gamma^a), \quad {\rm if } \quad d = 2n,
\nonumber\\
\tilde{\Gamma}^{(d)} :&=& (i)^{(d-1)/2}\; \prod_a \quad (\sqrt{\eta^{aa}} \tilde{\gamma}^a),
\quad {\rm if } \quad d = 2n\,.
\label{hand}
\end{eqnarray}
%
 We understand the product of $\gamma^a$'s in the ascending order with
respect to the index $a$: $\gamma^0 \gamma^1\cdots \gamma^d$.
It follows from the Hermiticity properties of $\gamma^a $ 
for any choice of the signature $\eta^{aa}$ that $\Gamma^{\dagger}= \Gamma,\;
\Gamma^2 = I. ($ Equivalent relations are valid for $\tilde{\Gamma}$.)
We also find that for $d$ even the handedness  anticommutes with the Clifford algebra objects
$\gamma^a$ ($\{\gamma^a, \Gamma \}_+ = 0$) (while for $d$ odd it commutes with
$\gamma^a$ ($\{\gamma^a, \Gamma \}_- = 0$)).

Taking into account the above equations it is easy to find a Weyl spinor irreducible representation
for $d$-dimensional space, with $d$ even or odd~\footnote{For $d$ odd
 the basic states are products of $(d-1)/2$ nilpotents and a factor $(1\pm \Gamma)$.}.
For $d$ even we simply make a starting state as a product of $d/2$, let us say, only nilpotents
$\stackrel{ab}{(k)}$, one for each $S^{ab}$ of the Cartan subalgebra elements
(Eqs.(\ref{choicecartan}, \ref{sabtildesab})), applying it on an (unimportant) vacuum
state.
Then the generators $S^{ab}$, which do not belong to the Cartan subalgebra, being applied on
the starting state from the left hand side,  generate all the members of one Weyl spinor.
\begin{eqnarray}
\stackrel{0d}{(k_{0d})} \stackrel{12}{(k_{12})} \stackrel{35}{(k_{35})}\cdots
\stackrel{d-1\;d-2}{(k_{d-1\;d-2})}
|\psi_0 \,>\nonumber\\
\stackrel{0d}{[-k_{0d}]} \stackrel{12}{[-k_{12}]} \stackrel{35}{(k_{35})}\cdots
\stackrel{d-1\;d-2}{(k_{d-1\;d-2})}
|\psi_0 \,>\nonumber\\
\stackrel{0d}{[-k_{0d}]} \stackrel{12}{(k_{12})} \stackrel{35}{[-k_{35}]}\cdots
\stackrel{d-1\;d-2}{(k_{d-1\;d-2})}
|\psi_0 \,>\nonumber\\
\vdots \nonumber\\
\stackrel{0d}{[-k_{0d}]} \stackrel{12}{(k_{12})} \stackrel{35}{(k_{35})}\cdots
\stackrel{d-1\;d-2}{[-k_{d-1\;d-2}]}
|\psi_0 \,>\nonumber\\
\stackrel{od}{(k_{0d})} \stackrel{12}{[-k_{12}]} \stackrel{35}{[-k_{35}]}\cdots
\stackrel{d-1\;d-2}{(k_{d-1\;d-2})}
|\psi_0\,> \nonumber\\
\vdots
\label{graphicd}
\end{eqnarray}
All the states have the same handedness $\Gamma $, since $\{ \Gamma, S^{ab}\}_{-} = 0$.
States, belonging to one multiplet  with respect to the group $SO(q,d-q)$, that is to one
irreducible representation of spinors (one Weyl spinor), can have any phase. We could make a choice
of the simplest one, taking all  phases equal to one. (In order to have the usual transformation
properties for spinors under the rotation of spin and under
${\cal C}_{{\cal N}}$ ${\cal P}_{{\cal N}}$,some of the states must be multiplied by $(-1)$.)

The above  representation demonstrates that for $d$ even
all the states of one irreducible Weyl representation of a definite handedness follow from a starting
state, which is, for example, a product of nilpotents $\stackrel{ab}{(k_{ab})}$, by transforming all
possible pairs of $\stackrel{ab}{(k_{ab})} \stackrel{mn}{(k_{mn})}$ into
 $\stackrel{ab}{[-k_{ab}]} \stackrel{mn}{[-k_{mn}]}$.
There are $S^{am}, S^{an}, S^{bm}, S^{bn}$, which do this.
The procedure gives $2^{(d/2-1)}$ states. A Clifford algebra object $\gamma^a$ being applied
from  the left hand side, transforms  a Weyl spinor of one handedness into a Weyl spinor of the
opposite handedness.

We shall speak about left handedness when $\Gamma= -1$ and about right
handedness when $\Gamma =1$
.

While $S^{ab}$, which do not belong to the Cartan subalgebra (Eq.~(\ref{choicecartan})), generate
all the states of one representation,  $\tilde{S}^{ab}$, which do not belong to the
Cartan subalgebra (Eq.~(\ref{choicecartan})),  generate the states of $2^{d/2-1}$ equivalent
 representations.

Making a choice of the Cartan subalgebra set (Eq.~(\ref{choicecartan}))
of the algebra $S^{ab}$ and
$\tilde{S}^{ab}$:
%
($S^{03}, S^{12}, S^{56}, S^{78}, S^{9 \;10}, S^{11\;12}, S^{13\; 14}\,$), 
($\tilde{S}^{03}, \tilde{S}^{12}, \tilde{S}^{56}, \tilde{S}^{78}, \tilde{S}^{9 \;10},
\tilde{S}^{11\;12}, \tilde{S}^{13\; 14}\,$),
%
a left handed ($\Gamma^{(13,1)} =-1$) eigenstate of all the members of the
Cartan  subalgebra, representing a weak chargeless  $u_{R}$-quark with spin up, hyper charge ($2/3$)
and  colour ($1/2\,,1/(2\sqrt{3})$), for example, can be written as 
\begin{eqnarray}
&& \stackrel{03}{(+i)}\stackrel{12}{(+)}|\stackrel{56}{(+)}\stackrel{78}{(+)}
||\stackrel{9 \;10}{(+)}\stackrel{11\;12}{[-]}\stackrel{13\;14}{[-]} |\psi_{0} \rangle = \nonumber\\
&&\frac{1}{2^7}
(\gamma^0 -\gamma^3)(\gamma^1 +i \gamma^2)| (\gamma^5 + i\gamma^6)(\gamma^7 +i \gamma^8)||
\nonumber\\
&& (\gamma^9 +i\gamma^{10})(1- i \gamma^{11}\gamma^{12})(1- i\gamma^{13}\gamma^{14})
|\psi_{0} \rangle \,.
\label{start}
\end{eqnarray}
This state is an eigenstate of all $S^{ab}$ and $\tilde{S}^{ab}$ which are members of the Cartan
subalgebra (Eq.~(\ref{choicecartan})).

The operators $ \tilde{S}^{ab}$, which do not belong to the Cartan subalgebra
 (Eq.~(\ref{choicecartan})), generate families from the starting $u_R$ quark, transforming the
$u_R$ quark from Eq.~(\ref{start}) to the $u_R$ of another family,  keeping all of the properties
with respect to $S^{ab}$ unchanged.
In particular, $\tilde{S}^{01}$ applied on a right handed $u_R$-quark
from Eq.~(\ref{start}) generates a
state which is again  a right handed $u_{R}$-quark,  weak chargeless,  with spin up,
hyper charge ($2/3$)
and the colour charge ($1/2\,,1/(2\sqrt{3})$)
\begin{eqnarray}
\label{tildesabfam}
\tilde{S}^{01}\;
\stackrel{03}{(+i)}\stackrel{12}{(+)}| \stackrel{56}{(+)} \stackrel{78}{(+)}||
\stackrel{9 10}{(+)} \stackrel{11 12}{[-]} \stackrel{13 14}{[-]}= -\frac{i}{2}\,
&&\stackrel{03}{[\,+i]} \stackrel{12}{[\,+\,]}| \stackrel{56}{(+)} \stackrel{78}{(+)}||
\stackrel{9 10}{(+)} \stackrel{11 12}{[-]} \stackrel{13 14}{[-]}\,.
\end{eqnarray}
One can find both states in Table~\ref{Table III.}, the first $u_{R}$ as $u_{R8}$ in the eighth line
of this table, the second one as $u_{R7}$ in the seventh line of this table.

Below some useful relations follow.
From Eq.(\ref{snmb:gammatildegamma}) one has
\begin{eqnarray}
\label{stildestrans}
S^{ac}\stackrel{ab}{(k)}\stackrel{cd}{(k)} &=& -\frac{i}{2} \eta^{aa} \eta^{cc}
\stackrel{ab}{[-k]}\stackrel{cd}{[-k]}\,,\,\quad\quad
\tilde{S}^{ac}\stackrel{ab}{(k)}\stackrel{cd}{(k)} = \frac{i}{2} \eta^{aa} \eta^{cc}
\stackrel{ab}{[k]}\stackrel{cd}{[k]}\,,\,\nonumber\\
S^{ac}\stackrel{ab}{[k]}\stackrel{cd}{[k]} &=& \frac{i}{2}
\stackrel{ab}{(-k)}\stackrel{cd}{(-k)}\,,\,\quad\quad
\tilde{S}^{ac}\stackrel{ab}{[k]}\stackrel{cd}{[k]} = -\frac{i}{2}
\stackrel{ab}{(k)}\stackrel{cd}{(k)}\,,\,\nonumber\\
S^{ac}\stackrel{ab}{(k)}\stackrel{cd}{[k]}  &=& -\frac{i}{2} \eta^{aa}
\stackrel{ab}{[-k]}\stackrel{cd}{(-k)}\,,\,\quad\quad
\tilde{S}^{ac}\stackrel{ab}{(k)}\stackrel{cd}{[k]} = -\frac{i}{2} \eta^{aa}
\stackrel{ab}{[k]}\stackrel{cd}{(k)}\,,\,\nonumber\\
S^{ac}\stackrel{ab}{[k]}\stackrel{cd}{(k)} &=& \frac{i}{2} \eta^{cc}
\stackrel{ab}{(-k)}\stackrel{cd}{[-k]}\,,\,\quad\quad
\tilde{S}^{ac}\stackrel{ab}{[k]}\stackrel{cd}{(k)} = \frac{i}{2} \eta^{cc}
\stackrel{ab}{(k)}\stackrel{cd}{[k]}\,.
\end{eqnarray}
We conclude from the above equation that $\tilde{S}^{ab}$ generate the
equivalent representations with respect to $S^{ab}$ and opposite.

We recognize in Eq.~(\ref{graphbinoms})
the demonstration of the nilpotent and the projector character of the Clifford algebra objects
$\stackrel{ab}{(k)}$ and $\stackrel{ab}{[k]}$, respectively.
\begin{eqnarray}
\stackrel{ab}{(k)}\stackrel{ab}{(k)}& =& 0\,, \quad \quad \stackrel{ab}{(k)}\stackrel{ab}{(-k)}
= \eta^{aa}  \stackrel{ab}{[k]}\,, \quad \stackrel{ab}{(-k)}\stackrel{ab}{(k)}=
\eta^{aa}   \stackrel{ab}{[-k]}\,,\quad
\stackrel{ab}{(-k)} \stackrel{ab}{(-k)} = 0\,, \nonumber\\
\stackrel{ab}{[k]}\stackrel{ab}{[k]}& =& \stackrel{ab}{[k]}\,, \quad \quad
\stackrel{ab}{[k]}\stackrel{ab}{[-k]}= 0\,, \;\;\quad \quad  \quad \stackrel{ab}{[-k]}\stackrel{ab}{[k]}=0\,,
 \;\;\quad \quad \quad \quad \stackrel{ab}{[-k]}\stackrel{ab}{[-k]} = \stackrel{ab}{[-k]}\,,
 \nonumber\\
\stackrel{ab}{(k)}\stackrel{ab}{[k]}& =& 0\,,\quad \quad \quad \stackrel{ab}{[k]}\stackrel{ab}{(k)}
=  \stackrel{ab}{(k)}\,, \quad \quad \quad \stackrel{ab}{(-k)}\stackrel{ab}{[k]}=
 \stackrel{ab}{(-k)}\,,\quad \quad \quad
\stackrel{ab}{(-k)}\stackrel{ab}{[-k]} = 0\,,
\nonumber\\
\stackrel{ab}{(k)}\stackrel{ab}{[-k]}& =&  \stackrel{ab}{(k)}\,,
\quad \quad \stackrel{ab}{[k]}\stackrel{ab}{(-k)} =0,  \quad \quad
\quad \stackrel{ab}{[-k]}\stackrel{ab}{(k)}= 0\,, \quad \quad \quad \quad
\stackrel{ab}{[-k]}\stackrel{ab}{(-k)} = \stackrel{ab}{(-k)}.
\label{graphbinoms}
\end{eqnarray}

Defining
\begin{eqnarray}
\stackrel{ab}{\tilde{(\pm i)}} =
\frac{1}{2} \, (\tilde{\gamma}^a \mp \tilde{\gamma}^b)\,, \quad
\stackrel{ab}{\tilde{(\pm 1)}} =
\frac{1}{2} \, (\tilde{\gamma}^a \pm i\tilde{\gamma}^b)\,,
\stackrel{ab}{\tilde{[\pm i]}} = \frac{1}{2} (1 \pm \tilde{\gamma}^a \tilde{\gamma}^b), \quad
\stackrel{ab}{\tilde{[\pm 1]}} = \frac{1}{2} (1 \pm i \tilde{\gamma}^a \tilde{\gamma}^b). \nonumber
\label{deftildefun}
\end{eqnarray}
one recognizes that
\begin{eqnarray}
\stackrel{ab}{\tilde{( k)}} \, \stackrel{ab}{(k)}& =& 0\,,
\quad \;
\stackrel{ab}{\tilde{(-k)}} \, \stackrel{ab}{(k)} = -i \eta^{aa}\,  \stackrel{ab}{[k]}\,,
\quad\;
\stackrel{ab}{\tilde{( k)}} \, \stackrel{ab}{[k]} = i\, \stackrel{ab}{(k)}\,,
\quad\;
\stackrel{ab}{\tilde{( k)}}\, \stackrel{ab}{[-k]} = 0\,.
\label{graphbinomsfamilies}
\end{eqnarray}
%

Below some more useful relations~\cite{pikanorma} are presented:
\begin{eqnarray}
\label{plusminus}
N^{\pm}_{+}         &=& N^{1}_{+} \pm i \,N^{2}_{+} =
 - \stackrel{03}{(\mp i)} \stackrel{12}{(\pm )}\,, \quad N^{\pm}_{-}= N^{1}_{-} \pm i\,N^{2}_{-} =
  \stackrel{03}{(\pm i)} \stackrel{12}{(\pm )}\,,\nonumber\\
\tilde{N}^{\pm}_{+} &=& - \stackrel{03}{\tilde{(\mp i)}} \stackrel{12}{\tilde{(\pm )}}\,, \quad
\tilde{N}^{\pm}_{-}= 
  \stackrel{03} {\tilde{(\pm i)}} \stackrel{12} {\tilde{(\pm )}}\,,\nonumber\\
\tau^{1\pm}         &=& (\mp)\, \stackrel{56}{(\pm )} \stackrel{78}{(\mp )} \,, \quad
\tau^{2\mp}=            (\mp)\, \stackrel{56}{(\mp )} \stackrel{78}{(\mp )} \,,\nonumber\\
\tilde{\tau}^{1\pm} &=& (\mp)\, \stackrel{56}{\tilde{(\pm )}} \stackrel{78}{\tilde{(\mp )}}\,,\quad
\tilde{\tau}^{2\mp}= (\mp)\, \stackrel{56}{\tilde{(\mp )}} \stackrel{78}{\tilde{(\mp )}}\,.
\end{eqnarray}
%

%
In Table~\ref{Table III.}~\cite{JMP2015} the eight families of the first member in
Table~\ref{Table so13+1.} (member number  $1$) of the eight-plet of quarks and the $25^{th}$
member  in Table~\ref{Table so13+1.} of the eight-plet of leptons are presented as an example.
The eight families of the right handed $u_{1R}$ quark  are
presented in the left column of  Table~\ref{Table III.}~\cite{JMP2015}. In the right column of the
same table the equivalent eight-plet of the right handed neutrinos $\nu_{1R}$ are presented.
All the other members of any of the eight families of quarks or leptons follow  from any member
of a particular family by the application of the  operators $N^{\pm}_{R,L}$ and
$\tau^{(2,1)\pm}$, Eq.~(\ref{plusminus}), on this particular member.


The eight-plets separate into two group of four families: One group  contains  doublets with respect
to $\vec{\tilde{N}}_{R}$ and  $\vec{\tilde{\tau}}^{2}$, these families are singlets with respect to
$\vec{\tilde{N}}_{L}$ and  $\vec{\tilde{\tau}}^{1}$. Another group of families contains  doublets
with respect to  $\vec{\tilde{N}}_{L}$ and  $\vec{\tilde{\tau}}^{1}$, these families are singlets
with respect to  $\vec{\tilde{N}}_{R}$ and  $\vec{\tilde{\tau}}^{2}$.

The scalar fields which are the gauge scalars  of  $\vec{\tilde{N}}_{R}$ and  $\vec{\tilde{\tau}}^{2}$
couple only to the four families  which are doublets with respect to these two groups.
The scalar fields which are the gauge scalars  of  $\vec{\tilde{N}}_{L}$ and  $\vec{\tilde{\tau}}^{1}$
couple only to the four families  which are doublets with respect to these last two groups.

After the electroweak phase transition, caused by the scalar fields with the space index $(7,8)$,
the two groups of four families become massive. The lowest of the two groups of four families
contains the observed three, while the fourth remains to be measured. The lowest of the upper
four families is the candidate  for the dark matter~\cite{IARD}.


\begin{thebibliography}{99}
\bibitem{NPLB} N.S. Manko\v c Bor\v stnik, "Do we have the explanation for
the Higgs and Yukawa couplings of the {\em standard model}", [arxiv:1212.3184,
arxiv:1011.5765].
\end{thebibliography}
\end{document}